\documentclass{PoS}

 \usepackage{amsmath}
 \usepackage{amsfonts}
 \usepackage{amssymb}

 \allowdisplaybreaks[4]

 \renewcommand{\normalfont}{\sffamily}

 \newcommand{\bp}[1]{\noindent{\bf #1}~~~}
 \newcommand{\nn}{\nonumber}
 \newcommand{\er}[1]{(\ref{#1})}
 
 \renewcommand{\d}{{\rm d}}     % Normally  \d{o}  gives an underdot.
 
 \newcommand{\td}[2]{\frac{{\rm d} {#1}}{{\rm d} {#2}}}
 \newcommand{\tdil}[2]{{\rm d} {#1} / {\rm d} {#2}}
 \newcommand{\tdt}[2]{\frac{{\rm d}^2 {#1}}{{\rm d} {#2}^2}}
 \newcommand{\tdtil}[2]{{\rm d}^2 {#1} / {\rm d} {#2}^2}
 \newcommand{\pd}[2]{\frac{\partial {#1}}{\partial {#2}}}
 \newcommand{\pdil}[2]{\partial {#1} / \partial {#2}}

 \newcommand{\Ra}{\Rightarrow}
 \newcommand{\LRa}{\Leftrightarrow}

 \newcommand{\LT}{Lema\^{\i}tre-Tolman}
 \renewcommand{\L}{Lema\^{\i}tre}

 \newcommand{\Rt}{\dot{R}}
 \newcommand{\Rtt}{\ddot{R}}
 \newcommand{\St}{\dot{S}}
 
 \newcommand{\Rh}{\hat{R}}
 \newcommand{\Rth}{\widehat{\dot{R}}}
 \newcommand{\Rrh}{\widehat{R'\,}}
 \newcommand{\Rtrh}{\widehat{\dot{R}'\,}}
 \newcommand{\Rrrh}{\widehat{R''\,}}
 \renewcommand{\th}{\hat{t}}
 \newcommand{\rh}{\hat{r}}
 
 \newcommand{\rhoh}{\hat{\rho}}
 \newcommand{\sgh}{\hat{\sigma}}

 \newcommand{\showlabel}[1]{
    \label{#1}}

\DeclareSymbolFont{operators}   {OT1}{cmr} {m}{n}  % These lines restore \epsilon to not look like \varepsilon
\DeclareSymbolFont{letters}     {OML}{cmm} {m}{it} % and \cal letters to not be twirly curly letters.
\DeclareSymbolFont{symbols}     {OMS}{cmsy}{m}{n}  % Provided by the PoS people when I enquired.
\DeclareSymbolFont{largesymbols}{OMX}{cmex}{m}{n} 

 \title{Modelling Inhomogeneity in the Universe}

 \ShortTitle{Modelling Inhomogeneity in the Universe}

 \author{Charles Hellaby%\thanks{A footnote may follow.}
         \\
         Department of Mathematics and Applied Mathematics, 
         University of Cape Town, Rondebosch 7701, South Africa\\
         E-mail: \email{Charles.Hellaby@uct.ac.za}}

 \abstract{
An overview of some recent developments in inhomogeneous models is presented.  

As the volume and precision of cosmological data improves, it will become more and more essential to understand the non-linear behaviour of the Einstein field equations.  This requires the study of exact inhomogeneous solutions, including their density distributions, their evolution, their geometry, and their causal structure.  Observations are strongly affected by the detailed geometry and evolution of a model, and therefore interpretation of observations depends on understanding them.

It is generally assumed the universe is homogeneous if averaged over large enough scales, but to actually prove this is so, will require the assumption to be relaxed, and a rigorous inhomogeneous approach to be applied.

Though the \LT\ metric has long been used for models of spherical inhomogeneities, there have been a number of new results, including a variety of methods for creating models with specific properties, and their application to cosmic structures on several different scales.

Interest in the Szekeres metrics is on the increase, and the quasi-spherical metric was recently used to model specific cosmic structures for the first time.  The quasi-planar and quasi-hyperspherical metrics have been hardly studied until recent work invesigated their physical and geometric properties.  There is enormous scope for work with these metrics.
 }

 \FullConference{5th International School on Field Theory and Gravitation,\\
		 April 20-24, 2009\\
		 Cuiab\'{a} city, Brazil}

 \tableofcontents

\begin{document}

 \section{Introduction}

   Why study inhomogeneous models?  The real universe is very lumpy.  To properly understand what we see, we should apply all possible methods: perturbation theory, $N$-body newtonian simulations, and exact inhomogeneous metrics --- each has its domain of validity.  Inhomogeneous metrics have the advantage that they are fully non-linear and relativistic solutions of the Einstein field equations (EFEs).

   The assumption of homogeneity has become so well established, that it has become all-pervasive.  But now, with so much data coming in, it's time to test homogeneity.  Cosmological data reduction relies heavily on the Robertson-Walker (RW) metric --- we need to beware of a circular argument.  It will be a significant challenge to check which of our well-known results actually depend on the assumption of homogeneity, and to re-derive them all without that assumption.
   
    Here I will present a selection of results in inhomogeneous cosmology, especially work done with Lu, McClure, Krasi\'{n}ski, Bolejko, C\'{e}l\'{e}rier, Alfedeel, Mustapha, Ellis and others, but I won't try to be comprehensive.  I'll attempt to provide the basics, and thereby promote the use of inhomogeneous metrics for the study of cosmological problems.

   Inhomogeneous metrics will become more important as the amount and accuracy of cosmological data increases, and more precise analysis is needed, so there are plenty of opportunities for good research.

 \section{The \LT\ Metric}

 The \LT\ (LT) metric was the first inhomogeneous non-vacuum metric to be discovered, and has probably been the most popular choice for modelling cosmic inhomogeneity ever since, certainly in recent decades.  It is a spherically-symmetric, inhomogeneous dust model, discovered by \L, rediscovered by Tolman, and studied by Bondi \cite{Lem33,Tol34,Bondi47}.  The metric is
 \begin{align}
   \d s^2 = - \d t^2 + \frac{(R')^2}{1 + f} \, \d r^2 + R^2 \d \Omega^2 ~,
   \showlabel{ds2LT}
 \end{align}
 where $\d\Omega^2 = \d\theta^2 + \sin^2\theta \, \d\phi^2$, $R(t,r)$ is the areal radius, and $R' = \pdil{R}{r}$.  The free function $f(r)$ determines the local geometry; it gives the ``embedding angle'' of constant $t, \theta$ surfaces in 3-d flat space \cite{Hell87}.  Also the Ricci scalar of the spatial 3-surfaces,
 \begin{align}
   {}^3\!{\cal R} = \frac{-2(R f' + f R')}{R^2 R'} ~,
 \end{align}
 is only zero (for all $r$) if both $f$ and $f'$ are zero.  The matter is a pressure-free perfect fluid, 
 \begin{align}
   T^{ab} = \rho u^a u^b ~,
 \end{align}
 that is comoving
 \begin{align}
   u^a = \delta^a_t ~.
 \end{align}
 From the EFEs we get
 \begin{align}
   \Rt^2 = \frac{2 M}{R} + f + \frac{\Lambda R^2}{3} ~, 
   \showlabel{RtSq}
 \end{align}
 where $\Rt = \pdil{R}{t}$, and
 \begin{align}
   \kappa \rho = \frac{2 M'}{R^2 R'} ~,
   \showlabel{rhoLT}
 \end{align}
 where $M(r)$ is a second free function that gives the gravitational mass within a comoving shell of radius $r$.  Here $f(r)$ also plays the role of twice the local energy per unit mass of the dust particles, so it's often written $f(r) = 2 E(r)$.  It follows from \er{RtSq} that
 \begin{align}
   \Rtt & = - \frac{M}{R^2} + \frac{\Lambda R}{3}
      \showlabel{Rtt} ~, \\
   \Rt \Rt' & = \frac{M'}{R} + W W' + \left( \frac{\Lambda R}{3} - \frac{M}{R^2} \right) R' ~.
      \showlabel{RtRtr}
 \end{align}

   When $\Lambda = 0$, the solutions of \er{RtSq}, in terms of parameter $\eta$, are
 \begin{align}
   & \Lambda = 0,~ f > 0:~~~~ & R & = \frac{M}{f} \, (\cosh \eta - 1) ~,~~~~ &
      (\sinh \eta - \eta) & = \frac{f^{3/2} (t - a)}{M} ~;
      \showlabel{HypEv} \\
   & \Lambda = 0,~ f = 0:~~~~ & R & = M \left( \frac{\eta^2}{2} \right) ~,~~~~ &
      \left( \frac{\eta^3}{6} \right) & = \frac{(t - a)}{M} ~;
      \showlabel{ParEv} \\
   & \Lambda = 0,~ f < 0:~~~~ & R & = \frac{M}{(-f)} \, (1 - \cos \eta) ~,~~~~ &
      (\eta - \sin \eta) & = \frac{(-f)^{3/2} (t - a)}{M} ~;
      \showlabel{EllEv}
 \end{align}
 for hyperbolic, parabolic, and elliptic evolution respectively.  (Near the origin, where $f \to 0$, the type of evolution is determined by the sign of $Rf/M$ or $f/M^{2/3}$.)  When $\Lambda \neq 0$ there is a very complicated solution in terms of elliptic integrals.  These solutions contain a third free function $a(r)$, which is the local time of the big bang, the time when $R = 0$ on each worldline.  In other words, the constant $r$ worldlines all emerge from the bang at different times, usually the outer spheres first and the origin last, as illustrated in the sketches below.  
 \\
 \includegraphics[scale=0.49]{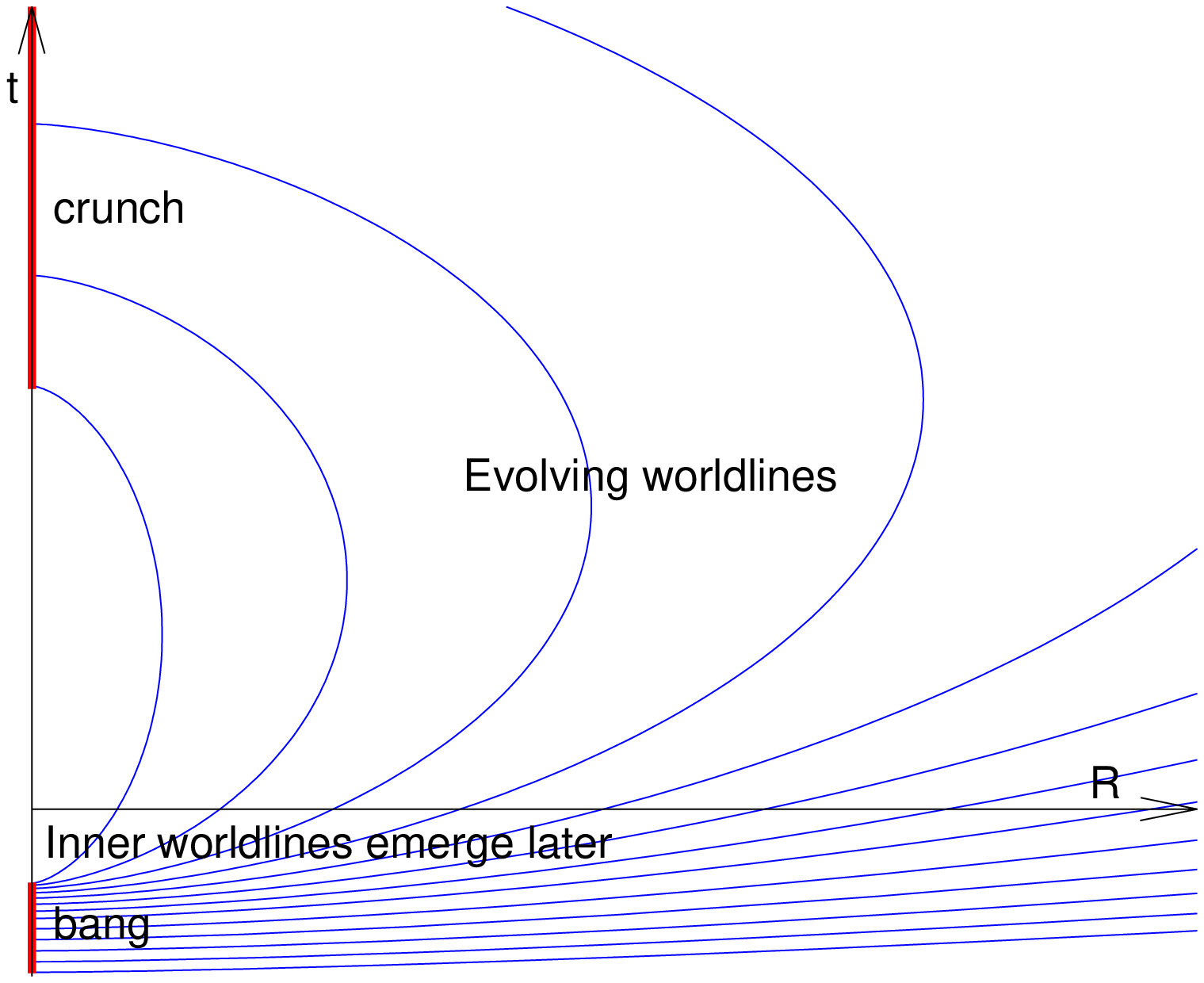}
 \includegraphics[scale=0.49]{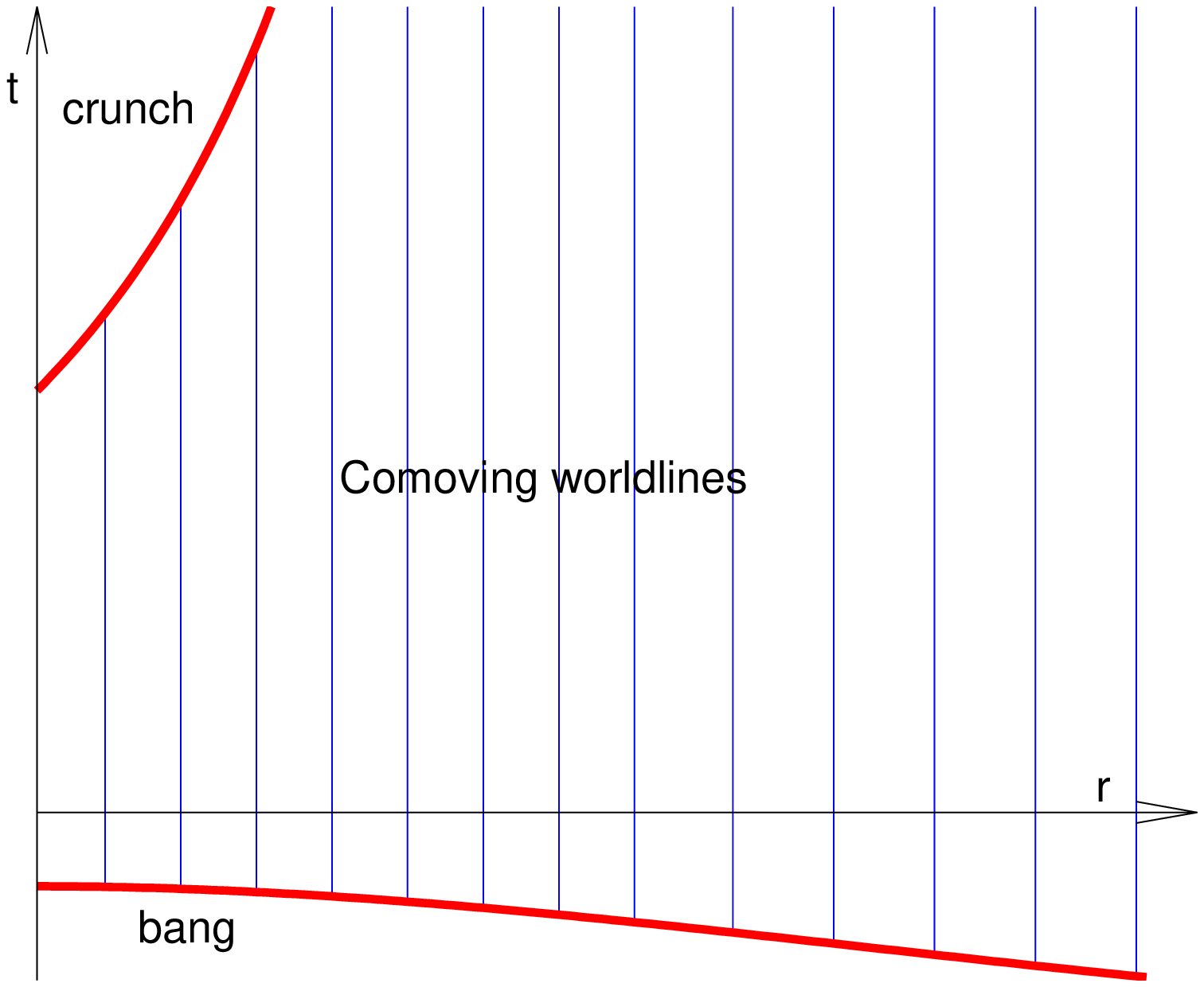}
 \\
 The above evolutions equation and solutions may also be written
 \begin{align}
   && \dot{\beta} & = \frac{2}{\alpha} + x + \frac{\Lambda \alpha^2}{3} ~, \\
   & \Lambda = 0,~ f > 0:~~ & t & = a + x^{-3/2} \left\{ \sqrt{(1 + x \alpha)^2 - 1}\;
      - {\rm arcosh} (1 + x \alpha) \right\} ~,
      \showlabel{tEvHX} \\
   & \Lambda = 0,~ f = 0:~~ & t & = a + \frac{\sqrt{2 \alpha^3}}{3} ~,
 % R & = \left( \frac{9 M (t - a)^2}{2} \right)^{1/3} ~,
      \showlabel{REvPX} \\
   & \Lambda = 0,~ f < 0,~ 0 \leq \eta \leq \pi:~~& t & = a + x^{-3/2} \left\{ \arccos(1 + x \alpha)
      - \sqrt{1 - (1 + x \alpha)^2}\; \right\} ~,
      \showlabel{tEvEX} \\
   & \Lambda = 0,~ f < 0,~ \pi \leq \eta \leq 2 \pi:~~& t & = a + x^{-3/2} \left\{ 2 \pi - \arccos (1 + x \alpha)
      + \sqrt{1 - (1 + x \alpha)^2}\; \right\} ~.
      \showlabel{tEvEC} \\
   & \mbox{where}~~~~~~ && \alpha = \frac{R}{M^{1/3}} ~,~~~~~~
      x = \frac{|f|}{M^{2/3}} ~,~~~~~~
      \beta = \frac{\Rt}{M^{1/3}} ~.
      \showlabel{alpha-x-beta-Def}
 \end{align}
 Naturally, the time reverses of these models, obtained by changing $(t - a)$ to $(a - t)$ and $a'$ to $-a'$, are also solutions.  It is quite possible to have adjacent elliptic and hyperbolic regions in one model --- for example, a re-collapsing dust cloud could be surrounded by an ever-expanding universe.  The two regions would have a parabolic shell at the boundary between them, but extended parabolic regions are also possible.  In practice, \er{HypEv}, \er{EllEv}, \er{tEvHX}, \er{tEvEX}, \er{tEvEC} are not good for calculating the evolution of worldlines that are close to parabolic, so a series expansion is used instead.  Similarly, near the bang or crunch, where the evolution is close to parabolic, one obtains better accuracy by using the same series expansion.

   It is sometimes useful to have an expression for $R'$.  When $\Lambda = 0$ it follows from \er{HypEv}-\er{EllEv}, that for all $f$ values one can write \cite{HelLak84}
 \begin{align}
    R' = \left( \frac{M'}{M} - \frac{f'}{f} \right) R - \left[ a' +
   \left( \frac{M'}{M} - \frac{3 f'}{2 f} \right) (t - a) \right] \dot{R} ~.
   \showlabel{R'gen}
 \end{align}
 Alternatively, one can write the parametric expressions
 \begin{align}
   f < 0:~~~~~~~~ \frac{R'}{R} & = \frac{M'}{M} \left( 1 - \phi_1 \right)
      + \frac{f'}{f} \left( \frac{3}{2} \phi_1 - 1 \right)
      - \frac{(-f)^{3/2} a'}{M} \phi_2 ~,
      \showlabel{R'Ell} \\
   & \phi_1(\eta) = \frac{\sin \eta (\eta - \sin \eta)}{(1 - \cos \eta)^2} ~,~~~~~~
      \phi_2(\eta) = \frac{\sin}{(1 - \cos \eta)^2} ~; \\
   f > 0:~~~~~~~~ \frac{R'}{R} & = \frac{M'}{M} \left( 1 - \phi_4 \right) + \frac{f'}{f} 
      \left( \frac{3}{2} \phi_4 - 1 \right) - \frac{f^{3/2} a'}{M} \phi_5 ~,
      \showlabel{R'Hyp} \\
   & \phi_4 = \frac{\sinh \eta (\sinh \eta - \eta)}{(\cosh \eta - 1)^2} ~,~~~~~~
      \phi_5 = \frac{\sinh \eta}{(\cosh \eta - 1)^2} ~.
 \end{align}

   A scale length and time may be defined by
 \begin{align}
   \tilde{R}(r) = \frac{M}{|f|} ~,~~~~~~ \tilde{T}(r) = \frac{M}{|f|^{3/2}} ~,
 \end{align}
 and for elliptic worldlines the maximum $R$ is $2\tilde{R}$, while the lifetime from bang to crunch is $2 \pi \tilde{T}$.

   By specifying $\Lambda$ and the three free functions --- $M(r)$, $f(r)$, and $a(r)$ --- an LT model is fully determined.  Between them they provide a radial co-ordinate freedom and two physical relationships, e.g. $M = M(r)$, $f = f(M)$ and $a = a(M)$, though it is normal to give all of them in terms of $r$.  It is not possible to give any kind of standard form for one of these functions that will cover all possibilities.  For example, the choice $M \propto r^3$ is common, but does not allow regions of vacuum where $M' = 0$; a standard choice for $f(r)$ cannot include both models in which $f$ changes sign, and those in which it doesn't; and similarly no choice of $a(r)$ can cover cases where $a$ is constant in some places and cases where it never is.

   See \cite{Kra97} for a survey of work done on inhomogeneous models up to 1997, \cite{PleKra06} for an introduction to some inhomogeneous models, and \cite{BoKrHeCe} for a summary of some recent developments.  A dynamical systems analysis is given in \cite{Suss08}.

 \subsection{Singularities}
 \showlabel{LTSings}

   Singularities occur where the density \er{rhoLT} or the curvature diverge.  The Kretschmann scalar is
 \begin{align}
   {\cal K} = R_{abcd} R^{abcd} = 
   \frac{48 M^2}{R^6} +  
   \frac{32 M M'}{R^5 R'} +  
   \frac{12 (M')^2}{R^4 (R')^2} ~.
 \end{align}

   \bp{Big Bang}
   At the big bang or the big crunch, we have $R = 0$, which occurs where $t = a$ or where $t = a + 2 \pi \tilde{T}$.  Here $R'$ diverges unless $a' = 0$.  The bang and crunch surfaces are spacelike \cite{Hell85,HelLak88}, except possibly at the origin.

   \bp{Shell Crossings}
   Shell crossings are timelike surfaces that occur where an inner spherical shell of matter collides with an adjacent outer shell, so that $R' = 0$.  These surfaces are timelike \cite{Hell85,HelLak85}, and have a different redshift structure from the bang \cite{HelLak84}.  Since the $r$ coordinate is comoving, it becomes degenerate at such loci.  Physically one might argue that non-zero pressure would develop before a shell crossing occurs, but for a ``fluid'' of many stars or galaxies that doesn't apply.  Clearly shell crossings represent a breakdown of the LT assumptions and for many purposes they are undesireable.  Shell crossings can be eliminated from an entire model, in the $\Lambda = 0$ case, by applying the conditions found in \cite{Hell85,HelLak85} to the 3 arbitrary functions.  These conditions were derived by writing $R'$ in terms of the parameter $\eta$ and looking at the early and late time behaviours.  They are important if you want your model to be everywhere well behaved.%
 \footnote{Shell crossings have been extensively investigated, e.g.\ \cite{Papa67,YoSeMu73,Nola03,BolLas08}}
 \\
 ${}$ \hfill
 \includegraphics[scale=0.33]{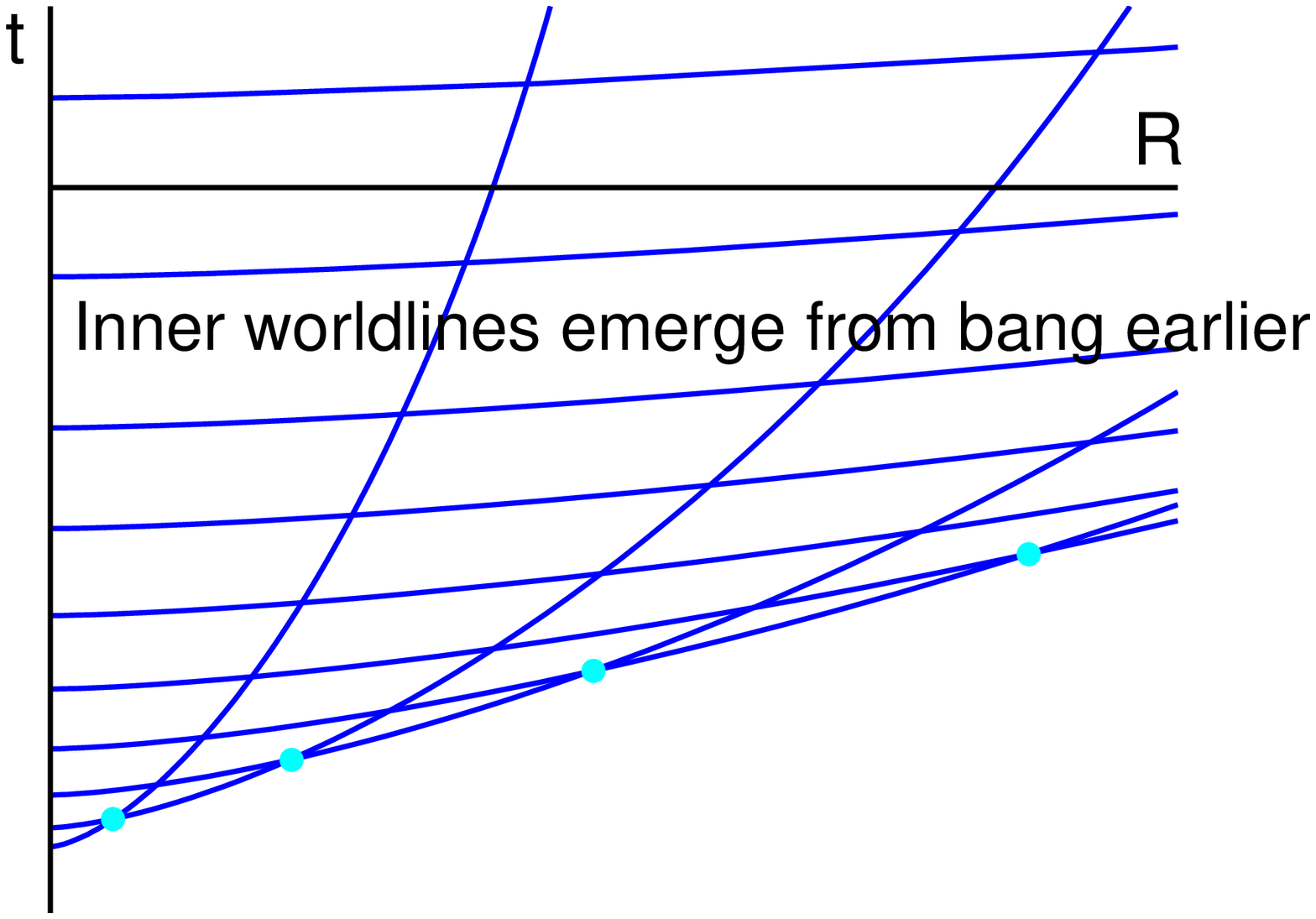}
 \hfill
 \includegraphics[scale=0.33]{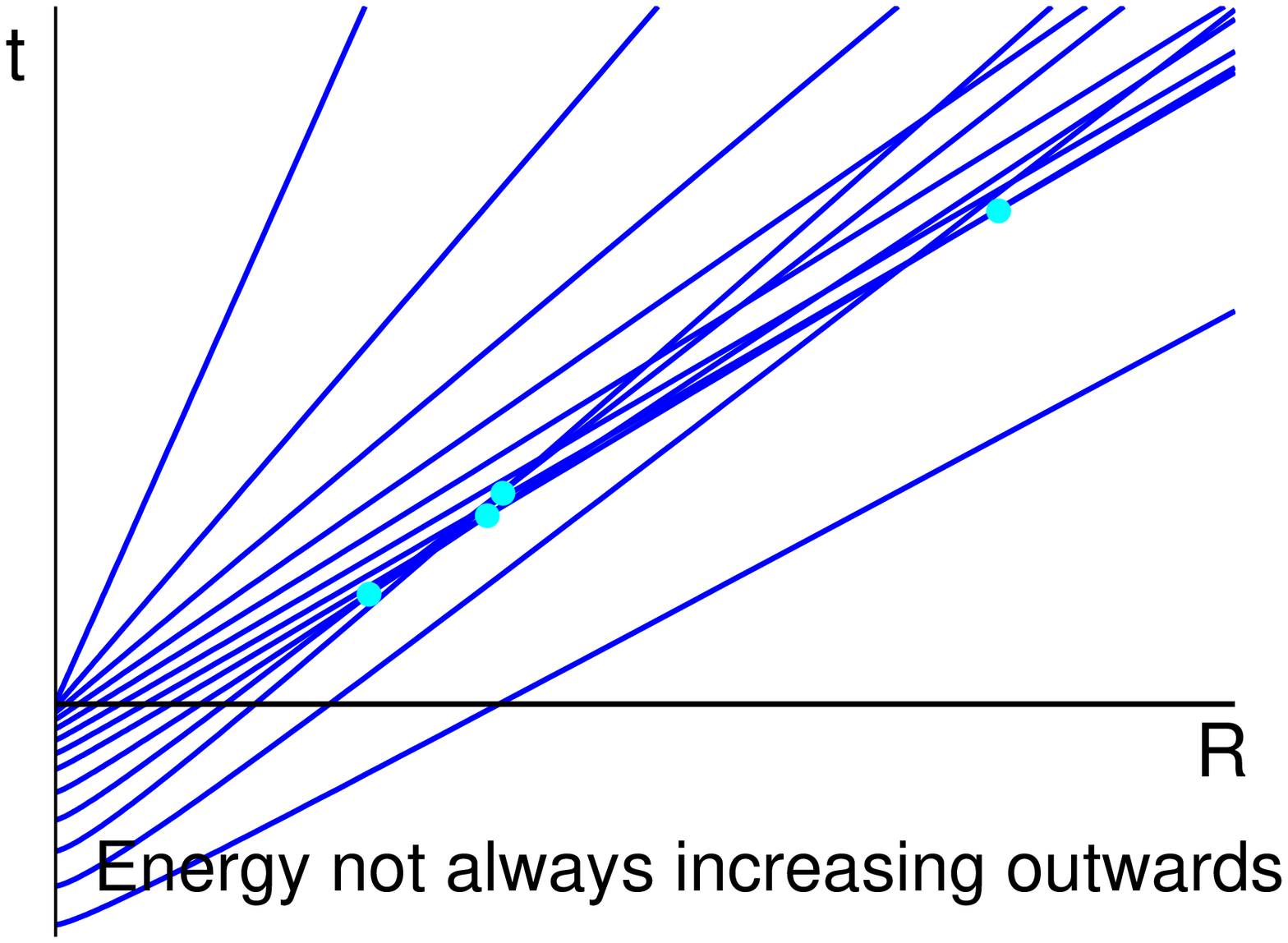}
 \hfill ${}$

   However, both $R = 0$ and $R' = 0$ can occur at non-singular locations as explained below.

   \bp{`Shell Focussing'}
   There are also ``shell focussing'' singularities, e.g. \cite{EarSma79,Chri84,Newm86,OriPir87,RajLak87,HelLak88,WauLak89,LakZan90,Gri91,Lemo91,Lak92,Josh93}.  For certain LT models, the first event of the big crunch to form, where the central worldline reaches the crunch surface, can emit many light rays, some of which may even reach infinity.  (So they might be better called ``light focussing'' singularities.)  The nature of the singularity is difficult to understand, and seems to depend on the path of approach to the singular point.

 \subsection{Regularity Conditions}
 \showlabel{LTRegCond}

   \bp{Regular signature}
   For the metric \er{ds2LT} to retain a Lorentzian signature, 
 \begin{align}
   f \geq -1
 \end{align}
 is required, the equality only occuring where $R' = 0$ --- see below.

   \bp{Regular Origins}
   An  origin of spherical coordinates is a locus $r_o$ where
 \begin{align}
   R(t, r_o) = 0 ~~~~~~\forall~~ t ~, 
   \showlabel{OrigDef}
 \end{align}
 so that $\Rt(t, r_o) = 0$, $\Rtt(t, r_o) = 0$, etc.  Obviously one usually wants an origin to be a normal timelike worldline.  The conditions for a regular origin are obtained by requiring that, in the limit as the origin is approached, the density and the curvature should not diverge, and the time evolution at the origin should be a smooth continuation of it's immediate neighbourhood.  See for example \cite{HuMaMa98,MusHel01}.  It is found that, away from the bang or crunch, on a constant $t$ slice,
 \begin{align}
   M \sim R^3 ~,~~~~~~ f \sim R^2 ~.
   \showlabel{OrigMf}
 \end{align}
 This may be realised by setting $f \propto M^{2/3}$, e.g. $M \sim r^3$, $f \sim r^2$, at the origin.  Variables $\alpha$, $x$ \& $\beta$ of \er{alpha-x-beta-Def} have the advantage that they are non-zero at the origin.  If in addition one wants the density to be smooth through the origin, i.e. to have zero gradient there, then there are further conditions \cite{MusHel01}, most notably
 \begin{align}
   a' \to 0 ~.
 \end{align}
 However, there is no singularity if this last one does not hold.  Thus, the locus $R = 0$ includes both the spacelike bang and crunch surfaces, and the timelike origins.  

 \noindent
 \hspace*{-10mm}
 \includegraphics[scale=0.55,angle=140]{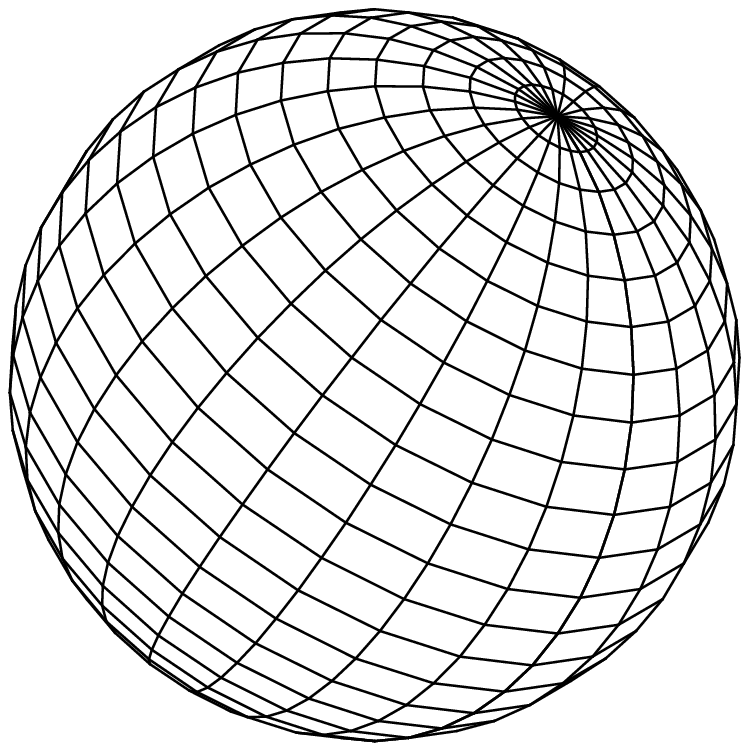}
 \hspace*{-9mm}
 \includegraphics[scale=0.55,angle=140]{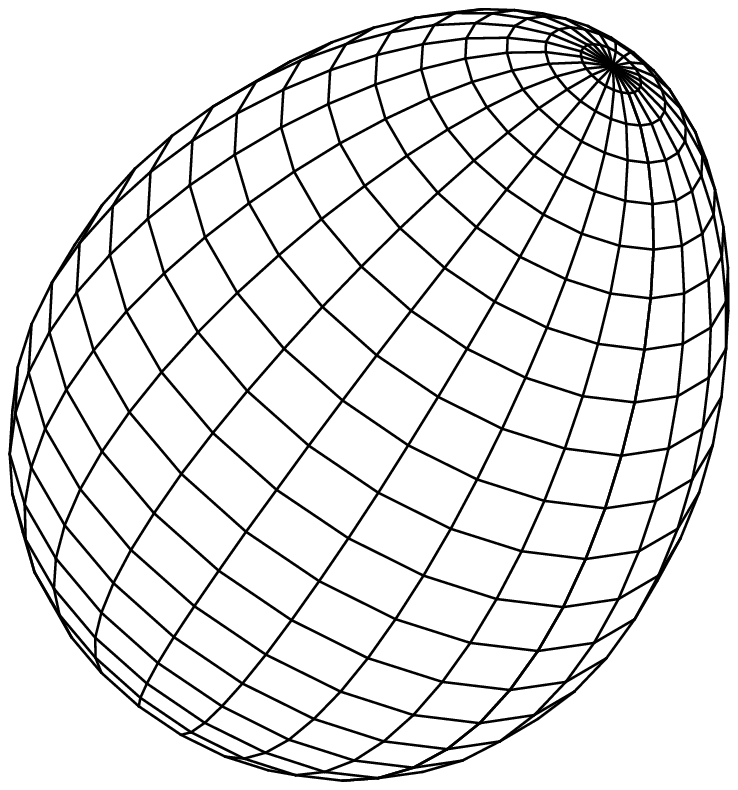}
 \hspace*{-10mm}
 \includegraphics[scale=0.55,angle=140]{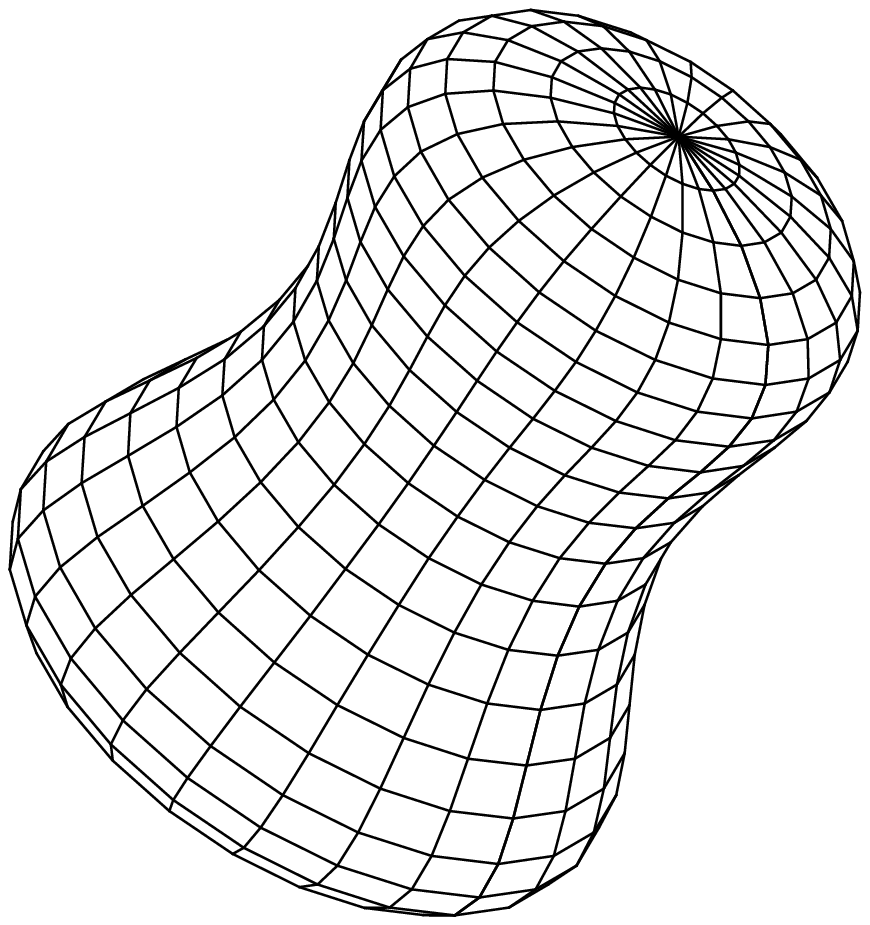} \\
   \bp{Regular Spatial Extrema}
   Similarly, $R' = 0$ includes regular loci as well as singular shell crossings.  As pointed out in \cite{ZelGri84}, any spherically symmetric model with closed sptial ($t =$~const.) sections, such as the $k = +1$ Freidmann-\L-Robertson-Walker (FLRW) model, has an origin, a maximum radius, and a second origin --- a north pole, an equator and a south pole.  At a spatial maximum we obviously have $R' = 0$, but we expect the density and curvature to be regular at such a locus.  This is only possible on a comoving shell, i.e.
 \begin{align}
   R'(t, r_m) = 0 ~~~~~~\forall~~ t ~.
 \end{align}
 The conditions for a regular maximum \cite{HelLak85,Bonn85} are that there is no shell crossing and no surface layer, i.e.
 \begin{align}
   M'(r_m) = 0 = f'(r_m) = a'(r_m) ~,~~ f(r_m) = -1 ~.
 \end{align}
 Therefore the LT model may have a number of interesting spatial topologies \cite{HelLak85,Hell87,KraHel04b}, such as a black hole in a cosmological background, or a sequence of maxima and minima --- ``bellies'' and ``necks''.  It is also possible to have an elliptic (recollapsing) model that is open.

 \subsection{Special Cases}

    \bp{Dust Robertson-Walker}
   The LT metric contains the dust RW metric as the special case
 \begin{align}
   f \propto M^{2/3} ~,~~~~~~ a' = 0 ~.   \showlabel{LTcondRW}
 \end{align}
 Putting this in \er{RtSq} and \er{rhoLT} makes $\Rt/M^{1/3}$ and $\rho$ independent of $r$.  In standard RW coordinates, $M = (\kappa \rho_0 S_0^3/6) r^3$, $f = - k r^2$, $a = 0$, $R = r S(t)$, so it is evident that $S(t)$ is the scale factor, and
 \begin{align}
   f_2 = -k ~,~~~~~~ M_3 = \frac{\kappa \rho_0 S_0^3}{6} ~.
   \showlabel{fM-RW}
 \end{align}
 Consequently one may write the LT arbitrary functions in a form that looks like RW plus perturbation, but is exact,
 \begin{align}
   M = M_3 r^3 (1 + \tilde{M}(r))   \showlabel{M-RW+} \\
   f = f_2 r^2 (1 + \tilde{f}(r))   \showlabel{f-RW+} \\
   a = a_0 (1 + \tilde{a}(r)) ~,   \showlabel{a-RW+}
 \end{align}
 where $\tilde{M}$, $\tilde{f}$ and $\tilde{a}$ may be set to zero at the origin, say.  In terms of the RW parameters of the `unperturbed' RW model that applies at the origin, we can write
 \begin{align}
   S = \frac{R}{r} ~,~~~~
      H & = \frac{\St}{S} ~,~~~~
      \Omega_m = \frac{2 M_3}{S^3 H^2} ~,~~~~
      \Omega_k = \frac{f_2}{S^2 H^2} ~,~~~~
      \Omega_\Lambda = \frac{\Lambda}{3 H^2} ~,   \showlabel{SHOmRW} \\
   \mbox{so that} ~~~~~~ f_2 & = - k ~~~~\to~~~~
      M_3 = \frac{\Omega_{m0}}{2 H_0^2 ( - k \Omega_{k0})^{3/2}} ~,~~~~
      a_0 = 0 ~,   \showlabel{fMaRW} \\
   \mbox{and of course} & ~~~~~~ S_0 = \frac{1}{H_0 \sqrt{- k \Omega_{k0}}\;} ~,~~~~
      \Lambda = 3 \Omega_\Lambda H_0^2 ~.   \showlabel{S0LamRW}
 \end{align}

   \bp{Schwarzschild}
   The spherical vacuum metric is obtained if $M' = 0$, and the different choices of $f(r)$ and $a(r)$ cover it with different families of geodesic coordinates.  But to get the full Scwarzschild-Kruskal-Szekeres (SKS) topology requires $f = -1$ and $a' = 0 = f'$ so that $R' = 0$ at the ``throat'' or ``neck'', and that $a$ decreases and $f$ increases on either side --- see \cite{Hell87,Hell96a} for the details and some plots.

 \noindent
 \hspace*{-2mm}
 \includegraphics[scale = 0.7, angle=-90]{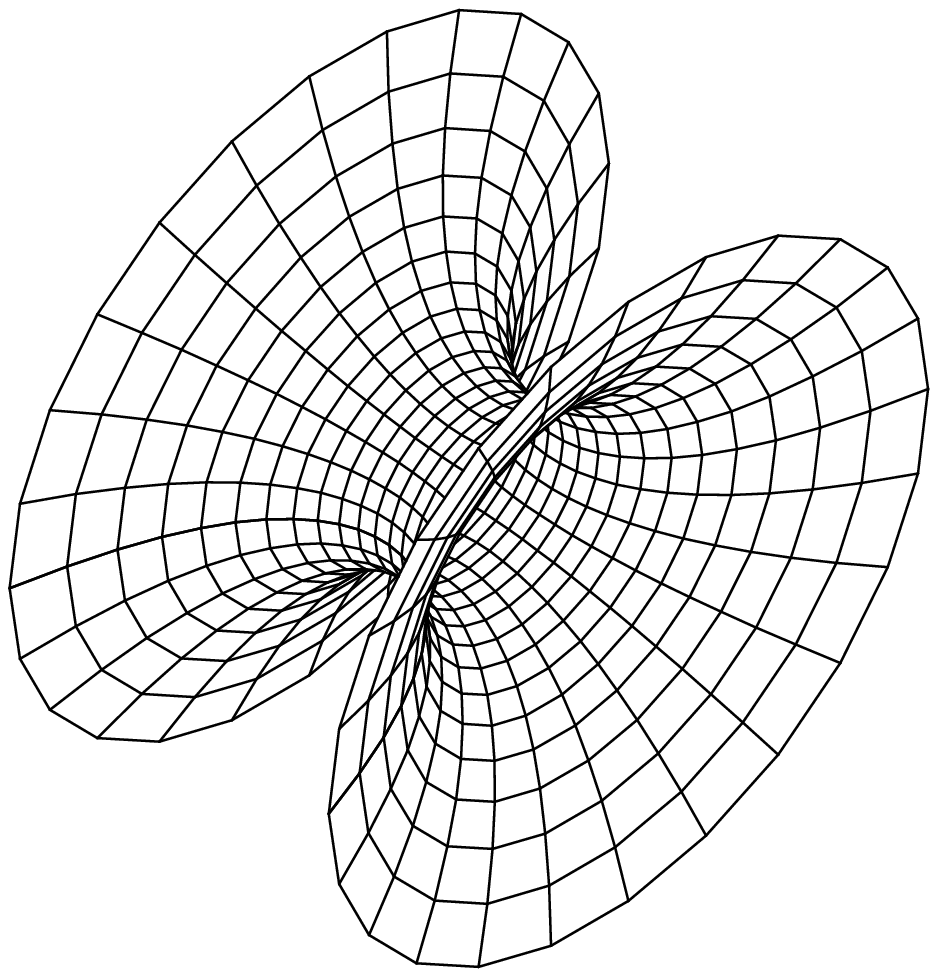}
 \hspace*{5mm}
 \includegraphics[scale = 0.4, angle=-90]{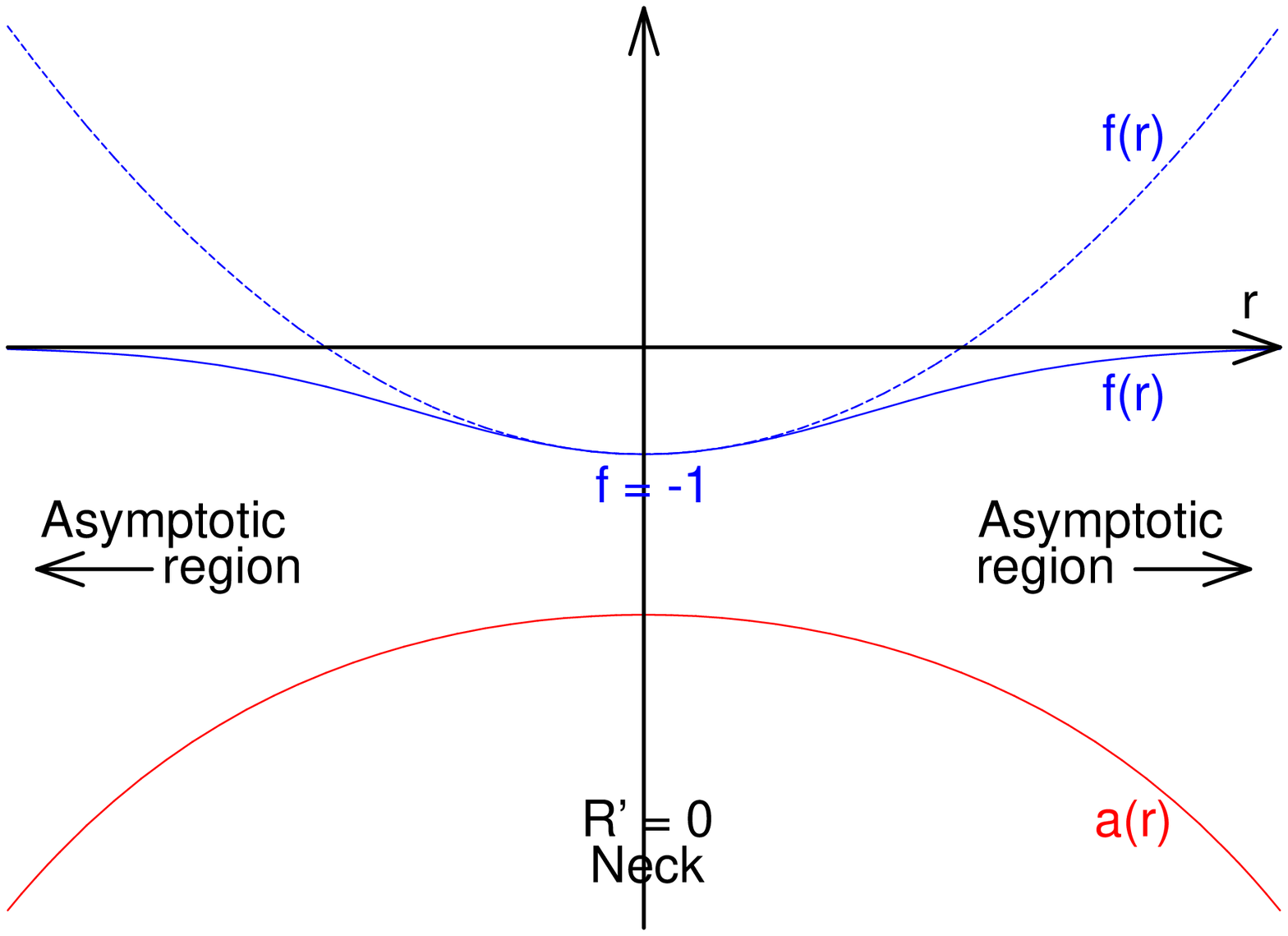}
 \\[1mm]

   \bp{Datt-Kantowski-Sachs}
   The Datt \cite{Datt38} models are inhomogeneous Kantowski-Sachs type models, and though often treated as a separate solution with $R' = 0$, they are in fact limits of LT models \cite{Hell96a}.  

   \bp{Vaidya}
   In the null limit, when $f \to \infty$ we get the Vaidya metric that represents incoherent radiation emanating from (or converging on) a spherical body \cite{Lemo92,Hell94}.  

 \subsection{Constructing Inhomogeneous Models}

   The most obvious way to construct an LT model is to choose the three arbitrary functions.  Choosing $f(r)$, for example, works quite well if one is interested in the geometry and topology of the model.  See \cite{HelLak85,Hell87}.  But for many situations, it is not always obvious what the density distribution and evolution will be, given $f$, $M$ and $a$.  In \cite{SusTru02}, for example, the use of the density $\rho_i(r)$, the 3-d Ricci scalar ${}^3\!{\cal R}_i(r)$ and the areal radius $R_i(r)$, on an initial surface at $t = t_i$ was advocated, and an appendix suggested how `lumps' and `voids' in the density and curvature could be prescribed on the initial surface.

   In place of $a(r)$, one may instead specify $R_i = R(t_i, r)$ at some initial time, set $\eta = 0$ at $t = t_i$ and re-write \er{EllEv} in the form
 \begin{align}
   R & = \frac{M}{(-f)} \, (1 - \cos\eta) + R_i \left( \cos\eta
      + \sqrt{\frac{2 M}{(-f) R_i} - 1}\; \sin\eta \right) ~, \nn \\
   t & = \frac{M}{(-f)^{3/2}} (\eta - \sin \eta) + \frac{R_i}{\sqrt{-f}\;} \left( \sin\eta
      + \sqrt{\frac{2 M}{(-f) R_i} - 1}\; (1 - \cos\eta) \right) ~;
      \showlabel{EllEvRi}
 \end{align}
 etc for the other cases, \er{ParEv} \& \er{HypEv}.

   If you prefer to think in terms of $R$, then choose $M(R_i)$, $f(R_i)$ and $a(R_i)$ on an initial surface $t = t_i$, and set $r = R_i$.  If $\rho = \rho_i(R_i)$ is given, then, again choosing $r = R_i$, 
 \begin{align}
   M(R_i) - M_0 = \int_{R_0}^{R_i} \frac{\kappa \rho_i(R) R^2}{2} \, dR ~.
 \end{align}

   If the expansion rate and radius $\Rt_i(M)$ and $R_i(M)$ are specified, then by \er{RtSq}
 \begin{align}
   f(M) & = \Rt^2_i - \frac{2 M}{R_i} - \frac{\Lambda R_i^2}{3} ~.
 \end{align}
 or if the $\Rt_i(M)$ and $f(M)$ are specified, then
 \begin{align}
   R(M) & = \frac{\Rt^2 - f + X^2}{\sqrt{\Lambda}\; X} ~,~~~~~~
      X = \left( \sqrt{9 M^2 \Lambda - (\Rt^2 - f)^3}\; - \sqrt{9 M^2 \Lambda}\; \right)^{1/3} ~, \\
   \mbox{or}~~~~~~ R(M) & = \sqrt{\frac{2 M}{\Rt^2 - f}}\; ~,~~~~~~ \mbox{when}~~ \Lambda = 0 ~.
 \end{align}

   \bp{Initial and Final Profiles}
   In \cite{KraHel02,KraHel04a,HelKra06} a number of very useful methods for $\Lambda = 0$ LT models were presented.  Since these have been summarised elsewhere \cite{BoKrHeCe} we will only outline the basic idea here.  Rather than specifying all the data on a single `initial' 3-surface, one may instead specify the density profile $\rho_1$ on one constant time surface $t = t_1$ and another density profile $\rho_2$ at a later time $t = t_2$.  There is a well-defined algorithm for finding the LT model that evolves from one to the other.

   Suppose, on the surfaces $t = t_1$ \& $t = t_2$, we specify the density to be $\rho = \rho_1(M)$ and $\rho = \rho_2(M)$, then%
 \footnote{In this case, though, \er{RofMint} would not be well defined if there were vacuum $\rho_i(M)$ anywhere, since the range of $R$ over which $M$ is constant could be anything.}
 from \er{rhoLT},
 \begin{align}
   R_i^3(M) - R_0^3 = \int_{M_0}^M \frac{6}{\kappa \rho_i(M)} \, dM ~,~~~~ i = 1,2
   \showlabel{RofMint}
 \end{align}
 and normally we would have $R_ 0 = 0 = M_0$.  We set the coordinate freedom via $r = M$.  Since $M$ is constant along each particle worldline, we now know $R_1$ and $R_2$ for each particle.  We consider a specific $M$, and we assume $\Rt(t_1, M) > 0$ and $R_2 > R_1$.  By the time $t_2$, the worldline is either hyperbolic and still expanding (HX), elliptic and still expanding (EX), or elliptic and already collapsing (EC).  In the HX case, we apply \er{tEvHX} and \er{alpha-x-beta-Def} at the two times and subtract them:
 \begin{align}
   & \sqrt{(1 + x \alpha_2)^2 - 1}\; - {\rm arcosh} (1 + x \alpha_2) \nn \\
   & - \sqrt{(1 + x \alpha_1)^2 - 1}\; + {\rm arcosh} (1 + x \alpha_1)
   - x^{-3/2} (t_2 - t_1) = \psi_{HX}(x) = 0 ~,
 \end{align}
 with similar expressions for the other cases.  This is solved numerically by the bisection method, and for this purpose, a pair of $x$ values that bracket the solution were found.  Having obtained $f = x M^{2/3}$, $a$ is found by using \er{tEvHX} again:
 \begin{align}
   a & = t_1 - x^{-3/2} \left\{ \sqrt{(1 + x \alpha_1)^2 - 1}\;
      - {\rm arcosh} (1 + x \alpha_1) \right\} ~.
 \end{align}
  Obviously it is important to know which case applies along each worldline.  In \cite{KraHel02} it was shown that
 \begin{align}
   t_2 - t_1 & > \left( \frac{\alpha_2}{2} \right)^{3/2} \left[
      \pi - \arccos\left(1 - \frac{2 \alpha_1}{\alpha_2} \right)
      + 2 \sqrt{\frac{\alpha_1}{\alpha_2} - \left( \frac{\alpha_1}{\alpha_2} \right)^2}\; \right]
      & & \to~~ EC \\
   t_2 - t_1 & < \frac{\sqrt{2}}{3} \left( \alpha_2^{3/2} - \alpha_1^{3/2} \right)
      & & \to~~ HX \\
   & \mbox{otherwise} & & \to~~ EX
 \end{align}
 The borderlines between these cases require careful treatment, see appendix B of \cite{HelKra06}.
 If $R_2 < R_1$ then there are 3 more solutions, including the collapsing hyperbolic model --- the time reverse of \er{tEvHX}.  

   A similar approach may be used if velocity profiles $\Rt_1(M)$ and $\Rt_2(M)$ are given at $t_1$ \& $t_2$, of if a density profile is given at one time and a velocity profile at another \cite{KraHel04a}.  There are quite a few other useful options, such as setting the late time density of velocity behaviour, specifying a simultaneous time of maximum expansion, specifying only growing or decaying modes, etc \cite{HelKra06}.

   Applications of these methods to model a galaxy cluster, a void, a galaxy with a central black hole, the Shapley Concentration and the Great Attractor, etc, can be found in \cite{KraHel02,KraHel04a,KraHel04b,BoKrHe05,Bole06b,BolHel08}.

 \section{Observations in {\LT} Models}

   The assumption that the universe is homogeneous, and thus well-represented by an FLRW model, has led to a very good understanding of its very large scale features and evolution.  But once the cosmological data are sufficiently accurate and complete over a large enough range of redshifts, this assumption should be checked.

   However, the assumption of homogeneity pervades so much theoretical and observational work so thoroughly, that there is a real danger of a circular argument.  Consequently, any proof of homogeneity must ensure it does not rely on results obtained using an assumption of homogeneity.  Clearly this will not be a simple task.  More precisely, the aim is not only to verify homogeneity but also to quantify it: how much fluctuation is there on each scale? 

   There are a several reasons why spherical symmetry is a good first step towards relaxing the homogeneity assumption: 
(a) we are at the centre of our past null cone, so it makes sense to consider spacetime in terms of spherical co-ordinates about the observer;
(b) the universe does seem close to isotropic on large scales, but radial homogeneity is not easy to verify because of the finite travel time of light and the miniscule duration over which cosmological observations have been made, so it is more urgent to determine the radial variation of the metric;
(c) there is no deep all-sky redshift survey at present, and the zone of avoidance is likely to be a gap in any survey for the foreseeable future;
(d) it keeps the theory and numerics tractable while the basics are sorted out.
Of course, in the long run, the assumption of spherical symmetry will be dropped.  

   Here we derive the observational relations that would be expected in an LT model with given arbitrary functions.  This is also known as the `forward problem'.  Below we focus on a central observer, though non-central observers have also been considered.  

 \subsection{Observables and Source Evolution}
 \showlabel{ObsSrcEv}

   The observables we shall use are those for which the dataset is already substantial and will in the near future become extensive, the redshift $z$, the number density in redshift space $n(z)$, the apparent luminosity and angular diameter $\ell(z)$ \& $\delta(z)$.  Connected with each of these is a source property, the peculiar velocity $\zeta$, the mass per source $\mu(z)$ the absolute luminosity $L(z)$ and the true diameter $D(z)$.

   The redshift $z$ is
 \begin{align}
   z = \frac{\lambda_o}{\lambda_e} - 1
   \showlabel{z}
 \end{align}
 where $\lambda_o$ and $\lambda_e$ are the observed and emitted wavelengths.  
 The diameter and luminosity%
 \footnote{In \cite{KriSac66} a ``corrected luminosity distance'' was defined to be the same 
as the diameter distance.  Some authors have called this latter the ``luminosity distance'', 
which perhaps has led to a confusion of terminology and sometimes to incorrect definitions.}
 distances are
 \begin{align}
   d_D = \frac{D}{\delta} ~,~~~~~~~~
   d_L = \sqrt{\frac{L}{\ell}}\; \, d_{10} = 10^{(m - \tilde{m})/5} d_{10} ~,
   \showlabel{dDdLDef}
 \end{align}
 where $\delta$ and $\ell$ are the angular diameter and apparent luminosity of a source, $D(z)$ and $L(z)$ are the corresponding true diameter and absolute luminosity, $m$ and $\tilde{m}$ are the apparent and absolute magnitudes, and $d_{10}$ is $10$~parsecs.  The two distances are related by the reciprocity theorem \cite{Eth33,Pen66,Ell71},
 \begin{align}
   (1 + z)^2 d_D = d_L ~.   \showlabel{reciprocity}
 \end{align}
 The Hubble and deceleration constants are obtained from the slope and concavity of the $d_L(z)$ plot at the origin,
 \begin{align}
   \left. \td{d_L}{z} \right|_{z=0} = \frac{1}{H_0} ~,   \showlabel{H0Def} \\
   \left. 1 - \frac{1}{H_0} \tdt{d_L}{z} \right|_{z=0} = q_0 ~,   \showlabel{q0Def}
 \end{align}
 and a common observational definition of $H(z)$ and $q(z)$, {\em based on the FLRW model}, is
 \begin{align}
   \frac{1}{H} & = \frac{1}{\sqrt{1 + \Omega_k (H_0 d_L/(1 + z))^2}\;}
      \td{}{z} \left( \frac{d_L}{(1 + z)} \right) ~,   \showlabel{HzRWDefObs} \\
   q & = \frac{(1 + z)}{H} \td{H}{z} - 1 ~.   \showlabel{qzRWDefObs}
 \end{align}
 For general non-homogeneous models, there is no obvious general definition of $H(z)$ or $q(z)$, and a number have been proposed.  In any case, what matters is the relation between the model and observations.

 If in a redshift survey of the sky, $dN$ sources are observed to lie between $z$ and $z + dz$ within solid angle $d\omega = \sin\theta \, d\theta \, d\phi$, then the redshift-space mass density is
 \begin{align}
   \frac{2 \sgh}{\kappa} = \mu n = \frac{\mu \, dN}{d\omega \, dz}
 \end{align}
 where $n$ is the redshift space number density and $\mu$ is the mean mass per source.  For a treatment which considers several different source types and observations at different wavelengths see \cite{Hel01}.

   A significant feature of these definitions is that each observable, $\delta$, $\ell$ and $n$, is associated with a source property, $D$, $L$ and $\mu$, which have certainly evolved over cosmological timescales.  The latter are much harder to determine observationally, and studies of their values and evolution invariably assume a homogeneous RW model in which to do the analysis.  However, if we eventually want to prove that the universe is homogeneous, it is imperative to avoid a circular argument.  The only way to be certain of the conclusion is to do the analysis without making the homogeneity assumption.

 \subsection{The Null Cone and the Observational Relations}
 \showlabel{NCOR}

   Light rays arriving at the central observer O follow $ds^2 = 0 = d \theta^2 = d \phi^2$, 
so from \er{ds2LT} the past null cone (PNC) of the observation event ($t = t_0, r = 0$) satisfies
 \begin{align}
   \td{t}{r} = - \frac{R'}{W} ~,~~~~~~~~ W = \sqrt{1 + 2 E}\; ~,
   \showlabel{dtdrNC}
 \end{align}
 and we write the solution $t = \th(r)$ or $r = \rh(t)$, defining the local time from the bang to O's PNC with
 \begin{align}
   \tau = \th - a ~.
   \showlabel{tau}
 \end{align}
 This radial null path is necessarily geodesic.  We denote a quantity evaluated on the observer's past null cone with a hat on top or as a subscript, for example $R(\th(r),r) \equiv \Rh$ or $[R]_\wedge$, though this will often be omitted where it is obvious from the context.  For a given LT model, equation \er{dtdrNC} must be solved numerically.  

   As is well known for the LT model (e.g. \cite{Bondi47,MuHeEl97,LuHel07}), the redshift of sources on the PNC observed at O obeys
 \begin{align}
   \frac{dz}{(1 + z)} = \frac{\Rtrh}{W} \, dr ~,
   \showlabel{dz(1+z)}
 \end{align}
 where $\Rt'$ is given by \er{RtRtr} and \er{RtSq}.

    The diameter distance is, by \er{dDdLDef}, the quantity that converts measured angular sizes of objects to their physical sizes at the time of emission.  It is evident from the metric \er{ds2LT} that this is the areal radius $R$, evaluated on the PNC,
 \begin{align}
   d_D = \Rh = R(\th(r), r) ~.
 \end{align}
 and of course $d_L$ follows from \er{reciprocity}.

   To convert the proper density of an LT model to the observed redshift space density, requires that we know how the locus of the PNC relates $z$ to comoving radius, i.e. $\rh(z)$.  Then the total mass contained in a small volume must be the same:
 \begin{align}
   \frac{2 \sgh}{\kappa} \, dz \, d\omega = \left[ \frac{\rho R' R^2}{W} \, dr \, d\omega \right]_\wedge
   ~~~~~~\to~~~~~~
   \kappa \rhoh {\Rh}^2 = 2 \sgh \td{z}{r} ~~~~\mbox{and}~~~~
   \sgh = \left[ \frac{M'}{W} \, \td{r}{z} \right]_\wedge
      \showlabel{n-rho}
 \end{align}
 where \er{rhoLT} has been used.

   In the numerical solution of these equations, \er{dtdrNC}, \er{dz(1+z)} \& \er{n-rho}, we need to evaluate $R$, $R'$ and $\Rt'$ at each new point along the PNC.  Along the constant $r$ worldline at each step we integrate
 \begin{align}
   \tau & = \int_0^R \frac{dR}{\Rt}
   \showlabel{tRint}
 \end{align}
 and
 \begin{align}
   \td{R'}{R} = \frac{\Rt \Rt'}{\Rt^2} ~,
   \showlabel{RrDE}
 \end{align}
 where $\Rt$, $\Rt^2$ and $\Rt \Rt'$ come from \er{RtSq} and \er{RtRtr}.  
 Equations \er{tRint} and \er{RrDE} are solved in one numerical integration for each step of integrating \er{dtdrNC}, \er{dz(1+z)} \& \er{n-rho}.%
 \footnote{
 The $\Lambda = 0$ special case is much easier, because \er{R'gen} allows one to integrate \er{dtdrNC} without solving \er{tRint} at every step.
 }

   The variation of the areal radius down the PNC is 
 \begin{align}
   \td{\Rh}{r} = \Rrh + \Rth \, \td{\th}{r}
   = \Rrh \left( 1 - \frac{\Rth}{W} \right) ~,
   \showlabel{dRhdr}
 \end{align}
 and its second derivative is
 \begin{align}
   \tdt{\Rh}{r} & = \left( \pd{}{r} + \td{\th}{r} \pd{}{t} \right) \td{\Rh}{r} \\
   & = \left[ \, \left( R'' - \frac{R' \Rt'}{W} \right) \left( 1 - \frac{\Rt}{W} \right)
      + \frac{R'}{W} \left( - \Rt' + \frac{\Rt W'}{W} + \frac{\Rtt R'}{W} \right) \, \right]_\wedge \\
   & = \left[ \, \left( 1 - \frac{\Rt}{W} \right) \left\{ R'' - \frac{R' \Rt'}{W}
      + \frac{R'^2}{W^2} \left( \frac{\Lambda R}{3} - \frac{M}{R^2} \right) \right\}
      + \frac{R' \Rt'}{W} \left( 1 - \frac{\Rt^2}{W^2} \right) - \frac{M' R' \Rt}{W^3 R} \, \right]_\wedge ~,
 \end{align}
 where $W'$ and $\Rtt$ were eliminated using \er{RtRtr} and \er{Rtt}.  It is important for later to note that $\Rh(r)$ may have a maximum value where $\tdil{\Rh}{r} = 0$, and at this locus we have
 \begin{align}
   \Rth = W ~~~~~~\LRa~~~~~~ \frac{2 M}{R} + f + \frac{\Lambda R^2}{3} = 1 + f ~,
   \showlabel{RtWAH}
 \end{align}
 and consequently, using \er{n-rho},
 \begin{align}
   \tdt{\Rh}{r} = - \frac{\sgh R'}{R W} \td{z}{r} ~.
   \showlabel{RhrrAH}
 \end{align}
 Thus the slope of the $d_L(z)$ curve is
 \begin{align}
   \td{d_L}{z} & = \td{(1 + z)^2 \Rh}{z} = 2 (1 + z) \Rh + (1 + z)^2 \frac{\tdil{\Rh}{r}}{\tdil{z}{r}} \nn \\
   & = (1 + z) \left( 2 \Rh + \frac{\Rrh}{\Rtrh} (W - \Rth) \right) ~,
 \end{align}
 and at the origin we have
 \begin{align}
   \frac{1}{H_0} & = \left. \frac{\Rrh}{\Rtrh} \right|_{z = 0} ~.
 \end{align}
 The definitions for the radial and tangential Hubble rates
 \begin{align}
   H_r = \frac{\Rt'}{R'} ~,~~~~~~ 
   H_t = \frac{\Rt}{R} ~,
 \end{align}
 represent the metric expansion rates in the radial and tangential directions, but one needs to be careful which of these, if any, relates to which observation.

   Near the origin --- the vertex of the PNC --- in addtion to \er{OrigDef} \& \er{OrigMf}, we have
 \begin{align}
   W \to 1 ~,~~~~~~ z \to 0 ~,~~~~~~
   \Rtrh \to H_0 \left[ \Rrh \right]_0 ~,~~~~~~
   \td{z}{r} \to H_0 \left[ \Rrh \right]_0 ~,~~~~~~
   \td{\Rh}{r} \to \left[ \Rrh \right]_0 ~.
   \showlabel{OrigPNC}
 \end{align}
 The origin value of $R'$ (and $\Rrh$) depends on the choice of arbitrary functions, but if $r \sim R$ there, then $R'$ is finite and non-zero, while $\tdtil{\Rh}{r} \to \left[ \Rrrh \right]_0$.

   The LT observational relations may be very different from the FLRW ones, especially near the maximum in $\Rh$ \cite{MBHE98}.  
   
   \bp{Inhomogeneous Models of SNIa Dimming}
   In recent years the LT model has seen quite a bit of use in investigations of whether the observed dimming of the supernovae can be explained as an effect of cosmic inhomogeneity, rather than invoking a `dark energy' whose magnitude and physical origin are obscure.

   For the case of an observer that is off-centre, \cite{HuMaMa97} calculated expressions for the angular variation of $d_L$, $H_0$, $q_0$, the source number count, and $\Delta T/T$, and showed the CMB dipole could be explained this way.

   It was first pointed out in Celerier \cite{Cel00} that the observed SNIa dimming can be explained by inhomogeneity.  That paper used a parabolic LT model and showed that a series expansion of $d_L(z)$ could easily manifest apparent `acceleration'.  This was generalised to non-parabolic LT models in \cite{TanNam07}.

   In \cite{AlAmGr06} the authors constructed an LT model that has a low density region (void) at the centre, and asymptotically approaches homogeneity.  Their functions $M$ \& $f$ have the form \er{M-RW+} with
 \begin{align}
   M_3 & = H_{\bot 0}^2 \alpha_0 ~,~~~~~~ &
      \tilde{M} & = \frac{\Delta \alpha}{2 \alpha_0}
      \left\{ 1 - \tanh\left(\frac{r - r_0}{2 \Delta r} \right) \right\} \\
   f_2 & = H_{\bot 0}^2 \beta_0 ~,~~~~~~ &
      \tilde{f} & = \frac{\Delta \beta}{2 \beta_0}
      \left\{ 1 - \tanh\left(\frac{r - r_0}{2 \Delta r} \right) \right\}
 \end{align}
 so that $M$ goes from $\sim H_{\bot 0}^2 (\alpha_0 - \Delta\alpha) r^3$ at the centre to $H_{\bot 0}^2 \alpha_0 r^3$ at large $r$, and similarly for $f$.  Their 3rd function was fixed via the hyperbolic version of \er{EllEvRi}, choosing $t_i$ to be recombination, and setting $R_i = a_* r$ where $a_*$ is the RW scale factor at recombination.  They then calculated the redshift and the apparent magnitude for a central observer, and found they could obtain good agreement with observations.  They also verified that they could retain the observed CMB power spectrum.  In \cite{AlnAma07} and \cite{AlnAma06} the authors investigated an off-centre observer in two versions of the void model.  They found a marginal improvement in the fit to the SNIa data is possible.  If the observed COBE dipole is due to this effect, it requires only a $15$~Mpc displacement from centre, but the corresponding quadrupole and octopole effects are then too small to match observations.

   In \cite{VaFlWa06} it was suggested that inhmogeneous models of supernova dimming have a `weak singularity' at the centre.  However, this is merely a conical point in the density profile, and not a singularity \cite{KrHeCeBo}.  Also, \cite{YoKaNa08} showed it is easy to smooth the central density without affecting the model much.

   Ref \cite{Bole08} considered a selection of LT models, with and without $\Lambda$ --- a central void model with $a = 0$, a uniform present-day density and a varying $H_t$, a varying $H_r$ with $a = 0$, both $\rho$ and $a$ varying.  All the models had small density oscillations imposed to represent smaller scale inhomogeneity.  It was argued that all of the $\Lambda = 0$ models considered are `peculiar'.

   In \cite{Garf06} the LT functions were chosen to be $2M = B r^3$, $a = 0$, $f = r^2/(1 + (c r)^2)$, $B$ was set from $\Omega_{m0}$ via \er{fMaRW}, and $c$ was adjusted to get the best fit to the SNIa magnitude-redshift data.  With $\Omega_m = 0.2$ the fit was better than that of $\Lambda$CDM.

   For observers near the centre of an overdensity, the model of \cite{ABTT06} agrees with $d_L$ observations for part of the $z$ range.

   \cite{EnqMat07} also compared two classes of LT models with the SNIa data, calculating $\chi^2$.  One model was fixed by $H_t = H + \Delta H e^{-r/r_0}$ and $\Omega_m(r) = 2 M/(H_{t0}^2 R_0^3) = \Omega_0$, where $H$, $\Delta H$, $r_0$ and $\Omega_0$ are constants.  The other had $a = 0$ and $\Omega_m$ varying.  They found that varying $H_t$ is very effective at fitting the data, but varying $\Omega_m$ is not.  The best fit LT model had slightly lower $\chi^2$ than the $\Lambda$CDM model, but including both inhomogeneous expansion and non-zero $\Lambda$ did not improve the fit.  
   
   In \cite{BolWyi08} two models were considered --- a local void model with a simultaneous bang time, and a `hubble bubble' model in which the expansion rate $H_t$ is higher locally than far away but the present-day density is uniform.  Each is a quite specific 2-parameter LT model.  They confronted their models with SNIa $d_L$ data, the BAO dilation scale, $d_V = [d_D^2 z / H_r]^{1/3}$, and the limit on $H_r$ set by the age of the oldest stars.  From $\chi^2$ calculations, they concluded that their best-fit hubble-bubble model fits the data almost as well as $\Lambda$CDM.

   In \cite{GarHau08}, void models with 4 or 5 parameters were considered, and it was shown that they can provide a good fit observations of the SNIa dimming, the CMB, and the BAO (within 1 $\sigma$) and a $\chi^2$ comparable to the $\Lambda$CDM model.
   In \cite{GarHau08b}, it was shown that observations of the kinematic Sunaev-Zeldovich effect already limit LT voids to $< 1.5$~Gpc, and future surveys will either put tighter limits on the size or constrain the density and expansion profiles.
   In \cite{GarHau08c}, the authors proposed the normalised cosmic shear as a test of inhomogeneity.  They also found that LT models still provide excellent agreement with updated SNIa and BAO data.

   A similar good fit with SNIa observations, i.e. a $\chi^2$ comparable to that of the $\Lambda$CDM model, was found in \cite{Enqv08}, which considered LT `bubble' models with decreasing $H(r)$ and constant $\Omega_m(r)$.  There was no improvement in the fit using a similar model with non-zero $\Lambda$.

   For a summary see \cite{Cel07a,Cel07b,BoKrHeCe}.  The important issue here is to highlight the difficulty of separating the effects on the null cone observations of the cosmic equation of state, of source evolution, and of cosmic inhomogeneity.  Whether or not $\sim$Gpc scale inhomogeneities are discovered, inhomogeneous models have to be taken seriously, firstly because inhomogeneities on many scales do exist, and secondly because we should rigorously verify homogeneity (instead of just assuming it), and such testing requires using an inhomogeneous model, so that the detection of inhomogeneity is a possible outcome.

   \bp{Differences between Dimming Models}
   Now the arbitrary functions of any given LT model determine not only a luminosity or diameter distance relation, $d_L(z)$ or $d_D(z)$, but also a redshift-space density relation $\sigma(z)$.  Each chosen model ``predicts'' a $\sigma(z)$ profile, and this will be important in distinguishing models.  Though number counts are not very complete or reliable today, the situation is likely to improve rapidly with future redshift surveys.  In fact, the different types of model predict very different $\sigma(z)$.

   If the LT arbitrary functions are written in the form of a central behaviour plus a variation,
 \begin{align}
   & M = M_3 r^3 (1 + \Delta \tilde{M}) ~,~~~~
      f = f_2 r^2 (1 + \Delta \tilde{f}) ~,~~~~
      a = a_0 + \Delta \tilde{a} ~,
 \end{align}
 then the leading terms may be related to the central cosmological parameters via
 \begin{align}
   & f_2 = {\rm sign}(H_0^2 (1 - 2 q_0)) ~,~~~~
      M_3 = q_0 H_0^2 \big( f_2/(H_0^2 (1 - 2 q_0)) \big)^{3/2} ~; \\
   & f_2 > 0 :~~~~ \eta_0 = {\rm arccosh}\left( 1 + \frac{f_2 H_0 \sqrt{f_2 (1 - 2 q_0)}\;}{q_0} \right) ~,~~~~
      \tau_0 = M_3 (\sinh\eta_0 - \eta_0)/f_2^{3/2} ~; \\
   & f_2 < 0 :~~~~ \eta_0 = \arccos\left( 1 + \frac{f_2 H_0 \sqrt{f_2 (1 - 2 q_0)}\;}{q_0} \right) ~,~~~~
      \tau_0 = M_3 (\eta_0 - \sin\eta_0)/(-f_2)^{3/2} ~; \\
   & a_0 = t_0 - \tau_0
 \end{align}
 which defines a `central RW model'.  A model with a pure bang time inhomogeneity, may be described by the functions
 \begin{align}
   % Model AAC1
   M & = M_3 r^3 ~,~~~~
      f = f_2 r^2 ~,~~~~
      a = a_0 + I (e^{-r/J} - 1) + K (e^{-r/J_2} - 1)  ~, \\
   H_0 & = 0.72 ~,~~~~ q_0 = 0.22 ~,~~~~ % \Omega_\Lambda = 0 ~,~~~~
      I = 0.8/H_0 ~,~~~~ J = 0.5 ~,~~~~ K = -0.7/H_0 ~,~~~~ J_2 = 0.7 ~,
 \end{align}
 and using this, we get good agreement with $d_L(z)$ from supernova data%
 \footnote{For a smoother density profile at the origin, the approach of \cite{YoKaNa08} may be used.}%
 .  It has become customary to compare the measured magnitudes with those expected in the Milne model, i.e. $\Delta m = m - m_\text{Milne} = 5 \log (L/L_\text{Milne})$.  The left plot below shows $\Delta m(z)$ against the supernova data of Kowalski et al. \cite{KowEtAl08}, with the blue line for the given LT model, and the red line for the RW model with the same central parameters; the middle plot shows the redshift-space density (number of sources per steradian per unit redshift interval times mean mass per source), with blue the LT model, and red the central RW model; the right plot shows $\rho(t_0, r)/\rho_\text{crit,0}$ the density as a multiple of the central critical density, on a constant time slice at the present day, against coordinate radius $r$.
 \\[1mm]
 \includegraphics[scale=0.27]{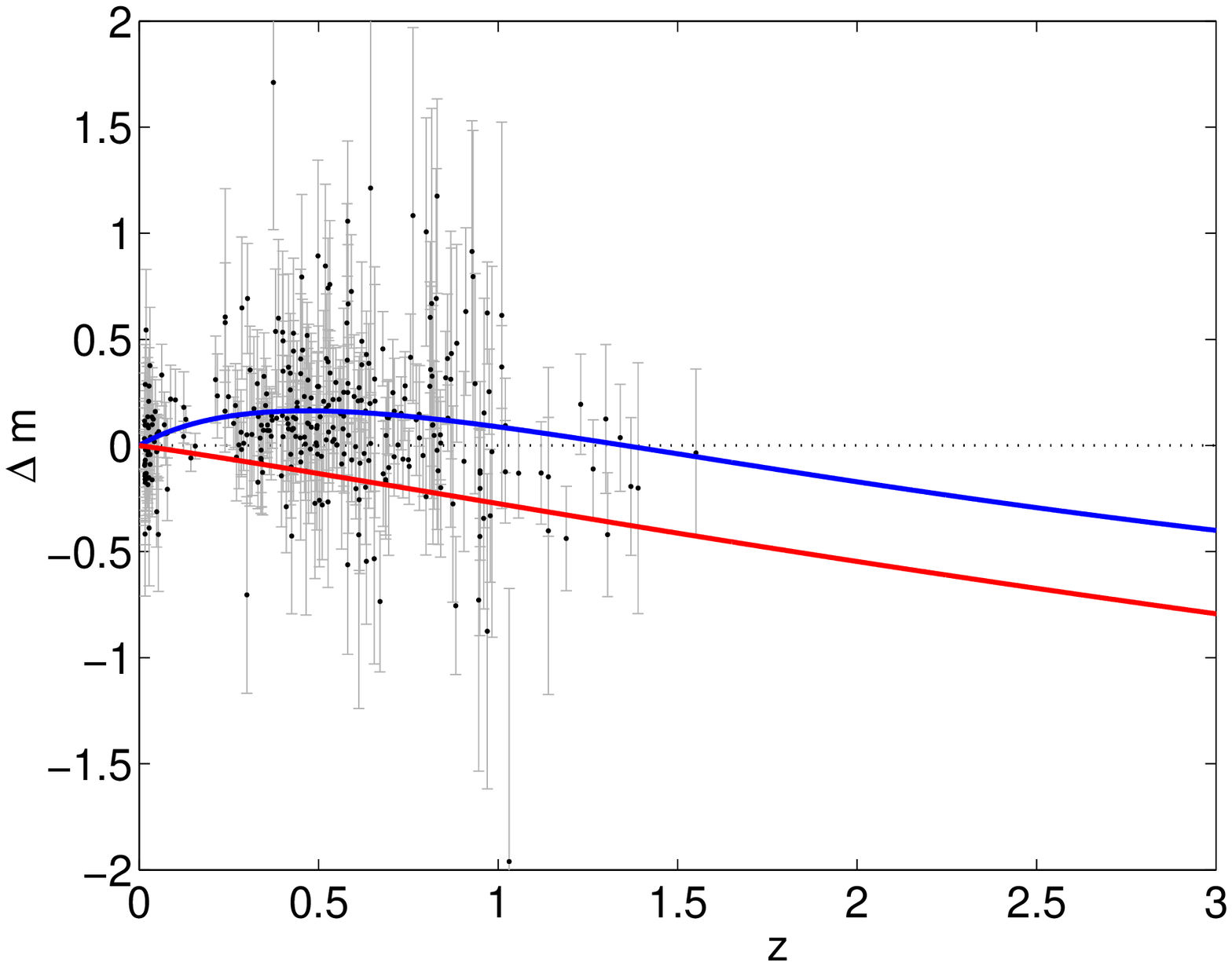}
 \includegraphics[scale=0.27]{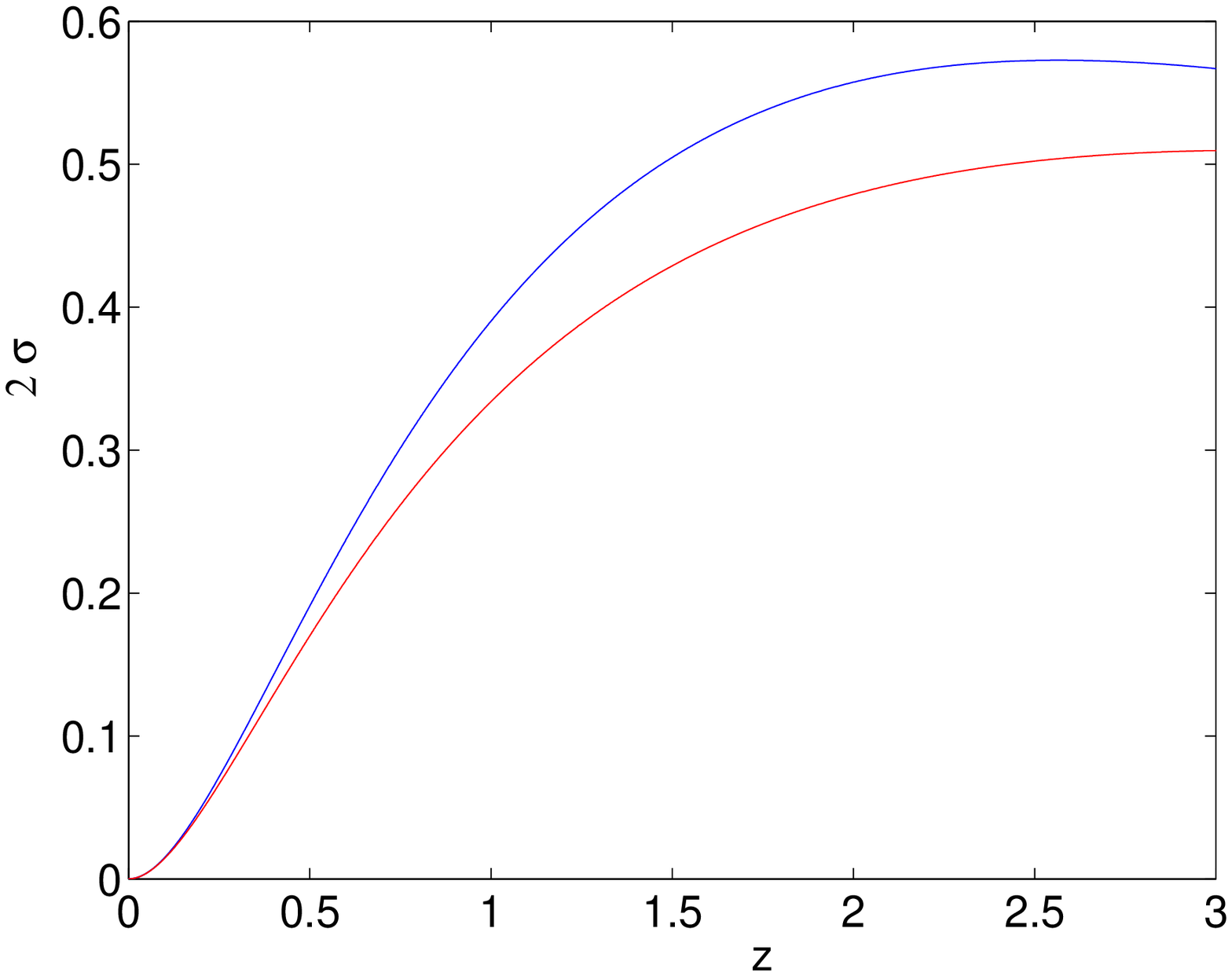}
 \includegraphics[scale=0.27]{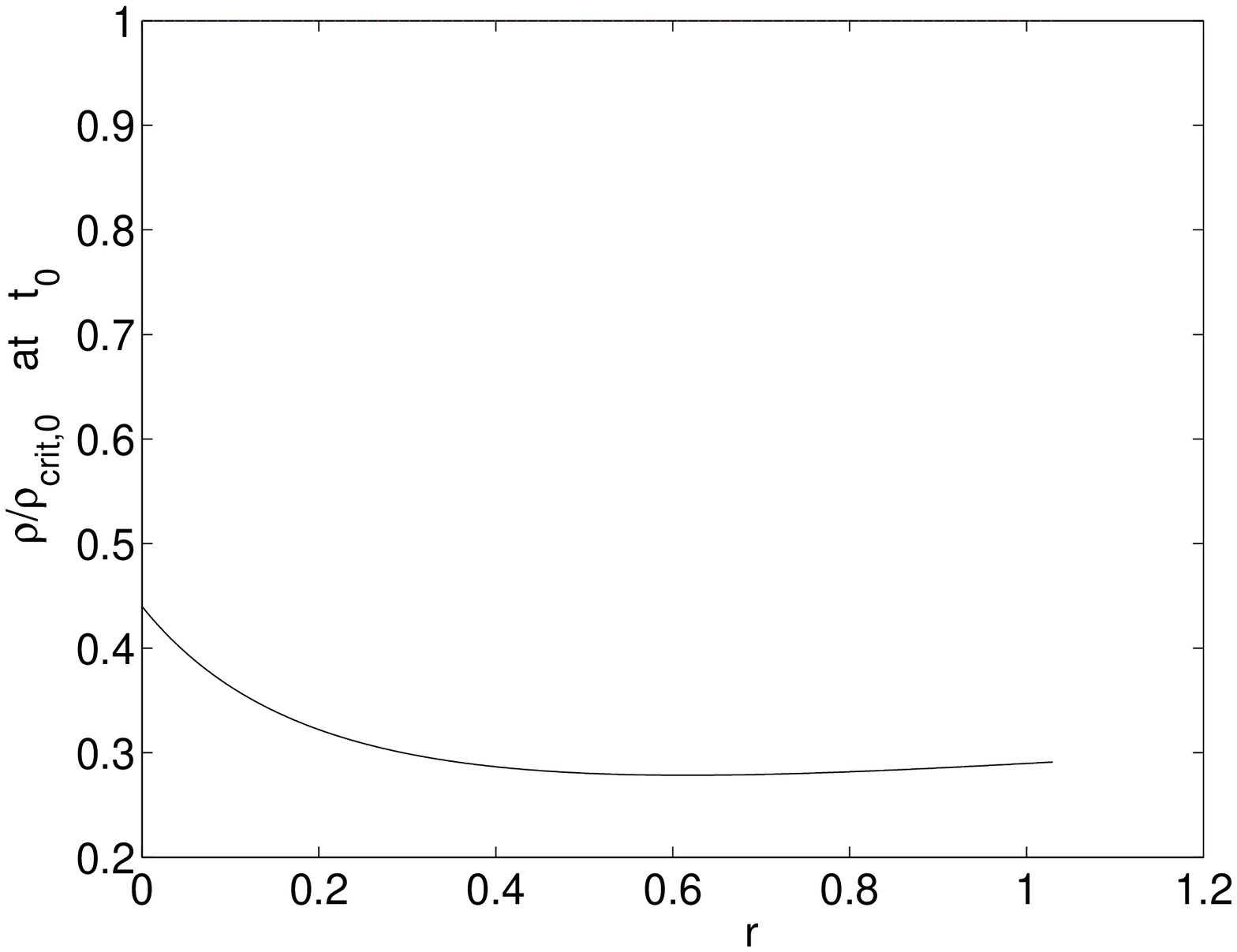}
 \\[1mm]
 For a pure mass-geometry-energy inhomogeneity, we can set the LT functions to
 \begin{align}
   % Model ADA1
   M & = M_3 r^3 ~,~~~~
      f = f_2 r^2 \big( 1 + E (e^{-r/F} - 1) + G (e^{r/F_2} - 1) \big) ~,~~~~
      a = a_0 ~, \\
   H_0 & = 0.72 ~,~~~~ q_0 = 0.09 ~,~~~~ % \Omega_\Lambda = 0 ~,~~~~
   E = 8.2 ~,~~~~ F = 0.4 ~,~~~~ G = -7.6 ~,~~~~ F_2 = 0.45 ~,
 \end{align}
 and the following plots show we also get good agreement with the SNIa data%
 \footnote{Both of these curves have a $\chi^2$ that rivals the least squares quadratic fit to the data.}%
 .
 \\[1mm]
 \includegraphics[scale=0.27]{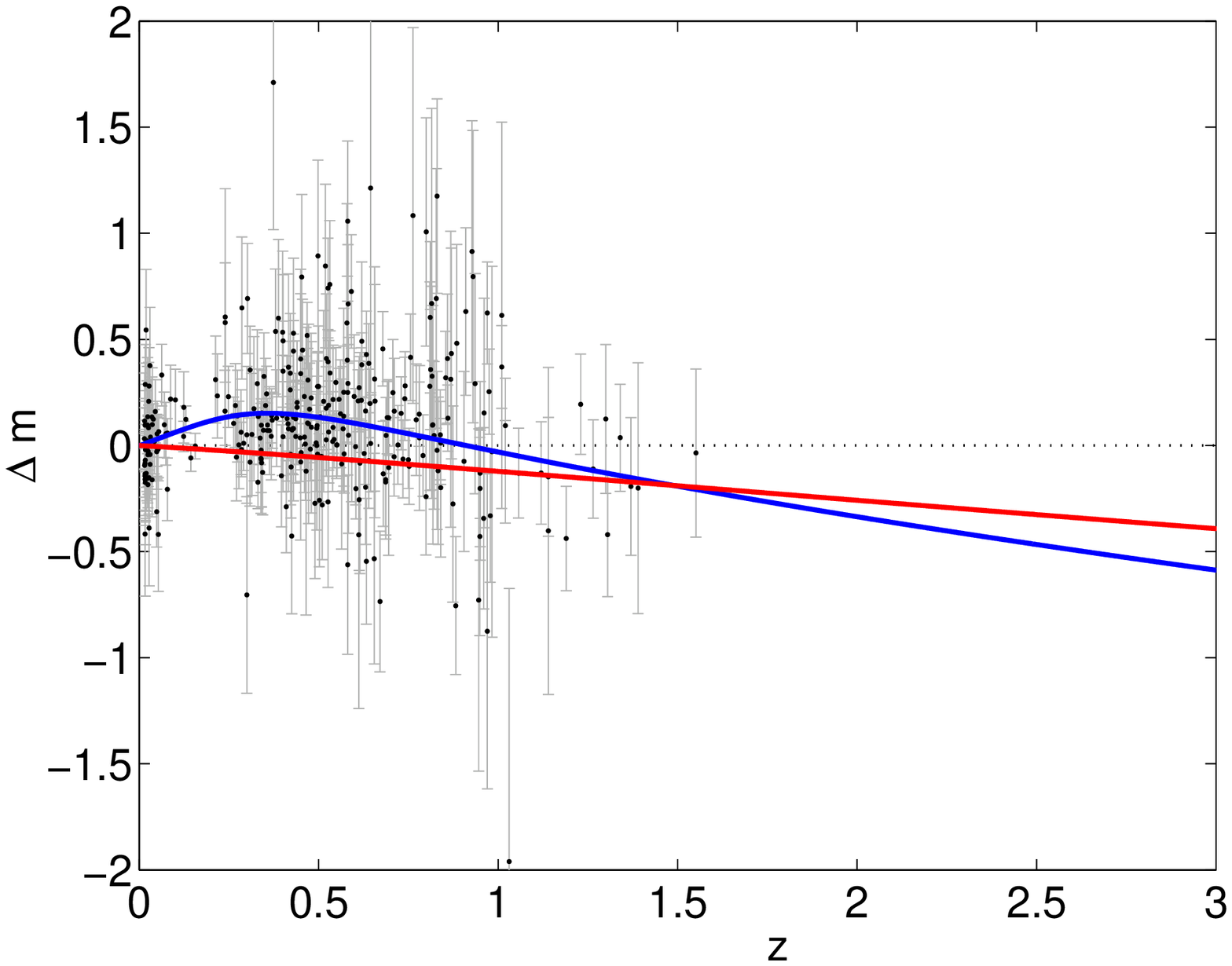}
 \includegraphics[scale=0.27]{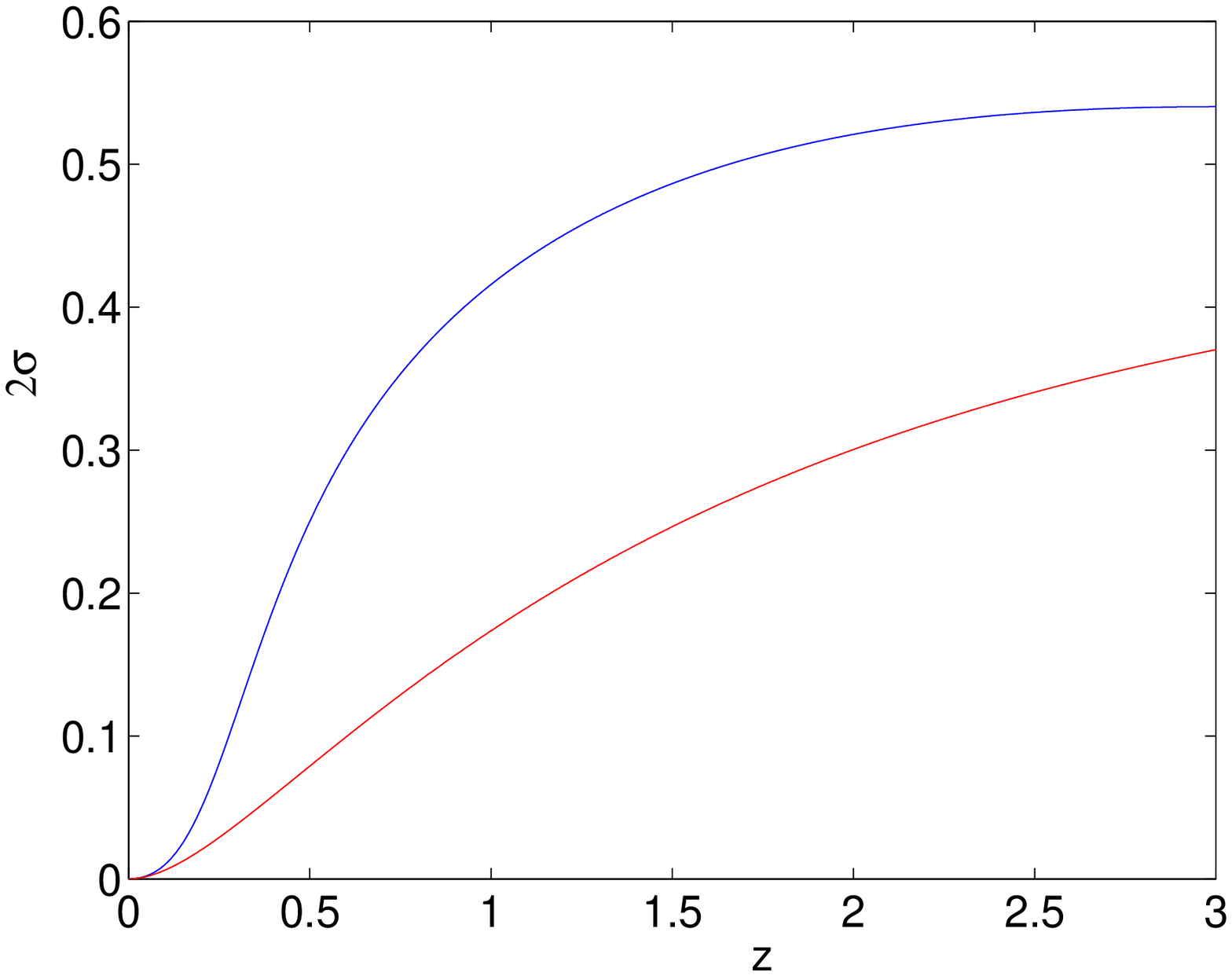}
 \includegraphics[scale=0.27]{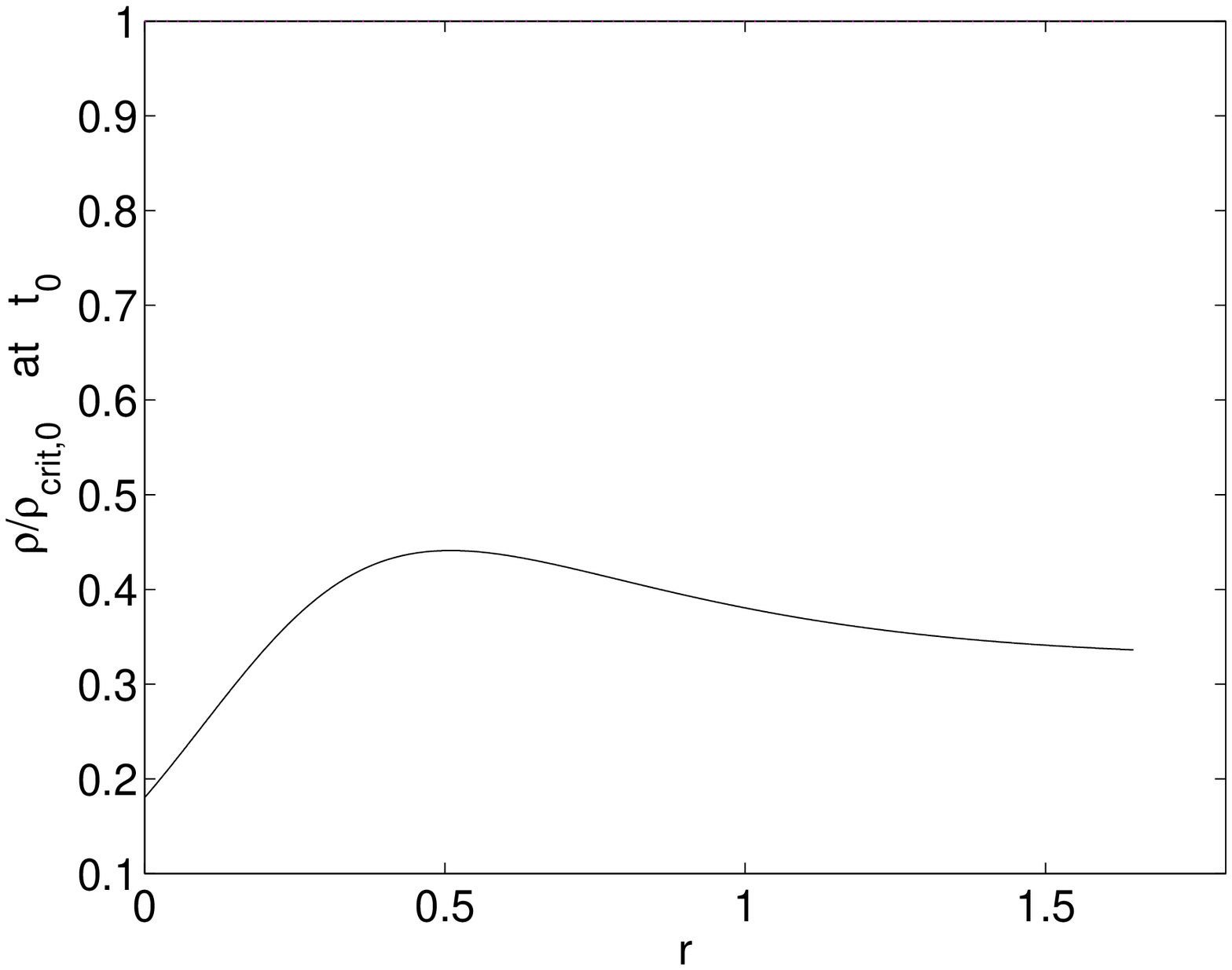}

   There is little difference between the $\Delta m(z)$ curves of these two and many other LT models, but the $\sigma(z)$ curves are quite distinct.  At present there are many uncertainties connected with estimating $\sigma$; the number count data is not sufficiently complete or accurate, the relation to total matter density is not well established, and the evolution of galaxy numbers and masses is an active area of research.  Still, one may hope this will improve dramatically with the next generation of redshift surveys.  Other ways of testing models, should be pursued.

   Now, observers are very likely to live on planets, which are highly likely to circle stars in galaxies, which have a good chance of being inside clusters within superclusters.  In other words, many observers are likely to be inside regions of higher density.  If the far universe is homogeneous (on average), then a model of an observer inside a density peak must have an intervening region of lower density to compensate the central overdensity.  The varying bang time models that reproduce the supernova dimming, also seem to generate a central overdensity quite well.  Of course, variations in both the bang time and the mass-geometry-energy functions will surely be needed to get the best fit to all the data.

 \subsection{Determining the Metric of the Cosmos}

   The Metric of the Cosmos project aims to determine the geometry of our universe from observational data.  This is an `inverse problem': given observations such as those described above, determine as much as possible about the spacetime metric.  In practice, one needs to make assumptions about the cosmic equation of state, etc, but the goal is to reduce them to a minimum.
    
   An important aim of this project is to determine the degree of homogeneity in the universe.  The large amounts of cosmological data now flowing in will soon make this a real possibility.  In order to do this, it is essential to remove the assumption of homogeneity, but since the use of a RW model is widespread in cosmological data analysis, many calculations will have to be carefully re-worked.  Now angular homogeneity --- that is isotropy --- is easy to check, and does not require us to know anything about the PNC.  Whatever the variation of observables with $z$ is, it must be the same in all directions.  But radial homogeneity is not at all easy to verify, since the $z$-dependence of observables depends on several things: the time evolution of the expansion (i.e. the equation of state), the source evolution, and whatever radial inhomogeneity is present.  So although a general treatment requires us to go beyond spherical symmetry, just pinning down the degree of radial variation would be a big step forward.  

   It was shown in \cite{MuHeEl97} that any reasonable `observational' functions $d_D(z)$ or $d_L(z)$ and $\sgh(z)$ can be reproduced by an LT model, and an algorithm for extracting the LT arbitrary functions was given.  This algorithm was implemented as a numerical procedure and clarified and extended significantly in \cite{LuHel07,Hel06,McCHel08}.

   For this project, we now need to invert the equations of the last section; we treat $\Rh(z)$ and $\sgh(z)$ as given, and we want to determine $f(r)$, $M(r)$ and $a(r)$.  We can use the radial coordinate freedom to choose
 \begin{align}
   \td{\th}{r} = -\beta(r) ~,~~~~\mbox{i.e.}~~~~ \Rrh = \beta W ~,
   \showlabel{RrPNC}
 \end{align}
 on the observer's past null cone, so that the solution to \er{dtdrNC} is
 \begin{align}
   \th = t_0 - \int_0^r \beta \, dr ~.
   \showlabel{rPNC}
 \end{align}
 In practice, $\beta = 1$ and $\th = t_0 - r$ is the obvious choice, providing there are no shell crossings.  Note that \er{RrPNC} and \er{rPNC} and much of the following only hold for the single null cone with apex $(t_0, 0)$.  

Putting this in \er{dRhdr} gives
 \begin{align}
   \td{\Rh}{r} = \beta (W - \Rth) ~,
   \showlabel{dRhdrNC}
 \end{align}
 and using \er{dRhdr}, \er{RrPNC} and \er{RtSq}, we find
 \begin{align}
   W = \frac{1}{2 \beta} \left( \td{\Rh}{r} \right) + 
   \frac{\beta \left( 1 - \frac{2 M}{\Rh} - \frac{\Lambda \Rh^2}{3} \right)}{2 \left( \td{\Rh}{r} \right)} ~,
   \showlabel{WPNC}
 \end{align}
 while \er{n-rho} and \er{RrPNC} give
 \begin{align}
   M' = \sgh W \td{z}{r} ~.
   \showlabel{MrPNC}
 \end{align}
 Combining \er{WPNC} with its derivative results in
 \begin{align}
   W' = \beta \left( \frac{M}{\Rh^2} - \frac{\Lambda R}{3} \right)
   - \frac{\beta M'}{\Rh \left( \td{\Rh}{r} \right)}
   - \left( \frac{\tdt{\Rh}{r}}{\td{\Rh}{r}} - \frac{\beta'}{\beta} \right)
   \left( W - \frac{1}{\beta} \td{\Rh}{r} \right) ~.
   \showlabel{WrPNC}
 \end{align}
 Putting \er{RrPNC} and \er{WrPNC} in \er{dz(1+z)} leads to
 \begin{align}
   \frac{dz}{(1 + z)} & = \frac{dr}{\Rth} \left( \frac{1}{\beta} \td{\Rh}{r} - W \right)
      \left( \frac{\beta M'}{\Rh \left( \td{\Rh}{r} \right) W}
      + \frac{\tdt{\Rh}{r}}{\td{\Rh}{r}} - \frac{\beta'}{\beta} \right) ~,
 \end{align}
 which simplifies, on substituting for $M'$ and $\Rth$ from \er{MrPNC} and \er{dRhdrNC}, to
 \begin{align}
   \td{z}{r} = - \frac{(1 + z) \left( \frac{\sgh \beta}{\Rh} \td{z}{r}
   + \tdt{\Rh}{r} + \frac{\beta'}{\beta} \right)}{\td{\Rh}{r}} ~.
   \showlabel{dzdrPNC}
 \end{align}
 Since the coordinate $r$ is not an observable, we convert all $r$ derivatives to $z$ derivatives using
 \begin{align}
   \td{\Rh}{r} = \td{\Rh}{z} \, \frac{1}{\varphi} ~,~~~~~~
   \tdt{\Rh}{r} = \tdt{\Rh}{z} \, \frac{1}{\varphi^2}
      - \td{\Rh}{z} \, \frac{1}{\varphi^3} \, \td{\varphi}{z} ~,
   \showlabel{dRhdzChain}
 \end{align}
 where
 \begin{align}
   \varphi = \td{r}{z}
   \showlabel{phiDef}
 \end{align}
 defines $\varphi$.  Then \er{MrPNC} and \er{WPNC} become
 \begin{align}
   \td{M}{z} & = \sgh W ~,
   \showlabel{dMdz} \\
   W & = \frac{1}{2 \beta \varphi} \left( \td{\Rh}{z} \right) + 
   \frac{\left( 1 - \frac{2 M}{\Rh} - \frac{\Lambda \Rh^2}{3} \right) \beta \varphi}{2 \left( \td{\Rh}{z} \right)} ~,
   \showlabel{WPNCz}
 \end{align}
 while putting \er{dRhdzChain} in \er{dzdrPNC} and solving for $\tdil{\varphi}{z}$, gives
 \begin{align}
   \td{\varphi}{z} = \varphi \left( \frac{1}{(1 + z)}
   + \frac{\frac{\sgh \beta \varphi}{\Rh} + \tdt{\Rh}{z}}{\td{\Rh}{z}} 
   - \frac{\beta_z}{\beta} \right) ~.
   \showlabel{dphidz}
 \end{align}
 where $\beta_z = \beta' \varphi$.  As stated, the obvious gauge choice is $\beta = 1$, $\beta_z = 0$. 
 Equations \er{phiDef}, \er{dphidz} and \er{dMdz} with \er{WPNCz} constitute the differential equations to be solved for $\varphi(z)$, $r(z)$, $M(z)$ and $W(z)$.  Then $\tau(z)$ and $a(z)$ follow from \er{HypEv}-\er{EllEv}, \er{tau} and \er{rPNC}.  Knowing $r(z)$, $M(z)$, $W(z)$, and $a(z)$ fully determines the LT metric that reproduces the given $\Rh(z)$ and $\sgh(z)$ data.  We note that \er{dphidz} is an independent DE, while \er{dMdz} and \er{phiDef} require $\varphi(z)$.  Also \er{dMdz} with \er{WPNCz} is a first order linear inhomogeneous ODE, for which the formal solution in known.  However it is easiest to solve all the DEs in parallel as one numerical procedure.

   The initial conditions for these DEs are set at the origin at the time of observation $t_0$.  The LT origin conditions applicable to these null cone equations were given in \cite{LuHel07,McCHel08}, and are reproduced and generalised in the appendix.

 \subsection{Apparent Horizon}
 \showlabel{ApHor}

   In an expanding decelerating model, there is a point on each PNC where the areal radius (i.e. $d_D$) is maximum, $\tdil{\Rh}{z} = 0$.%
 \footnote{This is evident in \cite{McC34, McC39}, but first stated explicitly in \cite{Hoyl61}.}
 We denote this point by $\Rh = R_m$, $z = z_m$, and the locus of all such points is the apparent horizon (AH) --- see \cite{KraHel04b,Hell87}.  

   But points where $\tdil{\Rh}{z} = 0$ make the DEs \er{dphidz} and \er{dMdz} with \er{WPNCz} singular.  However, in any given LT model $W$ is a fixed arbitrary function, so we don't expect any divergence on the right of \er{WPNCz}.  Further, it was shown in \er{RtWAH} that
 \begin{align}
   \td{\Rh}{z} = 0 ~~~~\Ra~~~~
   W - \Rth = 0 ~,
   \showlabel{Rhzm}
 \end{align}
 which open up the possibility that \er{WPNCz} is not actually singular on the AH.  Similarly, we don't expect $\tdil{z}{r}$ or $\tdtil{r}{z}$ to be divergent here in a general LT model with co-ordinate choice \er{RrPNC}, and in fact \er{RhrrAH} and \er{dRhdzChain} (with $\tdil{\Rh}{z} = 0$) show that
 \begin{align}
   \tdt{\Rh}{z} \Bigg|_m = \left[ \varphi^2 \tdt{\Rh}{r} \right]_m
      =  \left[ - \frac{\sgh \beta \varphi}{R} \right]_m ~.   \showlabel{Rhzzm}
 \end{align}
 Indeed, \er{Rhzm} and \er{Rhzzm} are exactly what happens at $\Rh_m$ in the FLRW case.  So although there are no divergencies at $\Rh_m$, the numerics break down.  In \cite{LuHel07} this was overcome by doing a series solution of the DEs \er{phiDef}-\er{dphidz} in $\Delta z = z - z_m$, and joining the numerical and series results at some $z$ value $z_j < z_m$ --- see sections 2.6, 3.3, and appendix B of that paper, and also appendix \ref{MaxSeries} below.  As pointed out in \cite{Hel06}, this phenomenon is not merely a cosmological curiosity.  At this locus, and no other, there is a simple relation between the diameter distance $d_D = \hat{R}$ and the gravitational mass $M_m$ that is independent of any inhomogeneity between the observer and sources at this distance:
 \begin{align}
   2 M_m = \Rh_m - \frac{\Lambda \Rh_m^3}{3} ~,   \showlabel{LamRhMEq}
 \end{align}
 or $R = 2 M$ if $\Lambda = 0$.  However, the redshift $z_m$ at which this occurs is not model independent.  Thus the maximum in $\Rh$ provides a new characterisation of our Cosmos --- the cosmic mass.  More practically, \er{LamRhMEq} and \er{Rhzzm} provide a cross-check on the numerical integration: the $M$ and $\varphi$ values obtained from the numerical integration must agree with those deduced from the measured $\Rh_m$ and $\sgh_m$ using \er{LamRhMEq} and \er{Rhzzm}.  This requirement enables systematic errors in the observational data to be spotted and at least partially corrected, as was done using \er{LamRhMEq} in \cite{McCHel08}.  In fact, the AH relation \er{LamRhMEq} generalises to the \L\ model, which has non-zero pressure \cite{AlfHel09}.

   Now \er{dphidz} may alternatively be written as
 \begin{align}
   \td{}{z} \left[ \td{\Rh}{z} \frac{(1 + z)}{\varphi \beta} \right]
      & = - \frac{\sgh}{\Rh} (1 + z) \\
   \left[ \td{\Rh}{z} \frac{(1 + z)}{\varphi \beta} \right]_0^z
      & = \frac{1}{\beta} \left( \td{\Rh}{z} \frac{(1 + z)}{\varphi} - 1 \right)
      = - \int_0^z \frac{\sgh}{\Rh} (1 + z) \, dz ~,
      \showlabel{dphidzint}
 \end{align}
 by \er{OrigPNC}, consequently giving
 \begin{align}
   r(z) = \int_0^z \varphi \, dz
   = \int_0^z \td{\Rh}{z} (1 + z) \left[ 1 - \beta \int_0^z \frac{\sgh}{\Rh} (1 + z) \, dz \right]^{-1} \, dz ~.
      \showlabel{rzint}
 \end{align}
 Although this appears to have no singularity at $\tdil{\Rh}{z} = 0$, in fact the term in square brackets in \er{rzint} goes to zero, as is evident from \er{dphidzint}.

    Some other attempts at solving a version of the cosmological inverse problem \cite{BisHai96,IgNaNa02,VaFlWa06} got stuck at this locus.  See also the discussion of the apparent horizon and the `critical points' in \cite{KrHeCeBo}.

 \subsection{Numerical Implementation}

    Now in reality, the observations do not provide smooth functions $\Rh(z)$ and $\sgh(z)$, they provide a set of discrete measurements of individual sources.  In order to proceed, the data must be collected into many redshift bins of width $\delta z$, and bin averages calculated.  Furthermore, the derivatives, $\tdil{\Rh}{z}$ and $\tdtil{\Rh}{z}$ must also be calculated.  So, for each of $\Rh(z)$ and $\sgh(z)$ it is necessary to fit a smooth function --- a polynomial say --- to a range of redshift bins, otherwise mild statistical variations in $\Rh(z)$ would create wild fluctuations in $\tdil{\Rh}{z}$ and especially $\tdtil{\Rh}{z}$.  The degree of smoothing is necessarily a compromise between eliminating statistical fluctuations and extracting inhomogeneity.  For example, in \cite{McCHel08} a quartic polynomial was fitted to 50 bins of width $\delta z = 0.001$ on either side of the bin in question.

   A second difficulty is that there is no data at the origin itself --- where initial conditions for the numerical integration are set.  The first bin extends from $z = 0$ to $z = \delta z$, so it provides average values at around $z = \delta z/2$.  This is resolved by fitting a near-origin series solution of the DEs to the first few data bins (see appendix \ref{OrigSeriesMoC}), and starting the numerical integration further out.

   As explained above, the maximum in $\Rh$ requires a series solution (see appendix \ref{MaxSeries}).  In addition, it has one undetermined coefficient, $M_1$.  The numerical and maximum-series results are joined at some redshift $z_j < z_m$ and this fixes $M_1$.  The numerical integration is resumed at $2 z_m - z_j$, using the series values there for ``initial'' numerical values.
   
   Thus the numerical integration has 4 regions --- the origin series, the pre-maximum numerical integration, the near maximum series, and the post-maximum numerical integration --- which must all be properly joined together.

    In \cite{LuHel07} the above programme was implemented as a numerical procedure, and tested using fake data generated from an LT model.  The fake observational data was exact, and contained no scatter or errors other than very small numerical errors.  Importantly, the numerics successfully reproduced the LT arbitrary functions of the various homogeneous and inhomogeneous models used to generate the data.  This validated the numerical procedure as viable in principle.

   In \cite{McCHel08}, the effects of statistical and systematic errors in the data were considered.  The numerical program was revised to output uncertainty estimates for each $f(z)$, $M(z)$, and $a(z)$.  It was shown how to use the data at the maximum of $\Rh(z)$, via \er{LamRhMEq}, to detect and correct for systematic errors in the observational data.  Several examples with fake data were given.  Lastly, the stability of the DEs \er{phiDef}-\er{dphidz} was analysed, and it was shown they are generally stable, except for the $\tdil{M}{z}$ DE which becomes unstable at redshifts larger than $z_m$.  This issue requires further attention.

   Application of this method to redshift survey data is under consideration.  However, at present, available data has a lot of scatter in $\ell(z)$, $\delta(z)$ and $n(z)$, and considerable uncertainty in the source properties $L(z)$, $D(z)$ and $\mu(z)$ at larger $z$ values.  It is particularly likely that studies of the source properties at large $z$ have assumed homogeneity, if not a particular FLRW model.  A method of testing source evolution theories, independently of possible inhomogeneity, was presented in \cite{Hel01}, which considered multiple source types and observations at several wavelengths.

   \bp{Combining Data with an Invariant Distance}   For many purposes, data at the same $z$ are grouped together and averaged, and it is assumed the errors cancel out.  However, as is well known the peculiar velocities of sources create a scatter in $z$ values, especially in clusters, and the observer's motion creates a dipole, so although redshift can be measured to high accuracy, it is not an ideal monotonic measure of distance.  According to Walker's and Etherington's argument \cite{Wal33,Eth33} the source area distance $d_S = (1 + z) d_D$ is independent of the observer's motion.  Since this distance is determined by the geometry of light rays emanating from the emission event, it is also independent of the source motion.  Therefore, if the data were good enough it would be more correct to combine data with the same $d_S = d_L/(1 + z)$.

 \subsection{Checking Homogeneity}

The LT model requires two physical functions of $r$ to be fully specified, so only models that satisfy {\em both} conditions \er{LTcondRW} are homogeneous.  As seen above, it is perfectly possible to reproduce one observational function, such as $d_L(z)$, with a variety of inhomogeneous models, so a one-function test is not sufficient.  Thus a two-function test is imperative for an unambiguous result; for example the redshift-space density $\sigma(z)$ can distinguish between models that fit the $d_L(z)$ data.  Clearly, then, the MoC procedure will provide an important test of homogeneity, when there is sufficient observational data of high precision and completeness.  If the procedure outputs LT functions that are close to the RW form \er{LTcondRW} (say within 1 sigma), then this is a strong indication of homogeneity.   Checking for homogeneity is so important that we should use all available tests.  Any deep $z$ test of homogeneity will depend on using the correct source evolution functions.  According to \cite{Hel01}, source evolution theories may possibly be tested with detailed multi-colour data.

 \subsection{Other Approaches}

   Although the above papers are the only ones that are seriously directed at eventually using real observational data, \cite{BisHai96} did develop a numerical code based on the characteristic initial value problem, and \cite{RinSus89,Isha04} also discussed the problem in broad terms.  There have been a number papers looking at restricted versions of the `inverse problem' \cite{IgNaNa02,ChuRom06,YoKaNa08} that only tried to fit the $d_L(z)$, and typically assumed this is identically the $\Lambda$CDM-FLRW curve.  Since this only fixed one of the LT physical freedoms, the other was fixed by the authors' choice.  As already noted, \cite{BisHai96,IgNaNa02} did not solve the apparent horizon (AH) problem.  In \cite{VaFlWa06} it was mistakenly suggested that it could not be solved using inhomogeneous models --- see \cite{KrHeCeBo} for corrections.
   
   In \cite{IgNaNa02} they chose $d_L(z)$ to be that of the $\Lambda$CDM-FLRW model with $\Omega_m = 0.3$, $\Omega_\Lambda = 0.7$, and set $M = M_0 r^3$.  They considered models with both $f = f_{RW}$, in which $a$ is not uniform, and $a = 0$, for which $f$ is not uniform.  They were able to find a variety of models that solved their inverse problem, and some of the varying bang-time models had quite reasonable redshift-space density.  They did however encounter difficulties at the AH.

   In \cite{ChuRom06} they assumed $d_L(z)$ has the $\Lambda$CDM form (i.e. that of an RW model with $\Omega_m = 0.3$, $\Omega_\Lambda = 0.7$, $\Omega_k = 0$), $a =$~constant, and $f = H_0 r^2 e^{-2 H_0 r}$, and they succeed in extracting $M(r)$ only up to $z = 0.4$.  Their comments below their eq (32) about the inversion method not probing the geometry or being unstable, actually originate from the not handling near-parabolic models appropriately, and from not identifying the $r$ coordinate freedom.  They also seemed unaware of earlier work on shell crossings in LT models.

   In \cite{YoKaNa08} they also assumed the $d_L$ of a $\Lambda$CDM universe, as well as $a = 0$, and they used the Dyer-Roeder equation to fix the coordinate freedom.  They did overcome the AH problem, though the details are unclear.  Their solution procedure involved integrating outwards from the centre and inwards from the AH, so joining the two parts up involved a `search' through multiple integrations to get a matching.  They tested different degrees of smoothness at the centre, but showed that the results in the outer regions were unaffected.  They also investigated the effect of a Dyer-Roeder clumpiness parameter that depends on $z$, and showed that this could reduce the amount of inhomogeneity needed to fit observations.

   \bp{Evolution of the Redshift}  Detecting how the redshift of sources evolves with time, may become an important method of distinguishing models of SNIa dimming \cite{EllTiv85,Lak07,UzClEl08,YoKaNa08}.

 \subsection{General case}

   The idea of deducing the metric of the universe from observations was first analysed in the classic paper \cite{KriSac66} by Kristian \& Sachs, and followed up in an important review by Ellis et al \cite{ENMSW85}.  Important early ideas appear in \cite{Eth33,Tem38}.  There has actaully been quite a lot of work on this problem \cite{StElNe92a,StNeEl92b,StNeEl92c,SNME92,MaaMat94,MHMS96,AraSto99,AABFS01,ArRoSt01,RibSto03,AlIrRiSt07,HelAlf09,ASAB08}, especially for the spherically symmetric case and its perturbations, though the general case is quite difficult and has not been developed very much.  See a summary in \cite{Hel03}.  There is much to be done here, especially turning the general case into a workable numerical procedure.

 \section{The Szekeres Metric}
 \showlabel{Szek}

   The Szekeres (S) metric \cite{Szek75a,Szek75b} is a very interesting and largely neglected inhomogeneous model.  Like LT, it is synchronous, comoving, and irrotational, with a dust equation of state.  The metric is: 
 \begin{align}
   ds^2 = - dt^2 + \frac{\left( R' - \frac{R E'}{E}\right)^2}{\epsilon + f} \, dr^2
   + \frac{R^2}{E^2} (dp^2 + dq^2) ~,
   \showlabel{ds2Sz}
 \end{align}
 where $\epsilon = -1, 0, +1$, $f = f(r)$ is an arbitrary function of $r$, $E = E(r, p, q)$, $R = R(t, r)$ and ${}' \equiv \partial/\partial r$, and an orthonormal basis for this metric is
 \begin{align}
   e^t{}_t = 1 ~,~~~~~~ e^r{}_r = \frac{(R' - R E'/E)}{\sqrt{\epsilon + f}\;} ~,~~~~~~
   e^p{}_p = \frac{R}{E} = e^q{}_q ~.
   \showlabel{ONBSz}
 \end{align}
 Applying the EFEs, the density and the Ricci and Kretschmann scalars are
 \begin{align}
   \kappa \rho & = \frac{2 (M' - 3 M E' / E)}{R^2 (R' - R E' / E)} ~,
      \showlabel{RhoSz} \\
   {\cal R} & = 4 \Lambda + \kappa \rho ~,
      \showlabel{RicScSz} \\
   {\cal K} & = \kappa^2 \left[ \frac{4}{3} \overline{\rho}^2
      - \frac{8}{3} \overline{\rho} \rho + 3 \rho^2 \right]
      + \frac{4 \Lambda}{3} \left[ 2 \Lambda + \kappa \rho \right] ~,
      \showlabel{KretschSz}
 \end{align}
 where
 \begin{align}
   8 \pi \overline{\rho} = \frac{6 M}{R^3} ~.
   \showlabel{rhoSz}
 \end{align}
 The function $R(t, r)$ has exactly the same dynamics \er{RtSq} and solution (e.g. \er{HypEv}-\er{tEvEC} for $\Lambda = 0$) as for LT.  The function $E$ is given by
 \begin{align}
   E(r, p, q) = \frac{S}{2} \left[ \left( \frac{p - P}{S} \right)^2
      + \left( \frac{q - Q}{S} \right)^2 + \epsilon \right]
      \showlabel{Edef}
 \end{align}
 where functions $S = S(r)$, $P = P(r)$ \& $Q = Q(r)$ are arbitrary.%  
 \footnote{ The function $E$ is often given in the form
 \begin{align}
   E(r,p,q) = A (p^2 + q^2) + 2 B_1 p + 2 B_2 q + C ,
 \end{align}
 where $A(r)$, $B_1(r)$, $B_2(r)$, and $C(r)$ must obey
 \begin{align}
   4(AC - B_1^2 - B_2^2) = \epsilon ~.
 \end{align}
 This last is automatically satisfied by \er{Edef}, so calculations are easier.  Also $S$, $P$, $Q$ have a natural interpretation in the Riemann projection given next.
 }
 The $p$-$q$ 2-surfaces and $E$ will be interpreted below; in brief the constant time 3-spaces are foliated by 2-surfaces of constant coordinate $r$, which have 2-metrics of spheres, planes or pseudo-spheres, depending on the value of $\epsilon$.  

   The $\epsilon = 0, -1$ cases have been ignored until recently \cite{HelKra08,Kra08}.  Although quantities like $r$, $M(r)$ etc do not have the same meaning as in spherically symmetric models, curves of constant $p$ \& $q$ will be called `radial', `p-radial' or `h-radial', and prefixes `p-' and `h-' will indicate quasi-planar and quasi-pseudospherical quantities.

   See \cite{Kra97} for a review of its known properties, or \cite{PleKra06} for an introduction.  See also \cite{HelKra02,HelKra08} for an analysis of the $\epsilon = +1$ and $\epsilon = 0, -1$ cases.  

   \bp{Free Functions}
   The S metric has 6 arbitrary functions $f$, $M$, $a$, $S$, $P$ and $Q$, which allow a rescaling of $r$ plus 5 functions to control the physical inhomogeneity.  For the $\epsilon = 0$ case, the mapping $(S, f, M) \to (S/F, f F^2, M F^3)$ for any $F(r)$ does not change the metric, the density or the evolution equations.  Thus $S(r)$ might as well be set to $1$ with $F = S$.

   \bp{Special Cases}
   The S model contains the LT model when $\epsilon = +1$ and $S$, $P$, $Q$ are all constant.  It therefore contains the same special cases, and has geometric possibilites at least as interesting as LT.  With $E' = 0$ it reduces to the Ellis metric \cite{Elli67}.  It also has a Kantowski-Sachs-type limit, and its null limit is a generalisation of the Kinnersley rocket metric \cite{Hell96b}.  

   \bp{No Killing Vectors}
   This metric has no Killing vectors \cite{BoSuTo77}, but that does not mean it is even close to a general inhomogeneous dust solution.  It is the $r$ dependence of $E$ that destroys any spherical, planar or pseudo-spherical symmetry.  Despite the inhomogeneity of the model, and the lack of Killing vectors, any surface of constant time $t$ is conformally flat \cite{BeEaOl77}.

   \bp{Matching to Vacuum}
   Also, any surface of constant coordinate `radius' $r$ can be joined to a symmetric vacuum metric with spherical, planar or pseudo-spherical symmetry \cite{Bonn76a,Bonn76b,HelKra08}.  This latter means that the S metric generates a symmetric gravitational field ``outside'' each and every constant $r$ shell.  

   \bp{Singularities}   
   The S model has the same singularities --- bang, crunch, shell crossings, shell focussings --- as discussed for the LT model in \S \ref{LTSings}.  The bang and crunch, where $R = 0$, still occur at $t = a$ and $t = a + 2 \pi \tilde{T}$.  Shell crossings however are more complicated, as they occur where $R' - R E'/E = 0$, provided $M' - 3 M E'/E$ and $\epsilon - f$ are not zero.
   
 \subsection{Riemann Projection}
 \showlabel{RiemProj}

   To understand the metric component $(dp^2 + dq^2)/E^2$, we note that the $p$-$q$ 2-surfaces can be transformed to 2-spheres, planes or pseudo-2-spheres by the Riemann projection:
 \begin{align}
   & \epsilon = -1~,~ E > 0:~~ &
         \frac{(p - P)}{S} & = \coth\left(\frac{\theta}{2}\right)
         \cos(\phi) ~, & \nn \\
      && \frac{(q - Q)}{S} & = \coth\left(\frac{\theta}{2}\right) \sin(\phi) ~, &
      \showlabel{Riemprojm} \\
   & \epsilon = -1~,~ E < 0:~~ &
         \frac{(p - P)}{S} & = \tanh\left(\frac{\theta}{2}\right)
         \cos(\phi)~, & \nn \\
      && \frac{(q - Q)}{S} & = \tanh\left(\frac{\theta}{2}\right) \sin(\phi) ~, &
      \showlabel{Riemprojm2} \\
   & \epsilon = ~0:~~ &
         \frac{(p - P)}{S} & = \left(\frac{2}{\theta}\right)
         \cos(\phi) ~, & \nn \\
      && \frac{(q - Q)}{S} & = \left(\frac{2}{\theta}\right) \sin(\phi) ~, &
      \showlabel{Riemproj0} \\
   & \epsilon = +1: ~~\mbox{either}~~ &
         \frac{(p - P)}{S} & = \cot\left(\frac{\theta}{2}\right)
         \cos(\phi) ~, & \nn \\
      && \frac{(q - Q)}{S} & = \cot\left(\frac{\theta}{2}\right) \sin(\phi) ~, &
      \showlabel{Riemprojp} \\
   & \mbox{or}~~~~~~~~~~~~~~~~~~~~ &
         \frac{(p - P)}{S} & = \tan\left(\frac{\theta}{2}\right)
         \cos(\phi)~, & \nn \\
      && \frac{(q - Q)}{S} & = \tan\left(\frac{\theta}{2}\right) \sin(\phi) ~. &
      \showlabel{Riemprojp2}
 \end{align}
 Notice that, with $\theta$ \& $\phi$ ranging over the whole sphere, {\em each} of the spherical transformations (\ref{Riemprojp}) \& (\ref{Riemprojp2}) covers the entire $p$-$q$ plane.   

   In contrast, {\em both} of the pseudospherical transformations (\ref{Riemprojm}) \& (\ref{Riemprojm2}), with $0 \leq \theta \leq \infty$, are required to cover the entire $p$-$q$ plane once; each transformation maps one of the hyperboloid sheets to the $p$-$q$ plane.  To distinguish the sheets, we choose $\theta$ to be negative on one and positive on the other.  Each constant $r$ ``shell'' seems to be a hyperboloid with two ``sheets'', but we shall determine whether both these sheets are needed, or even allowed.  In the planar case, the Riemann projection may be viewed as an inversion of the plane in a circle, or as a mapping of a semi-infinite cylinder to a plane.  These projections are illustrated below.

 \noindent
 \includegraphics[scale = 0.5]{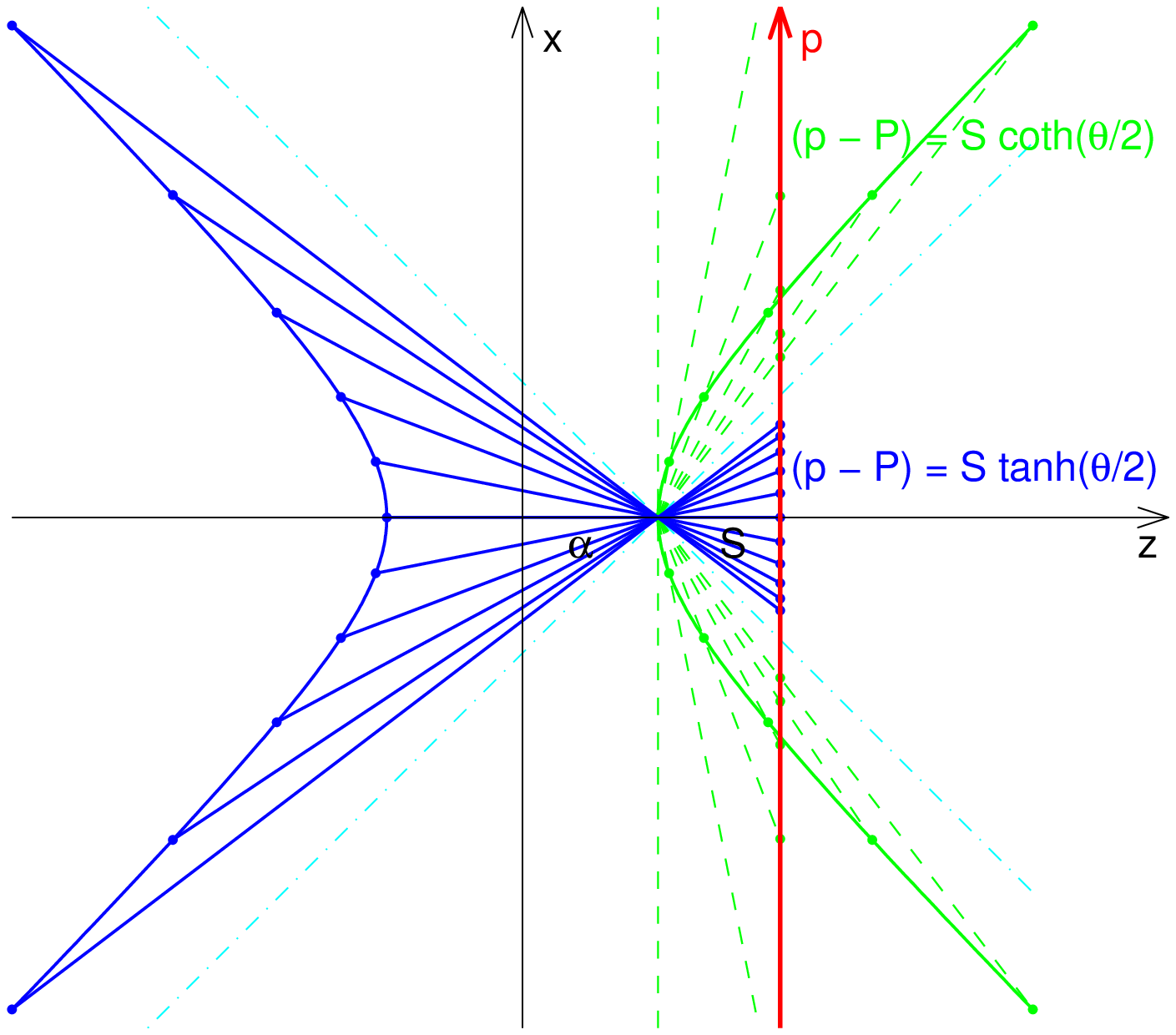}
 \hfill
 \includegraphics[scale = 0.5]{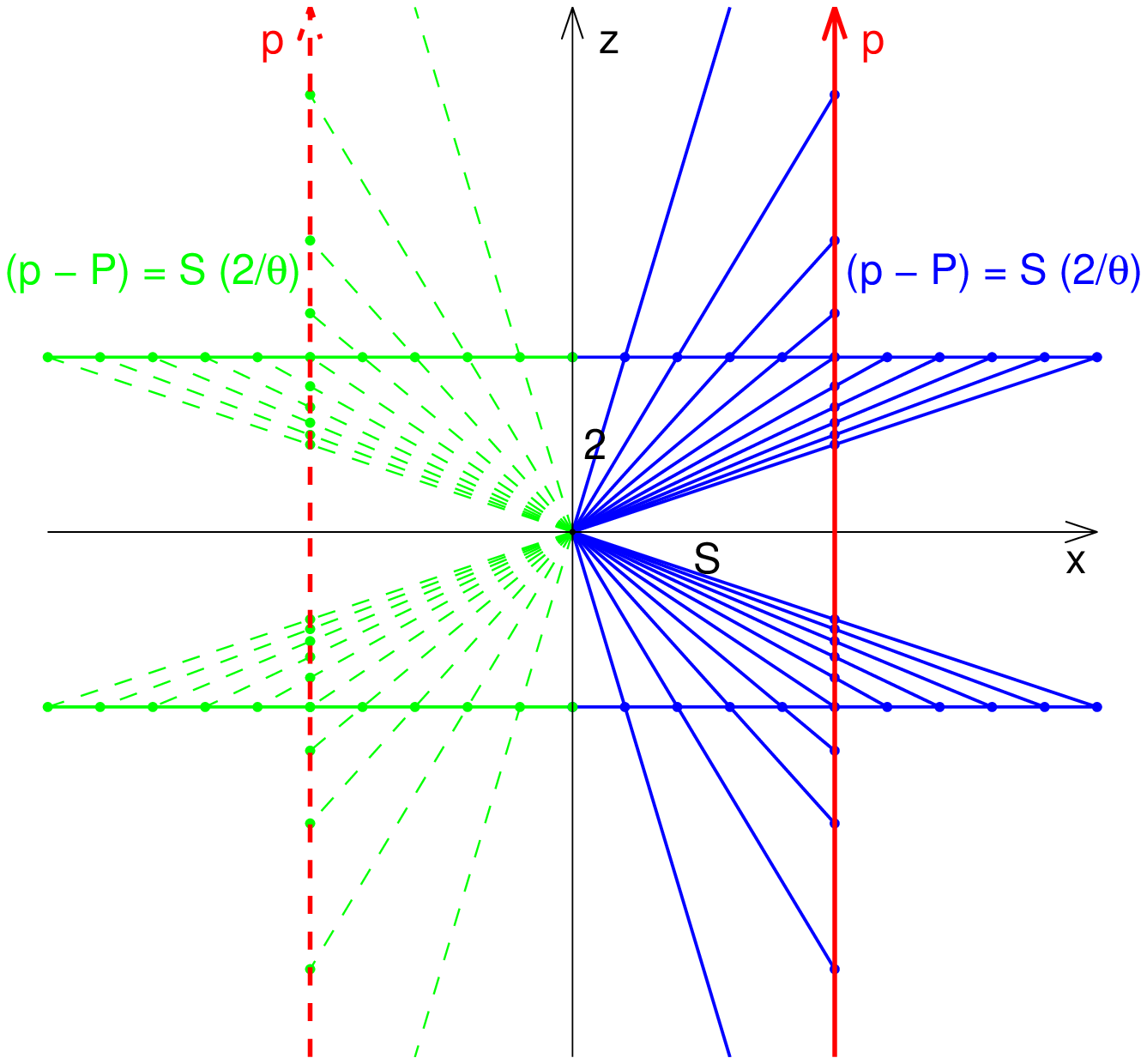} \\
 \includegraphics[scale = 0.55]{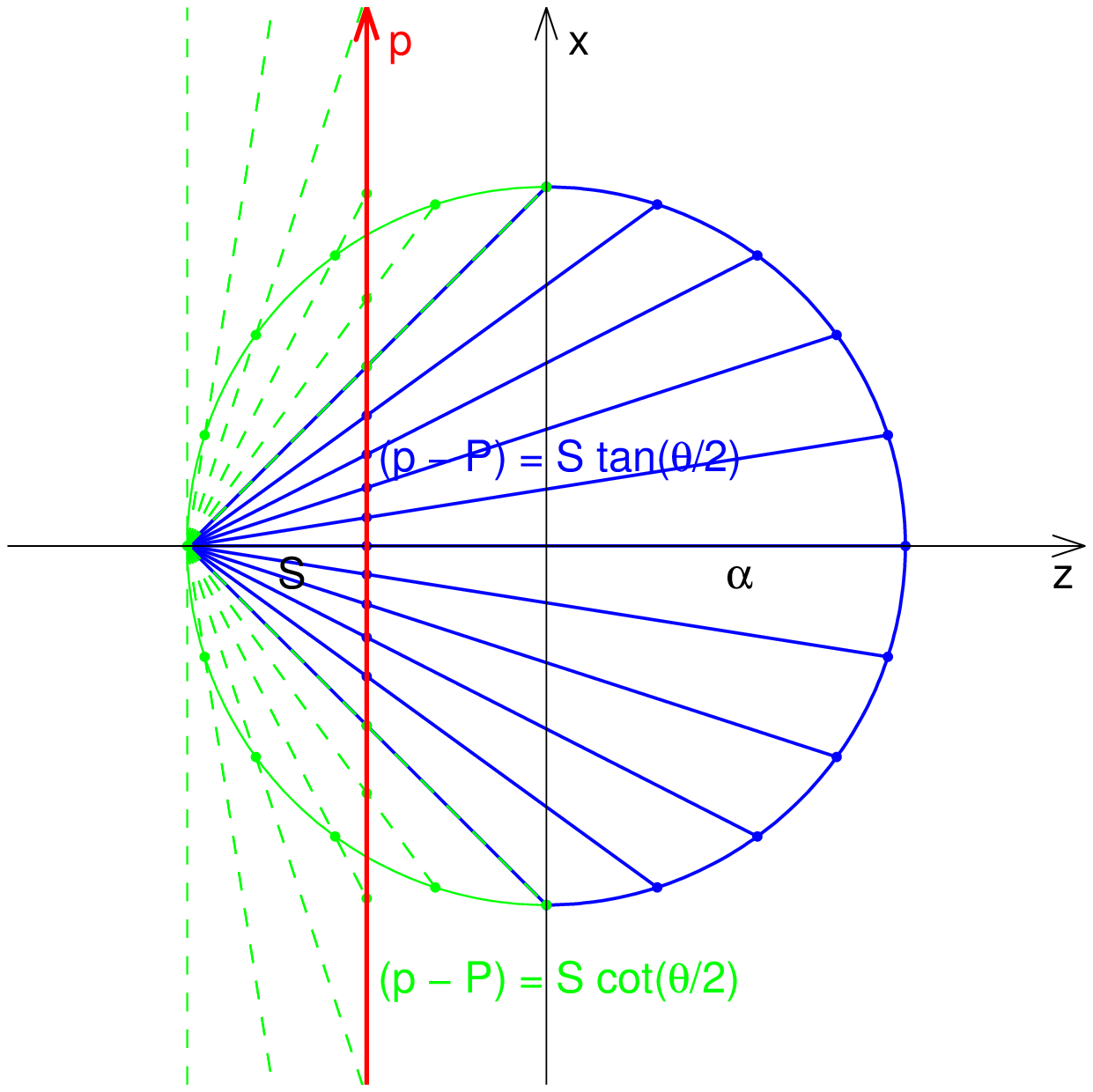}
 \hfill
 \includegraphics[scale = 0.45]{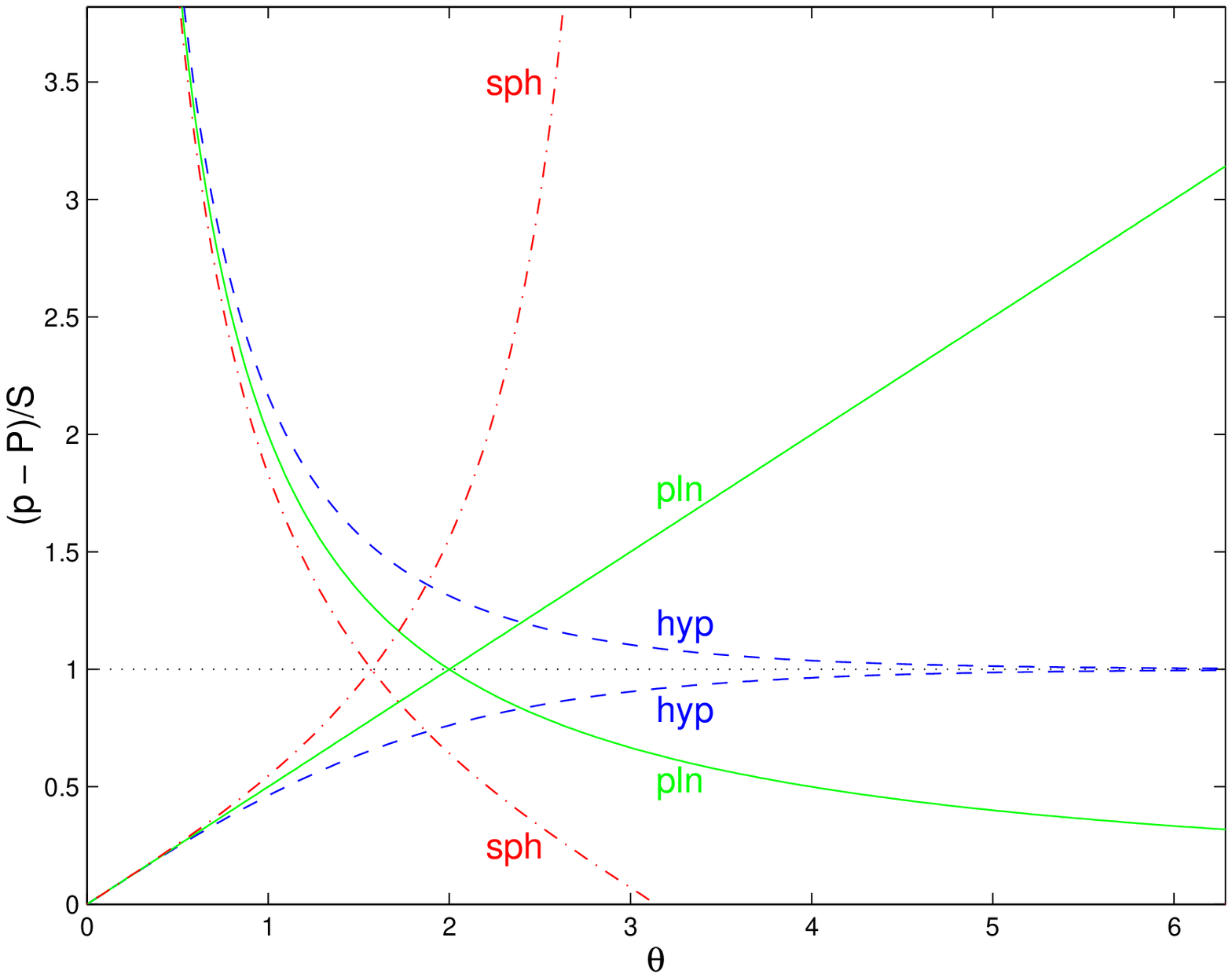}

 \noindent However, the transformations from $(p, q)$ to $(\theta, \phi)$ introduce cross terms in the metric, such as $dr \, d\theta$.  Of course, if $E' = 0$ everywhere, this transformation recovers the LT model.  

 \subsection{Properties of $E$}
 \showlabel{PropE}

   The function $E$ determines how the coordinates $(p,q)$ map onto the 2-d unit pseudosphere, plane or sphere at each value of $r$.  Each 2-surface is multiplied by factor $R = R(t,r)$ that is different for each $r$, and evolves with time. Thus the $r$-$p$-$q$ 3-surfaces are constructed out of a sequence of 2-dimensional spheres, pseudospheres, or planes that are not arranged symmetrically.  Obviously, for $\epsilon = 0,-1$ the area of the constant $t$ \& $r$ 2-surfaces could be infinite, but in the $\epsilon = +1$ case it is $4\pi R^2$.

   In the $p$-$q$ plane, $E$ has circular symmetry about the point $p = P$, $q = Q$, which is different for each $r$.  The $E = 0$ circle
 \begin{align}
   (p - P)^2 + (q - Q)^2 = - \epsilon S^2 ~,
   \showlabel{E=0circle}
 \end{align}
 has $E > 0$ on the outside, but becomes a point if $\epsilon = 0$, and does not exist if $\epsilon = +1$.  The divergence of the metric components $g_{pp}$ and $g_{qq}$ as $E \to 0$ has a
geometric significance that will be discussed below.  The $E' = 0$ locus is also a circle in the $p$-$q$ plane, which can be written
 \begin{align}
   \left( \frac{p - P}{S} + \frac{P'}{S'} \right)^2
      + \left( \frac{q - Q}{S} + \frac{Q'}{S'} \right)^2
      = \frac{(P')^2 + (Q')^2}{(S')^2} + \epsilon ~.
      \showlabel{E'=0circle}
 \end{align}
 For $\epsilon = 0, +1$, this locus always exists, and when $\epsilon = -1$ it
only exists if
 \begin{align}
   (S')^2 < (P')^2 + (Q')^2 ~,   \showlabel{hE'=0exists}
 \end{align}
 with the radius of this circle shrinking to zero as the equality is approached.  It can be shown these two circles always intersect, if they both exist.

To see how $E'/E$ affects the metric and the density, we write $x = E'/E$. Then
in the metric \er{ds2Sz}, $g_{rr}$ is a decreasing function of $x$ provided $x
> R'/R$, while for the density \er{rhoSz} we have
 \begin{align}
   8 \pi \rho & = \frac{6 M}{R^3} \, \frac{(M'/(3 M) - x)}{(R'/R - x)} ~,   \showlabel{Rho2}
 \intertext{so that}
   8 \pi \pd{\rho}{x} & = - \frac{6 M}{R^3} \frac{(R'/R - M'/(3M))}{(R'/R - x)^2} ~,
      \showlabel{drhodx}
 \intertext{and if $x \to \pm \infty$}
   8 \pi \rho & \to \frac{6 M}{R^3} ~.   \showlabel{rholim}
 \end{align}
Therefore at given $r$ and $t$ values, the density varies monotonically with $x = E'/E$.  (The sign of the numerator may possibly change as $R$ evolves.)  If $x$ can diverge, $\rho$ approaches a finite, positive limit.

 \subsection{Spatial 3- \& 2-Geometries}
 \showlabel{2,3Geom}

   It is apparent from the above that $\epsilon$ determines the shape of the constant $(t,r)$ 2-surfaces that foliate the spatial sections:
 \begin{align}
   \epsilon & = +1 && \rightarrow && \mbox{sequence of Riemann spheres} \nn \\
   \epsilon & = -1 && \rightarrow &&
      \mbox{sequence of Riemann hyperboloids} \\ 
   \epsilon & = 0 && \rightarrow && \mbox{sequence of Riemann planes} ~. \nn
 \end{align}
 This is confirmed by the curvature of the $p$-$q$ 2-spaces; the orthonormal basis components of the Riemann \& Ricci tensors and the Kretschmann and Ricci scalars are
 \begin{align}
   {}^2\!R_{(p)(q)(p)(q)} = \frac{\epsilon}{R^2} ~,~~~~~~
   {}^2\!R_{(p)(p)} = {}^2\!R_{(q)(q)} = \frac{\epsilon}{R^2} ~,~~~~~~
   {}^2{\cal K} = \frac{4 \epsilon^2}{R^4} ~,~~~~~~
   {}^2{\cal R} = \frac{2 \epsilon}{R^2} ~,
   \showlabel{pqCrvtrSz}
 \end{align}
 which also show $R$ is a scale length for the curvature.  In fact, it is quite possible to have the three types of foliation in one S model.  The original notation \cite{Szek75a,Szek75b} has a continuous function instead of $\epsilon$.  These 2-surfaces have area
 \begin{align}
   A = R^2 \int \int \frac{\d p \, \d q}{E^2} ~,
 \end{align}
  which is $4 \pi R^2$ when $\epsilon = +1$, but otherwise is infinite.

   Note that $g_{rr} \geq 0$ requires $\epsilon + f \geq 0$, to keep the metric Lorentzian, and so
 \begin{align}
        f & > 0 && \rightarrow & \epsilon & = +1, 0, -1 \nn \\
        f & = 0 && \rightarrow & \epsilon & = +1, 0 \\
   -1 < f & < 0 && \rightarrow & \epsilon & = +1 ~. \nn
 \end{align}
 Clearly, the 3-d geometry determined by $f$ may restrict the possible foliations.  For example, you can't foliate a positively curved space with hyperboloids, but you can foliate a negatively curved space with spheres.  

   Calculating the orthonormal basis components of the Riemann and Ricci tensors and scalars of the $r$-$p$-$q$ 3-spaces, we find
 \begin{align}
   {}^3\!R_{(r)(p)(r)(p)} & = {}^3\!R_{(r)(q)(r)(q)} = \frac{- 1}{R}
         \frac{(f'/2 - f E'/E)}{(R' - R E'/E)} ~,~~~~~~
      {}^3\!R_{(p)(q)(p)(q)} = \frac{- f}{R^2} ~,
      \showlabel{rpqRiemSz} \\
   {}^3\!R_{(p)(p)} & = {}^3\!R_{(q)(q)} = 
      \frac{-1}{R} \left( \frac{(f'/2 - f E'/E)}
         {(R' - R E'/E)} + \frac{f}{R} \right) ~'
      \showlabel{rpqRicciSz} \\
   {}^3{\cal K} & = 
      \frac{4}{R^2} \left( \frac{2 (f'/2 - f E'/E)^2}
         {(R' - R E'/E)^2} + \frac{f^2}{R^2} \right) ~,
      \showlabel{rpqKretschSz} \\
   {}^3{\cal R} & =
      \frac{-2}{R} \left( \frac{2 (f'/2 - f E'/E)}
         {(R' - R E'/E)} + \frac{f}{R} \right) ~.
      \showlabel{rpqRicScSz}
 \end{align}
%   Calculating the various curvature components for the constant $t$ spatial sections of the S metric, we find
% \begin{align}
%   {}^3\!R_{rprp} & = {}^3\!R_{rqrq} = \frac{- R^2 f}{E^2 (\epsilon + f)}
%         \left( \frac{R'}{R} - \frac{E'}{E} \right)
%         \left( \frac{f'}{2 f} - \frac{E'}{E} \right) \\
%   {}^3\!R_{pqpq} & = \frac{- R^2 f}{E^4} \\
%   {}^3\!R_{pp} & = {}^3\!R_{qq} = 
%      - \frac{f}{E^2} \left( \frac{(f'/2f - E'/E)}
%         {(R'/R - E'/E)} + 1 \right) \\
%   {}^3{\cal K} & = 
%      \frac{4 f^2}{R^4} \left( \frac{2 (f'/2f - E'/E)^2}
%         {(R'/R - E'/E)^2} + 1 \right) \\
%   {}^3\!R & =
%      - \frac{2 f}{R^2} \left( \frac{2 (f'/2f - E'/E)}
%         {(R'/R - E'/E)} + 1 \right) ~.
% \end{align}
 The flatness condition ${}^3\!R_{(a)(b)(c)(d)} = 0$ requires just%
 \footnote{If instead the coordinate dependent condition ${}^3\!R_{abcd} = 0$ is used, one gets a more complicated result \cite{HelKra08}.}
 \begin{align}
   f = 0 = f' ~.
      \showlabel{3FlatCondBasis}
 \end{align}
 This is not possible for $\epsilon = -1$.  When $\epsilon = 0$, \er{3FlatCondBasis} would make $g_{rr}$ diverge unless $R' - RE'/E = 0$, which in turn would make $\rho$ diverge unless $M' - 3ME'/E = 0$.  It will be shown this is only possible as an asymptotic limit.
% \begin{align}
%   & & \epsilon & \neq 0:~~ & & f = 0 = f' ~~\mbox{and}~~
%      \left| \frac{f'}{f} \right| < \infty ~,
%      \showlabel{3FlatCondepsnot0} \\
%   & & \epsilon & = 0:~~ & & f = 0 = E' = f' = R' ~~\mbox{and}~~
%      \left| \frac{(R' - R E'/E)}{f} \right| < \infty ~,
%      \showlabel{3FlatCondeps0}
% \end{align}
% and the latter is only possible as a limit, or as a Kantowski-Sachs type Szekeres model \cite{Hell96b}.  
% Note that $E' = 0$ implies all of $S' = 0$, $P' = 0$, $Q' = 0$.

   For the $r$-$p$ 2-spaces we find
 \begin{align}
   {}^2\!R_{(r)(p)(r)(p)} & = {}^2\!R_{(r)(r)} = {}^2\!R_{(p)(p)}
      = \frac{\sqrt{{}^2{\cal K}}\;}{2} = \frac{{}^2{\cal R}}{2} \nn \\
   & = \frac{1}{R}
      \left( \frac{E_q (E'_q - E' E_q / E) - (f'/2 - f E' / E)}{(R' - R E'/E)} \right) ~,
      \showlabel{rpCrvtrSz} \\
      \mbox{where}~~~~~~~~ & E_p = \pd{E}{p} ~,~~~~~~ E_q = \pd{E}{q} ~.
 \end{align}
% \begin{align}
%   {}^2\!R_{pqpq} & = \frac{R}{E^2 (\epsilon + f)} \left( R' - \frac{R E'}{E} \right)
%      \left( E_q \left( E_q' - \frac{E' E_q}{E} \right)
%      - \left( \frac{f'}{2} - \frac{f E'}{E} \right) \right) \\
%   {}^2\!R_{rr} & = \frac{1}{R (\epsilon + f)} \left( R' - \frac{R E'}{E} \right)
%      \left( E_q \left( E_q' - \frac{E' E_q}{E} \right)
%      - \left( \frac{f'}{2} - \frac{f E'}{E} \right) \right) \\
%   {}^2\!R_{pp} & = \frac{R}{E^2}
%      \left( \frac{E_q (E'_q - E' E_q / E) - (f'/2 - f E' / E)}{(R' - R E'/E)} \right) \\
%   {}^2{\cal K} & = \frac{4}{R^2}
%      \left( \frac{\left\{ E_q (E'_q - E' E_q / E) - (f'/2 - f E' / E) \right\}^2}{(R' - R E'/E)^2} \right) \\
%   {}^2\!R & = \frac{2}{R}
%      \left( \frac{E_q (E'_q - E' E_q / E) - (f'/2 - f E' / E)}{(R' - R E'/E)} \right) ~.
% \end{align}
 For these surfaces to be flat requires $E_q (E'_q - E' E_q / E) - (f'/2 - f E' / E) = 0$, and the only solution that can hold over an entire 2-surface is $f' = 0 = E'$.  This is because $E$ \& $E'$ depend on $p$, but $E_q$ \& $E'_q$ don't.  Note that $E' = 0$ implies all of $S' = 0 = P' = Q'$.  Obviously, these surfaces may be curved, even when the $r$-$p$-$q$ 3-space they foliate is flat.  

   \bp{RW in Szekeres Coordinates}
   Since the RW metrics can be written in the Szekeres form, it is useful to look at the transformations between Szekeres and standard RW coordinates --- see \cite{HelKra08} for a discussion.  The $k = -1$ case allows all three types of foliation, which are compared below in a constant $t$, $\phi = 0, \pi$ slice.  Blue curves are for the $\epsilon = +1$ case, red for $\epsilon = 0$ and green for $\epsilon = -1$.  Note that there's distortion, as a negatively curved 2-surface cannot be properly represented on a plane --- notably orthogonal lines do not look orthogonal.

 \centerline{%
 \includegraphics[scale = 1]{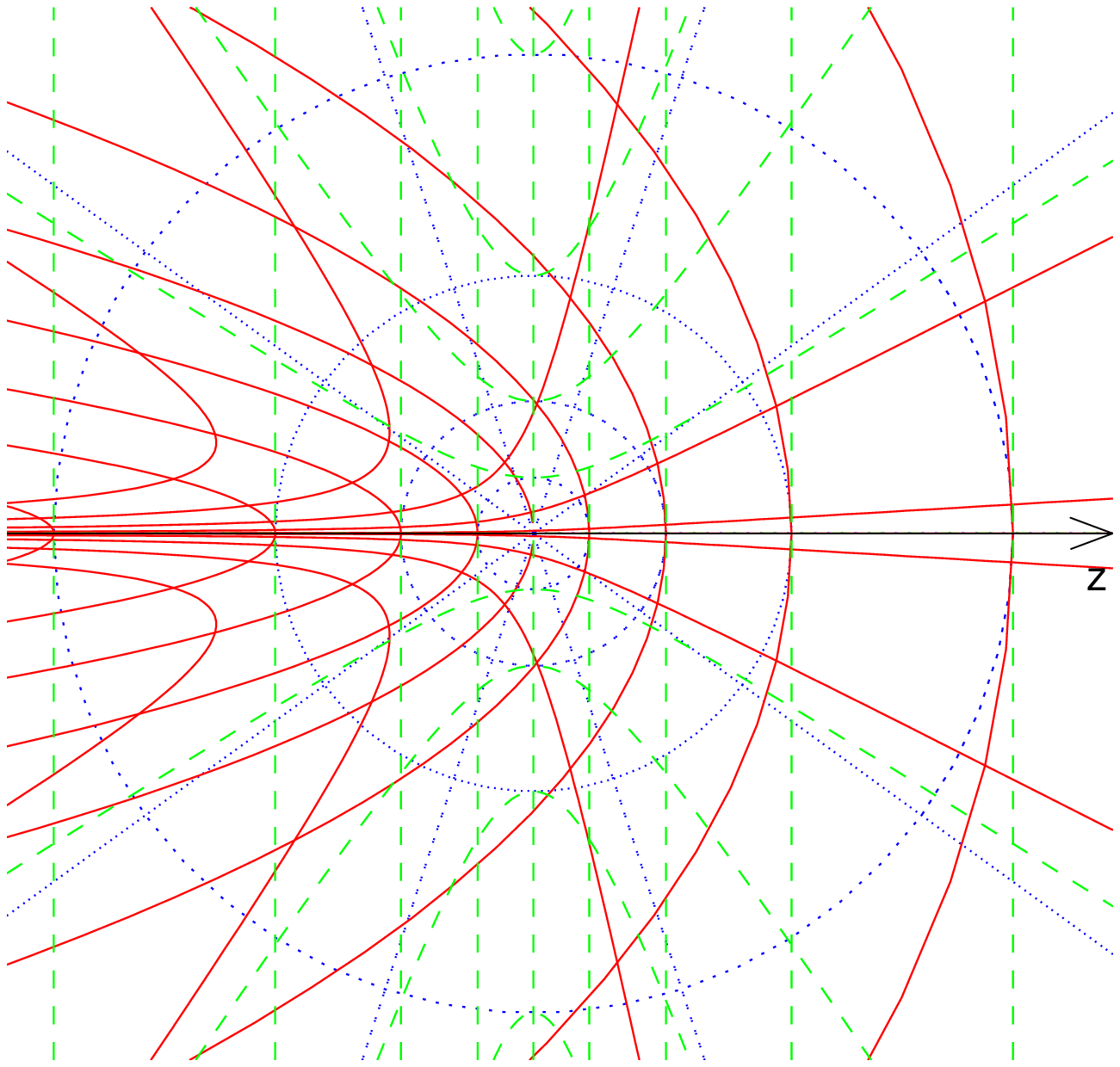}%
 }

 \subsection{Quasi Spherical Case}

   \bp{Dipole}
   The function $E$ describes a dipole distribution \cite{Szek75b,deS85,HelKra02} round the 2-sphere at each $r$ value, having $(E'/E)_{max} = - (E'/E)_{min}$ located at antipodal points, and $E' = 0$ on a great circle mid way inbetween.  From \er{Edef} and \er{Riemprojp}-\er{Riemprojp2} we find
 \begin{align}
   E & = \frac{S}{1 - \cos \theta},   \showlabel{E-theta} \\
   E' & = - \frac{S' \cos \theta + \sin \theta (P' \cos \phi + Q' \sin \phi)}
      {1 - \cos \theta},   \showlabel{E'-thetaphi}
 % \\
 %   E'' & = - \frac{S'' \cos \theta + \sin \theta (P'' \cos \phi + Q'' \sin \phi)}
 %         {(1 - \cos \theta)} \nn \\
 %      & + 2 \left( \frac{S'}{S} \right) \left( \frac{S' \cos \theta
 %         + \sin \theta (P' \cos \phi + Q' \sin \phi)}{(1 - \cos \theta)} \right) \nn \\
 %      & + \frac{(S')^2 + (P')^2 + (Q')^2}{S} ~.
 \end{align}
 so the locus $E' = 0$,
 \begin{equation}
   S' \cos \theta + P' \sin \theta \cos \phi + Q' \sin \theta \sin \phi = 0 ~,   \showlabel{E'0eq}
 \end{equation}
 is a great circle of the $\theta$-$\phi$ sphere.  The locations of the extrema of $E'/E$ are found by setting
 \begin{align}
   \frac{\partial (E'/E)}{\partial \phi} = 0 ~,~~~~~~
   \frac{\partial (E'/E)}{\partial \theta} = 0 ~,
 \end{align}
 which give
 \begin{align}
   \tan \phi_e & = \frac{Q'}{P'} ~~~~\Rightarrow~~~~
      \cos \phi_e = \epsilon_1 \frac{P'}{\sqrt{(P')^2 + (Q')^2}\;}, \qquad \epsilon_1 = \pm 1 ~,
      \showlabel{TanPhiX} \\
   \tan \theta_e & = \frac{P' \cos \phi_e + Q' \sin \phi_e}{S'}
      = \epsilon_1 \frac{\sqrt{(P')^2 + (Q')^2}\;}{S'}
      ~~~~\Rightarrow~~~~
      \showlabel{TanThetaX} \\
   \cos \theta_e & = \epsilon_2 \frac{S'}{\sqrt{(S')^2 + (P')^2 + (Q')^2}\;} ~,
   \qquad \epsilon_2 = \pm 1 ~,
 \end{align}
 where $\epsilon_1$ and $\epsilon_2$ are independent of $\epsilon$, and the extreme value is
 \begin{align}
   \left( \frac{E'}{E} \right)_{\rm extreme} =
      - \epsilon_2 \frac{\sqrt{(S')^2 + (P')^2 + (Q')^2}\;}{S} ~.
      \showlabel{E'EextremeQS}
 \end{align}
 These two points are symmetrically located relative to the $E' = 0$ circle, forming an equator and two poles --- a dipole.  Naturally, the dipole orientation varies with $r$.  By \er{ds2Sz} and \er{rhoSz} $E$ also creates a dipole variation in the $\sqrt{g_{rr}}\; dr$ metric interval and the density around each constant $t$-$r$ 2-sphere.  The distance between constant $r$ shells varies with $(p, q)$; $R E'/E$ is the correction to the radial separation $R'$ of neighbouring shells, $R S' / S$ is the forward $(\theta = 0)$ displacement, and $R P' / S$ \& $R Q' / S$ are the two sideways displacements $(\theta = \pi/2,~ \phi = 0)$ \& $(\theta = \pi/2,~ \phi = \pi/2)$.  
 \[
 \begin{tabular}{cccc}
   $E'/E$ &               & $g_{rr}$ & $\rho$ \\
   \hline
   max            & $\rightarrow$ & min      & min \\
   min            & $\rightarrow$ & max      & max
 \end{tabular}
 \]
The interpretation is that the Szekeres 3-spaces are constructed from a sequence of non-concentric 2-spheres, each having a density distribution that is exactly what's needed to generate a spherical field around a new centre.  Here we show a section through a set of spheres, the dipole on one 2-sphere, and a selection of possible $r$-$p$ surfaces at some moment in time, as well as the dipole on a single spherical shell, and some possible $r$-$\phi$ surfaces (of constant $t$ and $\theta = \pi/2$.

 \noindent
 ${}$ \hfill
 \includegraphics[scale = 0.4]{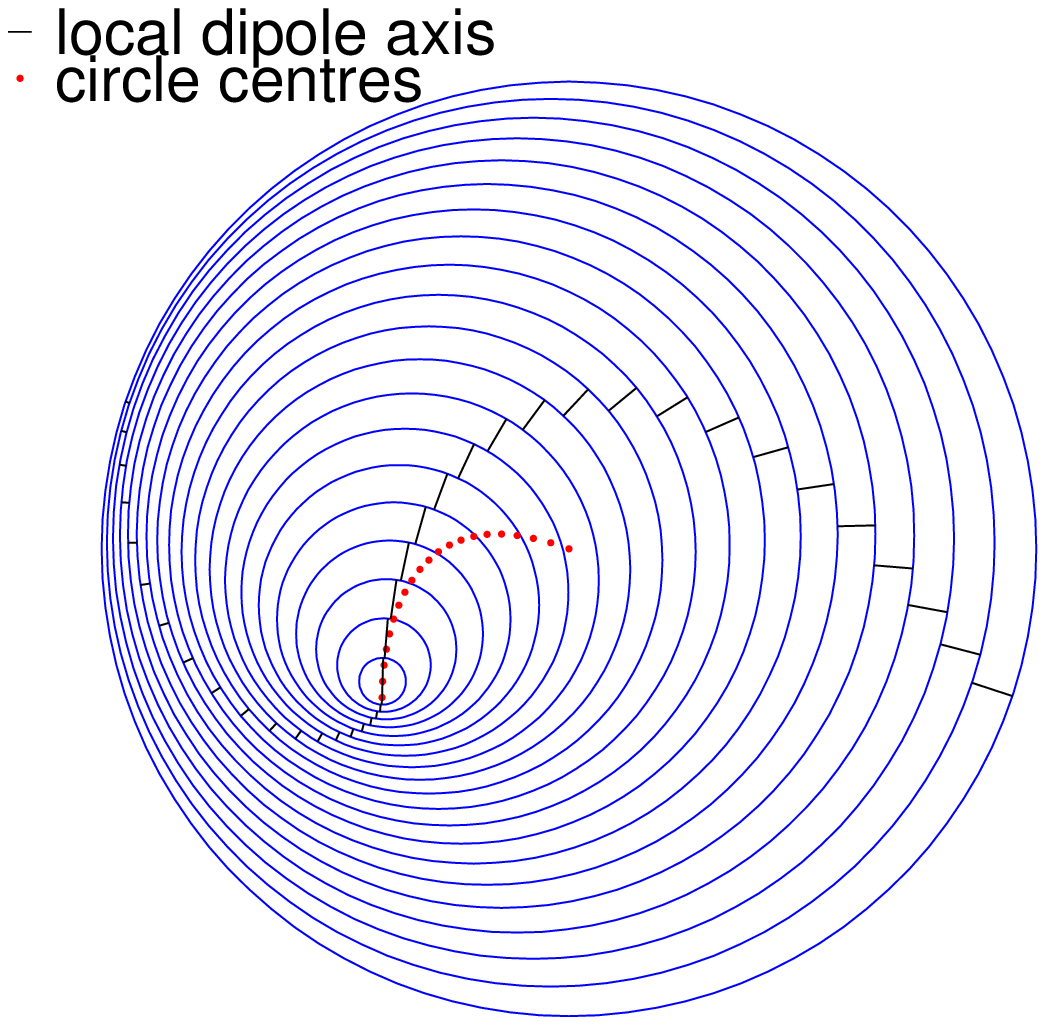}
 \hfill
 \includegraphics[scale = 0.5]{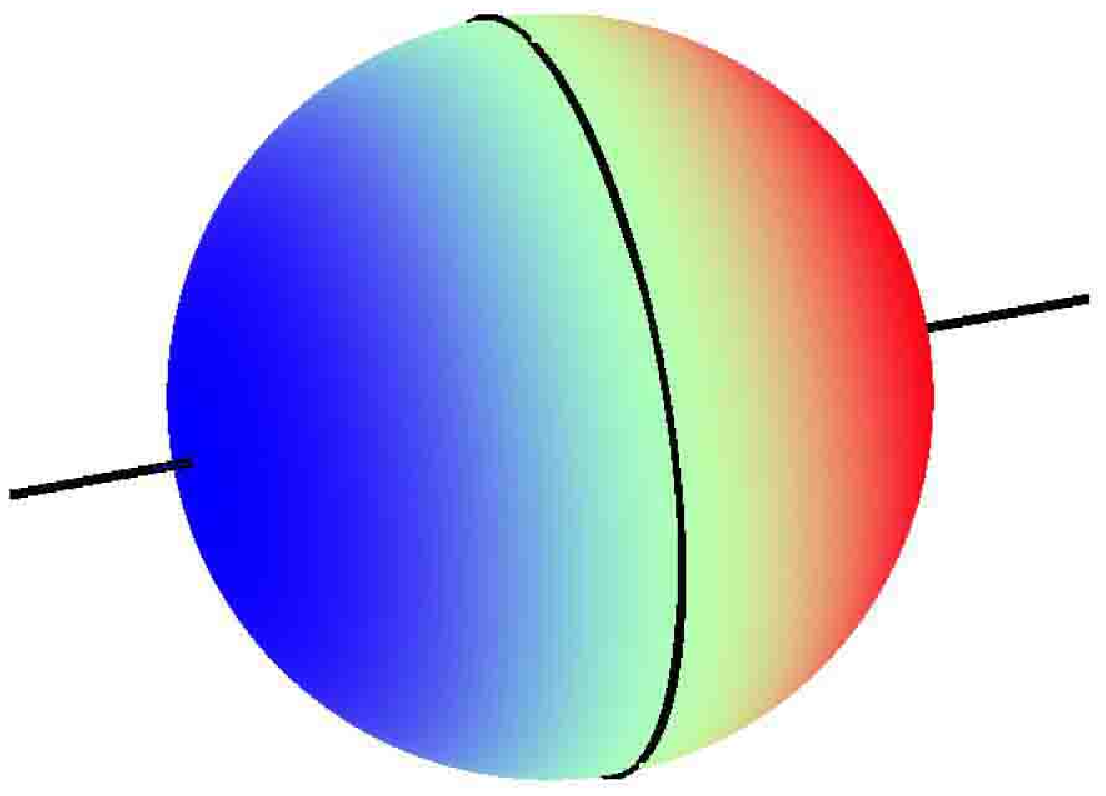}
% \resizebox{4mm}{!}{%
% \includegraphics[0, 0][20mm, 14.156mm]{SDipole.jpg}
% }
 \hfill ${}$

 \noindent 
 \includegraphics[scale = 0.4]{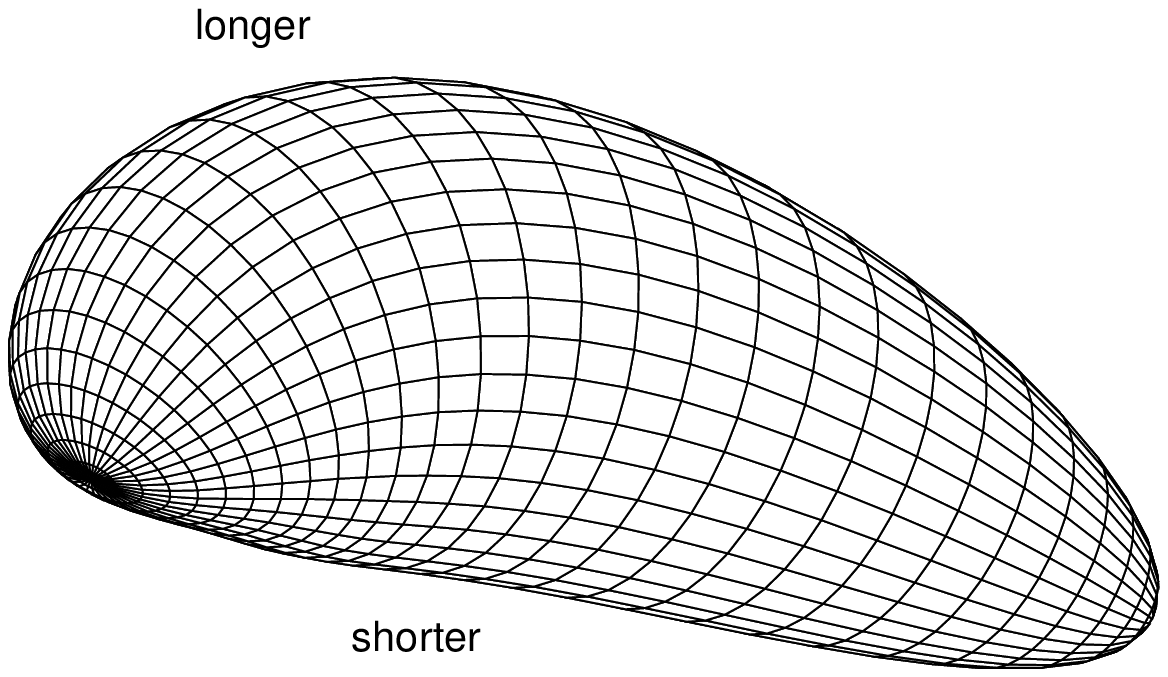}
 \hfill
 \includegraphics[scale = 0.4]{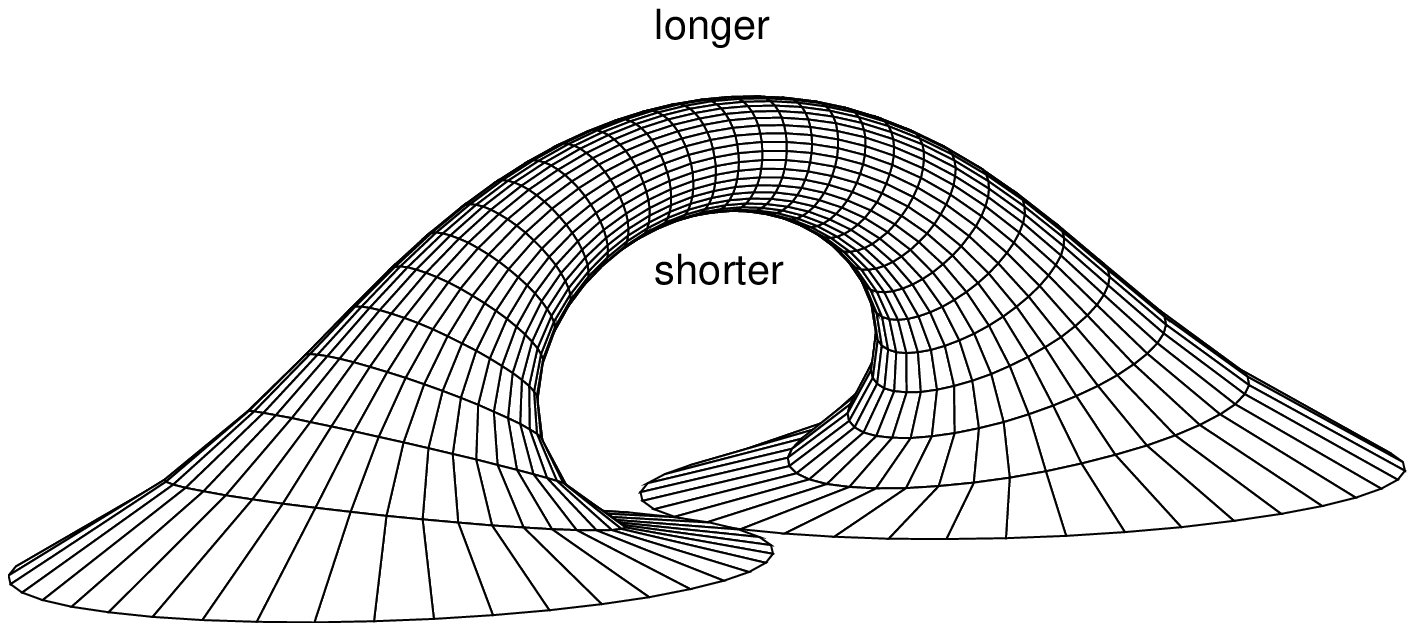}
 \hfill
 \includegraphics[scale = 0.27]{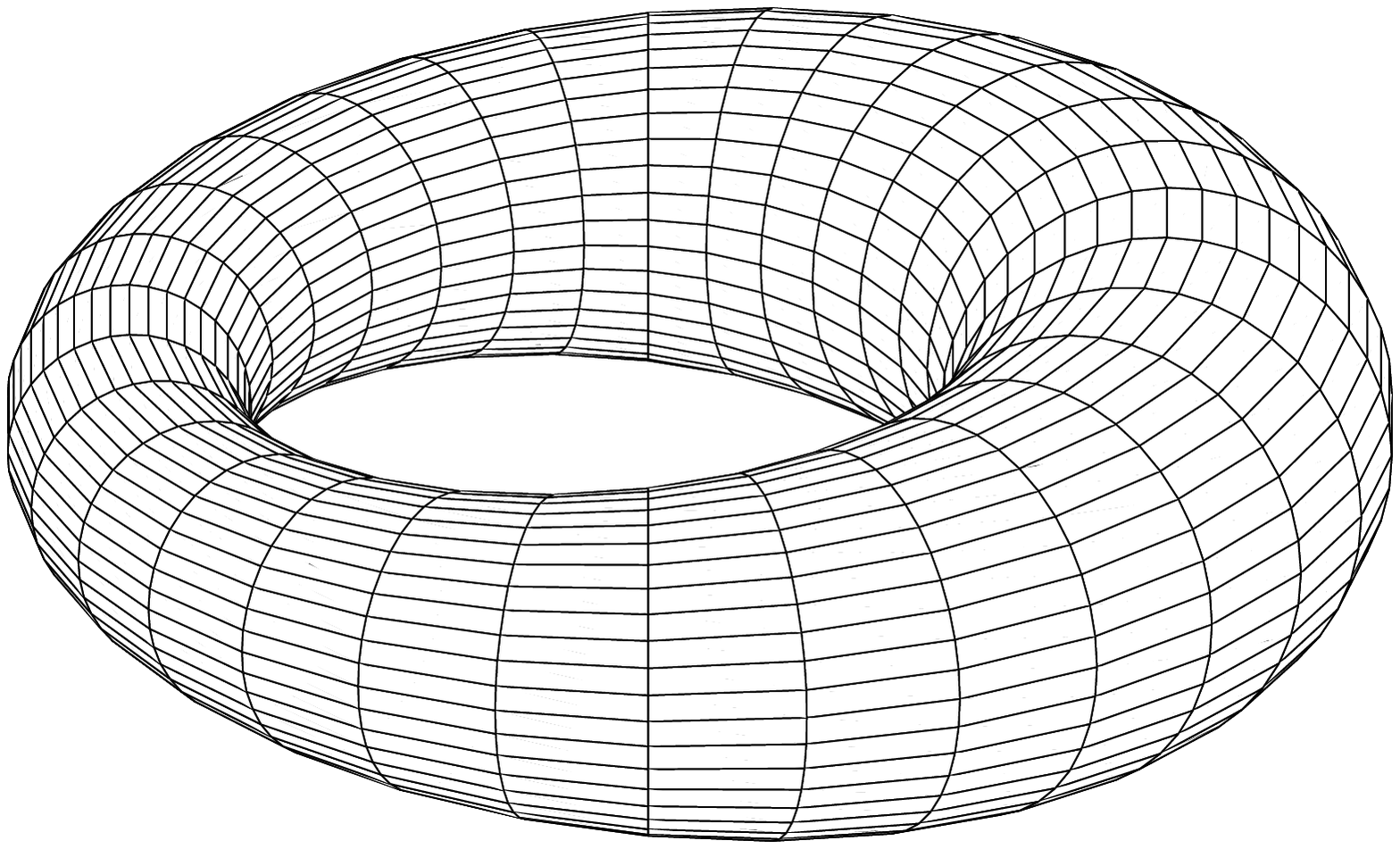}

   \bp{Regularity}   
   The conditions for regular origins, for regular spatial extrema, and for the avoidance of shell crossings are similar to those for LT models --- see section \ref{LTRegCond} --- except with further conditions on the new arbitrary functions $S$, $P$ and $Q$.  These are laid out in \cite{HelKra02}, and the case of non-zero $\Lambda$ is considered in \cite{ChaDeb08}.  Near an origin, $R \to 0$, regularity requires
 \begin{align}
   M \sim R^3 ~,~~~~
   f \sim R^2 ~,~~~~
   S \sim R^n ~,~~~~
   P \sim R^n ~,~~~~
   Q \sim R^n ~,~~~~
   0 \leq n \leq 1 ~.
   \showlabel{OrigCondSz}
 \end{align}

   The non-concentric nature of the constant $r$ shells means that shell crossings are more complex than in LT models.  Adjacent $r$-shells will first intersect at the point where $g_{rr}$ is minimum, and as time goes by the two will intersect on a circle parallel to the $E' = 0$ great circle --- i.e. aligned with the dipole.  See \cite{HelKra02}, and appendix \ref{NoShCr} for the conditions to avoid them.

   For a regular extremum we require
 \begin{align}
   f = - 1 ~,~~~~ \mbox{and}~~~~ M' = f' = a' = S' = P' = Q' = 0
 \end{align}
 and the conditions for no shell crossings have to be modified --- see \cite{HelKra02}.

   \bp{Apparent Horizons}
   According to the standard definition, surface $\Sigma$ is trapped if, for {\em any} null vector field $k^c$, $k^b k_b = 0$, we have
 \begin{align}
   \left. k^a{}_{;a} \right|_\Sigma < 0 ~,
 \end{align}
 and the apparent horizon (AH) is the boundary of the trapped region
 \begin{align}
   \left. k^a{}_{;a} \right|_{AH} = 0 .
 \end{align}
 Now null vectors that are momentarily `radial', $k^d  = K(t,r,p,q) \big( (R' - R E'/E), j W, 0, 0 \big)$, $j = \pm 1$, are also geodesic, $k^b \nabla_b k^a = 0$, if
 \begin{align}
   K' = - \frac{K (R' - R E'/E)'}{(R' - R E'/E)}
   - \frac{j}{W} \left( \dot{K} \left( R' - \frac{R E'}{E} \right)
   + 2 K \left( \Rt' - \frac{\Rt E'}{E} \right) \right) ~.
 \end{align}
 This together with \er{RtRtr} gives
 \begin{align}
   k^a{}_{;a} = \frac{2 K}{R} \left( R' - \frac{R E'}{E} \right) \left( \Rt + j W \right)
   = \frac{2 K}{R} \left( R' - \frac{R E'}{E} \right)
      \left( \ell \sqrt{ \frac{2 M}{R} + f + \frac{\Lambda R^2}{3}}\; + j \sqrt{1 + f}\; \right)
   ~,
 \end{align}
 so the expansion of these geodesics is zero when \er{LamRhMEq} holds and $\ell = - j$; that is, for incoming rays in an expanding region, or outgoing rays in a collapsing region.  Thus \cite{Szek75b} found $R = 2M$ is the apparent horizon when $\Lambda = 0$.

   The approach in \cite{HelKra02} was a bit different, as it was not required that $k^b$ be geodesic, and it was rather determined whether or not null paths were moving to shells of larger areal radius $R$.  It was established that, at any given point, the constant $r$ shells are traversed most rapidly by null vectors pointing `radially'.  (Null paths that stay radial are not geodesic in general.)  It was then found that the locus where $\tdil{R}{r} = 0$ along a radial null direction (geodesic or not) is not coincident with a constant $r$ shell, and is $p$-$q$ dependent.  This was called the `absolute apparent horizon' in \cite{BoKrHeCe}.  Not surprisingly, on any constant $r$ shell, the $\tdil{R}{r} = 0$ locus is (the history of) a circle aligned with the local dipole direction.

   \bp{Wormholes}
   We know light can't quite get through the Schwarzschild-Kruskal-Szekeres (SKS) `wormhole', and we know that dense LT `wormholes' are even less traversibile \cite{Hell87}.  But, if a dense Szekeres wormhole can be bent round round as shown above, so one side is shorter than the other, does that make it easier for light to get through on the shorter side?  In \cite{HelKra02} it was found light still can't get through, and ray paths were calculated and plotted for several models.  Shown here are some light paths (R) and apparent horizons (A) in a Szekeres `wormhole', showing fast (f) and slow (s) directions.

 \centerline{
 \includegraphics[scale = 0.55]{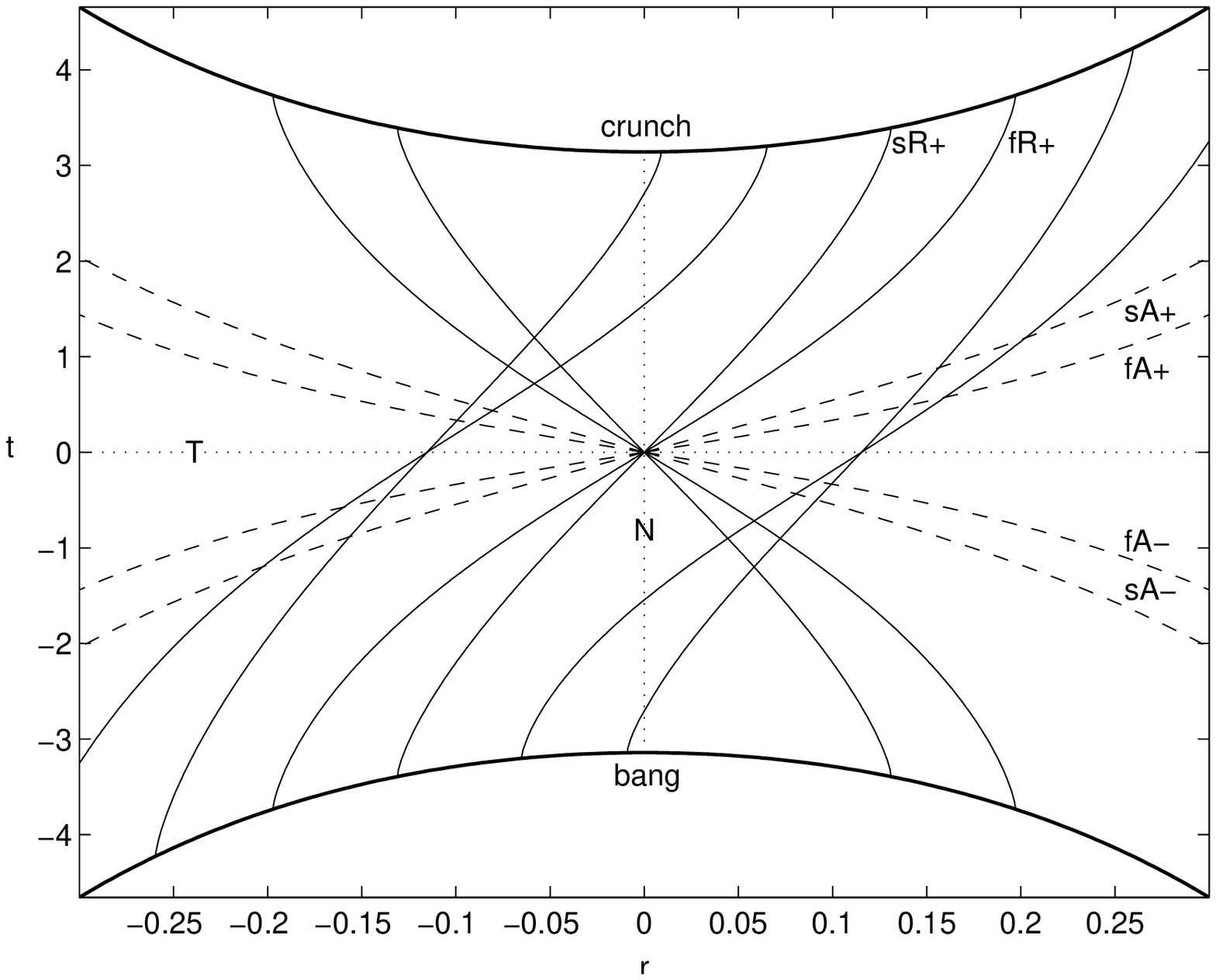}
 }

   Now since Szekeres spaces can bend round, this prompts the question of whether we make a handle topology by joining up the two sides across a boundary.  The Darmois matching conditions specify how to splice metrics together.  (Actually, we don't need the embedding to work, or the amount of bending to be sufficient, as long as the Darmois junction conditions are satisfied.)  However it was found in \cite{HelKra02} that the matching doesn't work, without creating surface layer.  This result includes the case of spherical vacuum --- so the idea of wormhole shortcuts --- commonly suggested in context of the SKS geometry --- is in fact impossible within Szekeres metrics.

 \subsection{Quasi-Pseudo-Spherical Case}

   When $\epsilon = -1$, the constant $r$ surfaces are not closed, and the physical and geometric meaning of $R$ and $M$ have to be re-thought.  We lay out some basic properties here, and attempt an interpretation later.
   
   \bp{h-Dipole}
   Recall that each shell of constant $t$ \& $r$ is a two-sheeted hyperboloid of revolution.  Using \er{Riemprojm} \& \er{Riemprojm2}, we can write
 \begin{align}
   E   & = \frac{\nu \, S}{\cosh \theta - \nu} ~,
      \showlabel{E-thetah}   \\ \nn \\
   E'  & = - \frac{S' \cosh \theta
      + \sinh \theta (P' \cos \phi + Q' \sin \phi)}{\cosh \theta - \nu} ~,
      \showlabel{E'-thetaphih} \\
 %   E'' & = - \frac{S'' \cosh \theta
 %         + \sinh \theta (P'' \cos \phi + Q'' \sin \phi)}{(\cosh \theta - \nu)}
 %         \nn \\ \nn \\
 %       &~~~~ + 2 \left( \frac{S'}{S} \right) \left( \frac{S' \cosh \theta
 %          + \sinh \theta (P' \cos \phi + Q' \sin \phi)}
 %          {(\cosh \theta - \nu)} \right)
 %         \nonumber \\ \nn \\
 %       &~~~~ - \frac{((S')^2 - (P')^2 - (Q')^2)}{S} ~,
 %         \showlabel{E''-thetaphih} \\
   \nu & = {\rm sign}(E) ~.
 \end{align}
 The $E= 0$ circle corresponds to $\theta \to \pm \infty$, and its neighbourhood represents the asymptotic regions of the two sheets of the hyperboloid of \er{Riemprojm} and \er{Riemprojm2}.  

   The locus $E' = 0$ is
 \begin{align}
   S' \cosh \theta + P' \sinh \theta \cos \phi + Q' \sinh \theta \sin \phi = 0 ~,
      \showlabel{E'=0h}
 \end{align}
 which only has a solution if (\ref{hE'=0exists}) holds.  It is a geodesic of the $p$-$q$ 2-space, and can be pictured as the intersection of a plane with a right hyperboloid.

   Writing 
 \begin{align}
   \frac{E'}{E} = - \nu \, \frac{S' \cosh \theta + \sinh \theta (P' \cos \phi
   + Q' \sin \phi)}{S}
   \showlabel{ErE-thetaphi} ~,
 \end{align}
 the extrema of $E'/E$ are 
 \begin{equation}
   \left( \frac{E'}{E} \right)_{\rm extreme} =
      - \epsilon_2 \,\nu \frac{\sqrt{(S')^2 - (P')^2 - (Q')^2}}{S} ~,
      \showlabel{E'EextremeQH}
 \end{equation}
where $\epsilon_2 = {\rm sign}(S')$.  These extrema only exist at finite $\theta$ if
 \begin{align}
   (S')^2 > (P')^2 + (Q')^2 ~,   \showlabel{maxE'/Eexists}
 \end{align}
 which is the opposite of (\ref{hE'=0exists}); so on a given constant $r$ shell, either $E' = 0$ exists, or the extrema of $E'/E$ exist, but {\em not both}.  

   It follows from (\ref{ErE-thetaphi}) that this extremum is a maximum where $E'/E$ is negative, and a minimum where $E'/E$ is positive.  Thus, for each constant $r$ hyperboloid, on the sheet with $E S' < 0$ (i.e. $\nu \epsilon_2 = -1$), $E'/E$ has a positive minimum and goes to $+\infty$ as $|\theta| \to \infty$, while on the sheet with $E S' > 0$, $E'/E$ has a negative maximum and goes to $-\infty$.  We now specify that $\theta < 0$ on the $E < 0$ sheet.  From the foregoing considerations, if \er{maxE'/Eexists} holds, then $E'/E$ is the pseudospherical equivalent of a dipole, having a negative maximum on one sheet and a positive minimum on the other, but diverging in the asymptotic regions of each sheet near $E = 0$.

   We see in the metric \er{ds2Sz} that $R E'/E$ is the correction to the separation $R'$, along the $r$ curves, of neighbouring constant $r$ shells, meaning that the hyperboloids are centered differently and are ``non concentric'', as sketched below.  In particular $R S' / S$ is the forward displacement, and $R P' / S$ \& $R Q' / S$ are the two sideways displacements.  The shortest h-radial distance is where $E'/E$ is maximum.  

 \centerline{\includegraphics[scale = 0.65]{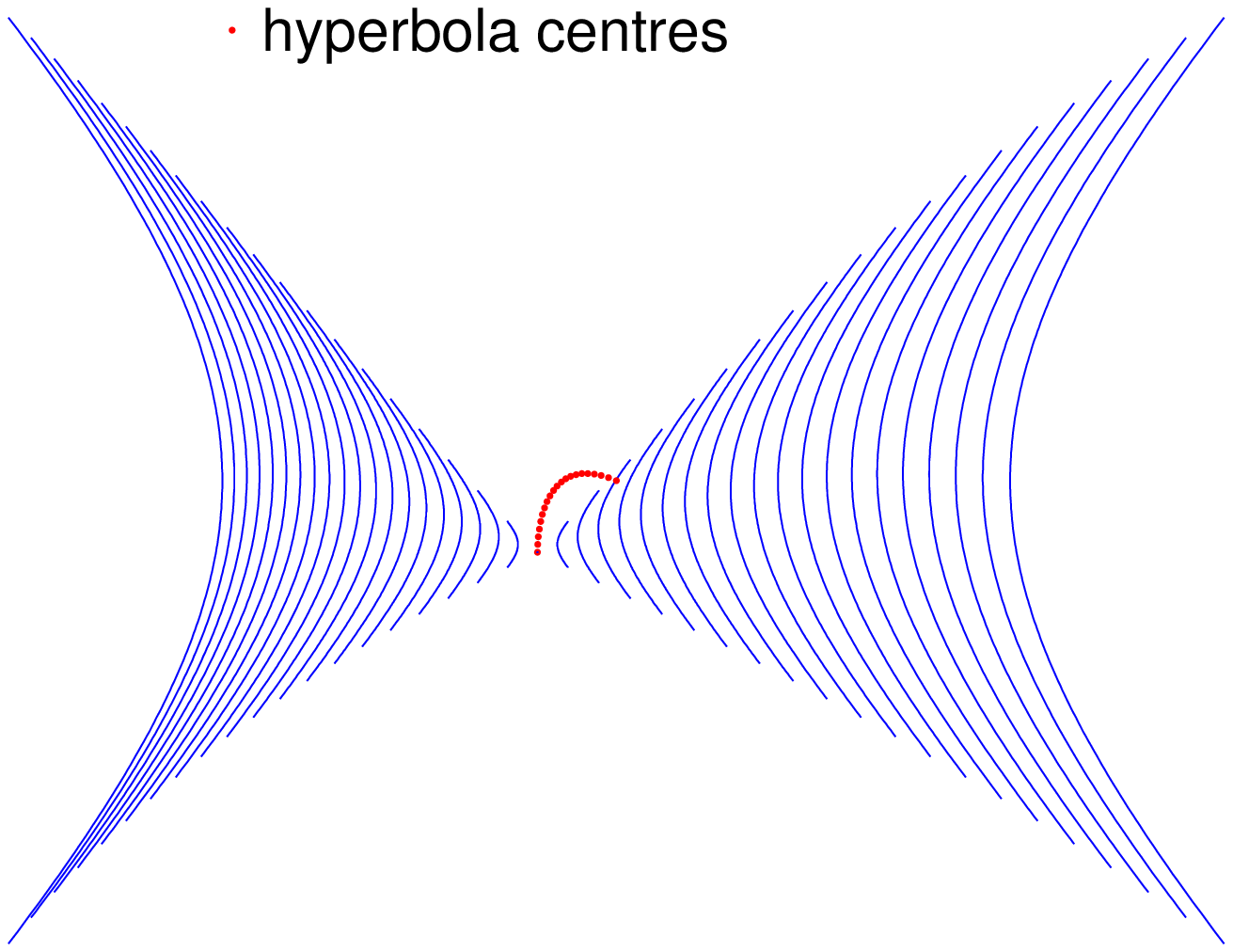}}

   \bp{Regularity}
   Can $\epsilon = -1$ regions have $R(t, r_o) = 0$ for some $r_o$?  The derivation of the `origin' conditions \er{OrigCondSz} does not depend on $\epsilon$, but when $\epsilon = -1$, $f \to 0$ is not allowed, since $f \geq 1$, so `origins' are not possible.

   The analysis of \cite{HelKra08} shows that, to be free of shell crossings, \er{maxE'/Eexists} must hold.  Even then only one sheet of the two-sheeted hyperboloid at each $r$ --- the one with $0 \geq {(E'/E)_\text{max}} \geq (E'/E) > -\infty$ --- can be free of shell crossings.  However on that sheet, the conditions are weaker than in LT --- see appendix \ref{NoShCr}.  As with the LT case, these are obtained by studying the evolution of $R'/R$.  An important conclusion is that only {\em one} sheet of the Riemann hyperboloid should be used to construct regular models, which means not all of the $p$-$q$ plane is used.

   Regular extrema, $R'(t, r_m) = 0)$, are indeed possible.  The calculations in \cite{HelKra08} lead to
 \begin{align}
   f = 1 ~,~~~~ \mbox{and}~~~~ M' = f' = a' = S' = P' = Q' = 0 ~,
 \end{align}
 but note that the conditions for no shell crossings are more subtle at such a locus.  This is not an obscure possibility --- the $k = -1$ RW metric in pseudo-spherical coordinates has a spatial minimum in $R$.

 \subsection{Quasi-Planar Case}

   \bp{No dipole}
   For the $\epsilon = 0$ case, we find
 \begin{align}
   E   & = \frac{2 S}{\theta^2} ~,
      \showlabel{E-thetap} \\
   E'  & = - \frac{2(S' + \theta (P' \cos \phi + Q' \sin \phi))}{\theta^2} ~,
      \showlabel{E'-thetaphip} \\
 %   E'' & = - \frac{2(S'' + \theta (P'' \cos \phi + Q'' \sin \phi))}{\theta^2} \nn \\
 %       & + 4 \left( \frac{S'}{S} \right) \left( \frac{S'
 %          + \theta (P' \cos \phi + Q' \sin \phi)}{\theta^2} \right) \nn \\
 %       & + \frac{(P')^2 - (Q')^2}{S}.
 %         \showlabel{E''-thetaphip}
   \frac{E'}{E} & = - \frac{S' + \theta (P' \cos \phi + Q' \sin \phi)}{S}
      \showlabel{ErE-thetaphip} ~,
 \end{align}
 and though the $E = 0$ locus has shrunk to the point $p = P$, $q = Q$, it still corresponds to the asymptotic regions of the plane, $\theta = \infty$.  The locus $E' = 0$,
 \begin{align}
   S' + P' \theta \cos \phi + Q' \theta \sin \phi = 0 ~,
      \showlabel{E'=0p}
 % \\   \tanh \theta = \frac{- S'}{P' \cos \phi + Q' \sin \phi} ~.
 \end{align}
 is obviously a geodesic of the $p$-$q$ 2-space, and it exists provided
 \begin{align}
   S' \neq 0 ~~~~\mbox{and}~~~~ (P' \neq 0 ~~\mbox{or}~~ Q' \neq 0) ~.
      \showlabel{pE'=0exists}
 \end{align}
 There are no extrema of $E'/E$, and it its value extends to both $\pm \infty$.  We interpret the above as showing that adjacent $r$-shells are planes tilted relative to each other, with $\tan\phi_0 = Q'/P'$ being the direction of maximum tilt, but if $E' = 0$ they are parallel.

   The behaviour found here cannot really be termed a dipole.

   \bp{Regularity}
   The question of whether an $\epsilon = 0$ model may have an `origin', $R \to 0$, is a little tricky.  The origin conditions \er{OrigCondSz} require $f \to 0$.  But if the metric and the 3-curvature is to be regular, we expect 
 \begin{align}
   \lim_{r \to r_o} g_{rr} = \lim_{r \to r_o}
   \frac{ \left\{ R' \left[ 1 - \frac{R E'}{R' E} \right] \right\}^2 }{f}
 \end{align}
 to be finite and non-zero.  Since $R E'/(R' E)$ is not divergent, this implies
 \begin{align}
   R' \sim \sqrt{f} \sim R ~~~~\Rightarrow~~~~
   R \sim e^{br} ~,~~~~ b~\mbox{constant,}
   \showlabel{R=0condp}
 \end{align}
 while the p-radial distance is
 \begin{align}
   s = \int \sqrt{g_{rr}}\; \, dr \sim r ~.
 \end{align}
 In other words, $R$, $M$ and $f$ may only approach zero asymptotically, and the scale of the planar foliations becomes ever smaller.  This is what happens in the planar foliation of the $k = -1$ RW metric.  

    Since the 3-spaces of a completely quasi-planar model consist of planes tilted relative to each other, they inevitably intersect somwhere, unless
 \begin{align}
   S' = P' = Q' = E' = 0 ~,   \showlabel{S'P'Q'E'Condp}
 \end{align}
 and the no-shell crossing conditions for $f \geq 0$ LT models also hold.  This reduces the model to planar symmetry --- an Ellis model \cite{Elli67}.

    As with the other foliations, regular extrema require no shell crossings and $f \to - \epsilon$, as given in appendix \ref{NoShCr}.  But, by \er{rpqRiemSz}-\er{rpqRicScSz} and \er{rhoSz}, $f \to 0$ also requires $R' - R E'/E \to 0$ and $M' - 3 M E'/E \to 0$, which is the Kantowski-Sachs type limit for this case.  Alternatively, $f \to 0$ also occurs in the origin requirement $R \to 0$ above, which can only be approached asymptotically.

    \bp{Quasi-Planar Limit}
    A S model may have both quasi-spherical and quasi-pseudo-spherical regions, and the boundary surface between them is a quasi-planar timelike 3-surface.  It was verified in \cite{HelKra08} that the $\epsilon = 0$ case and projection are suitable limits of both the $\epsilon = \pm 1$ cases.

 \subsection{Physical Discussion of the $\epsilon = -1,0$ Cases}

   \bp{Role of $R$}

   In the metric \er{ds2Sz} and in the area integral, $A = R^2 \int 1/E^2 \, dp \, dq$, the factor $R^2$ multiplies the unit sphere or pseudosphere, and therefore determines the magnitude of the curvature of the constant $(t,r)$ surfaces \er{pqCrvtrSz}.  By \er{rpqRiemSz}-\er{rpqRicScSz}, it is also a major factor in the curvature of the constant $t$ 3-spaces.  Therefore we view it as an ``areal factor'' or a ``curvature scale''. However, when $\epsilon \leq 0$, it is not at all like a spherical radius.  We note that when $\epsilon = -1$, there can be no origin, but $R$ can have maxima and minima as $r$ varies, while in the $\epsilon = 0$ case, $R$ cannot have extrema, and it can only approach zero asymptotically.

   \bp{Role of $M$}
   In \er{RtSq}, $M$ looks like a mass in the gravitational potential energy term of the evolution equation, while in \er{Rtt} $M$ determines the deceleration of $R$.  For $\epsilon = +1$, where the surfaces of constant $r$ are 
spheres enclosing a finite amount of matter, the function $M(r)$ does play the role of the gravitational mass contained within a comoving ``radius'' $r$.  For $\epsilon \leq 0$ however, $R$ is not the spherical radius that is an important part of these ideas in their original form, and $M$ is not a total gravitational mass, since the constant $t$ \& $r$ surfaces are not closed.  Consequently these ideas need revising.

   In fact, the impossibility of an ``origin'' or locus where $M$ and $R$ go to zero when $\epsilon = -1$ means that $M$ must have a global minimum, and indeed regular extrema in $R$ and $M$ are possible.  Therefore the local $M$ value is not independent of its value elsewhere, and integrals of the density over a region always have a boundary term, suggesting the value of $M$ (rather than its change between two shells) is more than can be associated with any finite part of the mass distribution.  

In $\epsilon = 0$ models, an asymptotic ``origin'' is possible, but not required, regular maxima are not possible, and regular ``minima'' are actually asymptotic origins.  So, with an asymptotic origin (as occurs in the planar foliation of $k = -1$ RW), the boundary term could be set to zero.

Nevertheless, the central roles of $R$ and $M$ are confirmed by the fact that the 3 types of Szekeres model can be joined smoothly to vacuum across a constant $r$ surface at which the values of $R$ and $M$ must match. The vacuum metric ``generated'' by the Szekeres dust distribution must have spherical, planar, or pseudospherical symmetry, and in each, $M$ is the sole parameter, while $R$ is an areal factor.

We note that, even in the Poisson equation, the gravitational potential does not need to be associated with a particular body of matter, and indeed it is not uniquely defined for a given density distribution.

Therefore we find that $M$ is a mass-like factor in the gravitational potential energy.

   \bp{Role of $f$}
   As shown in \S \ref{2,3Geom}, and as is apparent from the metric \er{ds2Sz}, the function $f$ determines sign of the curvature of the 3-space $t=$~const, as well as being a factor in its magnitude.  In the case, $\epsilon = +1$, this 3-space becomes flat (represented in unusual coordinates) when $f = 0$.  In the quasi-pseudospherical case, with $f = 0$ it becomes `flat' if the signature is made pseudoeuclidean, $(- + +)$.  In the quasi-planar case, $f = 0$ is possible as a Kantowski-Sachs type limit.  

   As with LT, $f$ appears in the gravitational energy equation \er{RtSq} as twice the total energy per unit mass of the matter particles, and we do not need to revise this interpretation.  Therefore, this variable has the same role as in quasi-spherical and spherically symmetric models.

   \bp{Role of $E$}
   As we have seen, for $\epsilon = +1$, $E'/E$ is the factor that determines the dipole nature of the constant $r$ shells, and for $\epsilon = -1$, it is the pseudospherical equivalent of a dipole, except that the two sheets of the hyperboloid each contain half the dipole, and only one of them can be free of shell crossings.  The shell separation (along the $r$ lines) decreases monotonically as $E'/E$ increases.  If $E' = 0$, it is uniform, otherwise it is minimum at some location and diverges outwards.  For $\epsilon = 0$, the effect of $E'/E$ is merely to tilt adjacent shells relative to each other, but only the zero tilt case ($E' = 0$) is free of shell crossings.

   \bp{Density Distribution}
   For $\epsilon = +1$ models, the density has a dipole variation around each constant $r$ sphere, though the strength an orientation of the dipole varies with $r$.  For $\epsilon = -1$ models, which must have $f \geq 1$, it was found that, if $f'/(2f) \geq M'/(3M)$ and there are no shell crossings, the density is at all times monotonically decreasing with $E'/E$, but asymptotically approaches a finite value as $E'/E$ diverges.  Therefore the density distribution on each shell is that of a void, but the void centres on successive shells can be at different $(p, q)$ or $(\theta, \phi)$ positions, in other words, the void has a snake-like or wiggly cylinder shape.  The minimum density is only zero if $M'/(3M) = -(E'/E)_\text{max}$.  Far from the void, at large $\theta$, the density is asymptotically uniform with $p$ \& $q$ (i.e. with $\phi$), but can vary with $r$.  However, where $f'/(2f) < M'/(3M)$ everywhere, an initial void in a constant $r$ shell can evolve into an overdensity.  

   The no shell crossings conditions imply limits on how far the location of the minimum density can be displaced between shells with different $r$.

 \subsection{Applications of the Szekeres Metric}

   The Szekeres metric was not used to model cosmological structures until very recently.

   In \cite{Bole06a}, models of voids next to superclusters were constructed, and it was found that the growth of the supercluster is strongly enhanced, relative to the linear perturbation approach.

   In \cite{Bole07}, it was found small voids surrounded by large overdensities evolve more slowly than isolated voids, while large voids enhance the evolution of adjacent superclusters.

   A swiss-cheese model based on Szekeres inhomogeneities was used in \cite{Bole08b} to investigate the effect of non-linear inhomogeneities on the CMB.  While compensated inhomogeneities have a tiny Rees-Sciama effect, the effect of uncompensated inhomogeneities is around $\sim 10^{-3}$ and so could be responsible for the low multipoles in the CMB.

   The effect of volume averaging was considered in \cite{Bole08c} and it was found the results are the same as in the LT case, c.f.\ \cite{Hell88}.

    A generalisation of the LT void models for SNIa dimming of \S\ref{NCOR} was given in \cite{IRGWNS08}.  They used a quasi-spherical S model with quite restricted arbitrary functions and few parameters.  This allowed some angular variation in $d_L(z)$.  It was shown that the model fits the data almost as well as the $\Lambda$CDM model, even though the potential of the S model has hardly been explored.

    Shell crossings in certain specific examples of higher dimensional quasi-spherical models were considered in \cite{ChaDeb08}.  These authors have also investigated generalised quasi-spherical models, including collapse and the occurrence of ``shell focussing'' naked singularities, often in higher dimensions and involving non-zero pressure or heat flux.

 \section{Conclusion}

    The universe is of course very inhomogeneous on many scales.  To fully understand how these structures evolve, and properly analyse our observations will require the non-linearity of exact inhomogeneous metrics.
    
    Up to now, homogeneity has been assumed, and was key to making progess.  In the age of precision cosmology, we should thoroughly test this assumption and quantify how good an approximation it is on each scale.  Nearly all data analysis assumes the RW metric.  To be sure we avoid circular arguments, there is an urgent need to re-do all calculations in a general non-homogeneous metric.  The methods of inhomogeneous cosmology will be an essential component of this endeavour.  

   \LT\ models have produced a wide variety of interesting results, and the investigations are far from exhausted.

   The Szekeres models have a lot of flexibility, and can be used to model quite complex structures --- but they have been very little investigated.

   There are plenty of opportunities for good research.

 \appendix

 \section{Near Bang and Near Parabolic Series}
 \showlabel{BangSeries}

    Near the bang, where $R = 0$, \er{RtSq} is dominated by $2M/R$, and the exact solutions \er{HypEv}-\er{tEvEC} for $\Lambda = 0$ involve the cancellation of nearly identical terms, thus generating large numerical errors.  Taking our cue from the $f = 0$, $\Lambda = 0$ solution, \er{ParEv}, i.e. $R = (9 M (t - a)^2/2)^{1/3}$, we write $R$ as a series,
 \begin{align}
   R & = \sum_{i = 1}^{\infty} R_i s^i ~,~~~~~~~~ s = \tau^{1/3} = (t - a)^{1/3} ~,
 \end{align}
 and put it into \er{RtSq} in the form
 \begin{align}
   3 R \Rt^2 = 6 M + 3 f R + \Lambda R^3 ~.
 \end{align}
 Solving for each power of $s$ in turn we find
 \begin{align}
   R & = R_2 s^2 \Bigg( 1 + V - \frac{3 V^2}{7} + \frac{23 V^3}{63} - \frac{U}{4}
      - \frac{1894 V^4}{4851} + \frac{V U}{11} + \frac{3293 V^5}{7007} + \frac{45 V^2 U}{2002}
      \cdots \Bigg) ~, \\
   & \mbox{where}~~~~~~~~ R_2 = \left( \frac{9 M}{2} \right)^{1/3} ~,~~~~~~ 
      V = \frac{9 f s^2}{20 R_2^2} = \frac{f s^2}{10} \left( \frac{9}{2 M^2} \right)^{1/3} ~,~~~~~~
      U = \frac{\Lambda s^6}{3} ~.
 \end{align}
 In the case $\Lambda = 0$, this is also the near-parabolic series for small $f$ (and $s$ not small).  When $\Lambda \neq 0$, and  \er{RtSq} is integrated numerically, small $f$ is not problematic.

   It is a good idea to have more terms in the series than the bare minimum, so that there is a range where both the series and numerical solutions are accurate, and each calculation provides a check on the coding of the other.

 \section{Near Origin Series for Observational Relations}
 \showlabel{OrigSeriesObs}

   When calculating $\th$, $\Rh$, $z$ and $\sgh$ for an LT model with given $f(r)$, $M(r)$ and $a(r)$, the origin, where all but $\th$ and $a$ go to zero, requires special numerical treatment.  Therefore it is useful to have a series expansion for the null cone quantities in the neighbourhood of the origin.  Writing 
 \begin{align}
   R = \sum_{i=1}^\infty \sum_{j=0}^\infty R_{ij} \, r^i \, \delta t^j ~,~~~~~~~~ \delta t = t - t_0 ~,
 \end{align}
 we can solve the evolution equation \er{RtSq} for the coefficients $R_{ij}$, when $j \neq 0$,
 \begin{align}
   R_{11} & = \sqrt{\frac{2 M_3}{R_{10}} + f_2 + \frac{\Lambda R_{10}^2}{3}}\; ~,~~~~
      R_{12} = - \frac{M_3}{2 R_{10}^2} + \frac{R_{10} \Lambda}{6} ~,~~~~ \nn \\
   R_{21} & = \left( \frac{M_4}{R_{10}} + \frac{f_3}{2} - \frac{M_3 R_{20}}{R_{10}^2} + \frac{R_{10} R_{20} \Lambda}{3} \right) \frac{1}{R_{11}} ~,~~~~
      R_{22} = - \frac{M_4}{2 R_{10}^2} + \frac{M_3 R_{20}}{R_{10}^3} + \frac{R_{20} \lambda}{6} ~,
 \end{align}
 but the $R_{i0}$, or equivalently the origin values of $R'$, $R''$, $R'''$, etc, must be found by numerical integration, e.g.\ using \er{RtRtr} \& \er{RrDE}.  Then using 
 \begin{align}
   M & = \sum_{i=3}^\infty M_i r^i ~,~~~~~~~~
      f = \sum_{i=2}^\infty f_i r^i ~,~~~~~~~~
      a = \sum_{i=0}^\infty a_i r^i ~,~~~~~~~~ \nn \\
   \th & = t_0 + \sum_{i=1}^\infty t_i r^i ~,~~~~~~~~
      z = \sum_{i=1}^\infty z_i r^i ~,~~~~~~~~
      \Rh = \sum_{i=1}^\infty \Rh_i r^i ~,~~~~~~~~
      \sgh = \sum_{i=2}^\infty K_i r^i ~,
 \end{align}
 and solving \er{dtdrNC}, \er{dz(1+z)}, \er{dRhdr} and \er{n-rho} power by power, leads to
 \begin{align}
   t_1 & = - R_{10} ~,~~~~
      t_2 = - R_{20} + \frac{R_{10} R_{11}}{2} ~,~~~~ \\
   z_1 & = R_{11} ~,~~~~
      z_2 = \frac{3 M_3}{2 R_{10}} + \frac{f_2}{2} + \left( \frac{M_4}{R_{10}} + \frac{f_3}{2} - \frac{M_3 R_{20}}{R_{10}^2} + \frac{R_{10} R_{20} \Lambda}{3} \right) \frac{1}{R_{11}} ~,~~~~ \\
   \Rh_1 & = R_{10} ~,~~~~
      \Rh_2 = R_{20} - R_{10} R_{11} ~,~~~~ \\
   K_2 & = \frac{3 M_3}{R_{11}} ~,~~~~
      K_3 = - \frac{3 M_3}{R_{11}^2} \left( \frac{3 M_3}{R_{10}} + f_2 \right) \nn \\
   &~~~~ + \left(  - 3 M_3 f_3 + 4 M_4 f_2 + \frac{2 M_3 M_4}{R_{10}} + \frac{6 M_3^2 R_{20}}{R_{10}^2} - 2 M_3 R_{10} R_{20} \Lambda + \frac{4 R_{10}^2 M_4 \Lambda}{3} \right) \frac{1}{R_{11}^3} ~.
 \end{align}

 \section{Near Origin Series for the Metric of the Cosmos}
 \showlabel{OrigSeriesMoC}

   Not only do $f(z)$, $M(z)$, $\sgh(z)$ and $\Rh(z)$ all go to zero at the origin, $\rh = 0$, $z = 0$, but more importantly we don't actually have any observational data at the origin.  Therefore, we fit a series solution to the data from the first few data bins.  We write the LT arbitrary functions as Taylor series in powers of $z$, 
 \begin{align}
   \Rh & = z \hat{\cal S}  = \sum_{i=1}^\infty R_i z^i ~,~~~~~~~~
      \sgh = \sum_{i=2}^\infty K_i z^i ~, \\
   \rh & = \sum_{i=1}^\infty r_i z^i ~,~~~~~~~~
      M = \sum_{i=3}^\infty M_i z^i ~,~~~~~~~~
      f = \sum_{i=2}^\infty f_i z^i ~.
 \end{align}
 The coefficients in the $\Rh$ and $\sgh$ series are determined by fitting polynomials to the observational data near the origin.  The null cone DEs, \er{phiDef}, \er{dphidz}, \er{dMdz}, and \er{WPNCz}, with $\beta = 1$, are then solved power by power, from which we find the coefficients of the $\rh$, $M$, and $f$ series.  The results of a Maple calculation are 
 \begin{align}
   \rh & = R_1 z
      + \left( R_2 + \frac{R_1}{2} \right) z^2
      + \left( R_3 + \frac{2 R_2}{3} + \frac{K_2}{6} \right) z^3
      + \left( R_4 + \frac{3 R_3}{4} + \frac{5 K_2}{24} + \frac{K_3}{12}
         + \frac{K_2 R_2}{6 R_1} \right) z^4 \nn \\
   &~~~~ + \left( R_5 + \frac{4 R_4}{5} + \frac{K_4}{20} + \frac{7 K_3}{60} + \frac{K_2}{15}
      + \left\{ K_3 + \frac{13 K_2}{5} \right\} \frac{R_2}{12 R_1}
      + \frac{K_2 R_3}{4 R_1} + \frac{K_2^2}{20 R_1}
      - \frac{K_2 {R_2}^2}{12 {R_1}^2} \right) z^5 \cdots ~, \\
   \frac{M}{z^3} & = \frac{K_2}{3} + \frac{K_3}{4} z
      + \left( \frac{K_4}{5}
      + \left\{ 1 - \frac{\Lambda R_1^2}{3} \right\} \frac{K_2}{10}
      - \frac{K_2^2}{15 R_1} \right) z^2 \nn \\
   &~~~~ + \left( \frac{K_5}{6} + \left\{ 1 - \frac{\Lambda R_1^2}{3} \right\} \frac{K_3}{12}
      - \left\{ \frac{1}{2} + \frac{\Lambda R_1^2}{6}
      + \frac{\Lambda R_1 R_2}{3}
      + \frac{7 K_3}{12 R_1} \right\} \frac{K_2}{6}
      + \left\{ \frac{1}{2} + \frac{R_2}{R_1} \right\} \frac{K_2^2}{18 R_1} \right) z^3 \cdots ~, \\
   \frac{f}{z^2} & = \left( 1 - \frac{\Lambda R_1^2}{3} - \frac{2 K_2}{3 R_1} \right)
      - \left( 1 + \frac{\Lambda R_1^2}{3} + \frac{2 \Lambda R_1 R_2}{3} - \frac{K_2}{3 R_1}
         + \frac{K_3}{2 R_1} - \frac{2 K_2 R_2}{3 R_1^2} \right) z \nn \\
   &~~~~ + \Bigg( \frac{5}{4} - \frac{\Lambda R_1^2}{6} - \frac{\Lambda R_2^2}{3}
      - \left[ 2 R_2 + 2 R_3 - \frac{K_2}{30} \right] \frac{\Lambda R_1}{3}
      + \frac{\Lambda^2 R_1^4}{36} \nn \\
   &~~~~~~~~~~ + \frac{K_3}{6 R_1} - \frac{11 K_2}{30 R_1} - \frac{2 K_4}{5 R_1}
      + \frac{2 K_2 R_3}{3 R_1^2} + \frac{K_3 R_2}{2 R_1^2} - \frac{2 K_2 R_2^2}{3 R_1^3}
      + \frac{29 K_2^2}{180 R_1^2} \Bigg) z^2 \nn \\
   &~~~~ + \Bigg( - \frac{3}{2}
      - \left\{ R_2 + 2 R_3 + 2 R_4 + \frac{3 K_2}{10} - \frac{K_3}{12} \right\} \frac{\Lambda R_1}{3}
      - \left\{ R_2 + 2 R_3 + \frac{K_2}{5} \right\} \frac{\Lambda R_2}{3} \nn \\
   &~~~~~~~~~~ + \left\{ \frac{R_1}{2} + R_2 \right\} \frac{\Lambda^2 R_1^3}{9}
      + \frac{19 K_2}{20 R_1} - \frac{K_3}{4 R_1} + \frac{K_4}{10 R_1} - \frac{K_5}{3 R_1} \nn \\
   &~~~~~~~~~~ + \left\{ \frac{11 K_2}{6} + 2 K_4 \right\} \frac{R_2}{5 R_1^2}
      + \left\{ \frac{K_2}{3} + K_3 \right\} \frac{R_3}{2 R_1^2}
      + \frac{2 K_2 R_4}{3 R_1^2} - \frac{K_2^2}{180 R_1^2} + \frac{2 K_2 K_3}{9 R_1^2} \nn \\
   &~~~~~~~~~~ - \left\{ \frac{K_2}{3} + K_3 \right\} \frac{R_2^2}{2 R_1^3}
      - \left\{ \frac{11 K_2^2}{15} + 4 K_2 R_3 \right\} \frac{R_2}{3 R_1^3} + \frac{2 K_2 R_2^3}{3 R_1^4}
      \Bigg) z^3 \cdots ~.
 \end{align}
 Then $a$ is found from a numerical integration of \er{RtSq}, using \er{tau}, in the form
 \begin{align}
   a = t_0 - \rh - \tau ~,~~~~
   \tau = \int_0^{\hat{\cal S}} \frac{\d {\cal S}}{\dot{\cal S}} ~,~~~\mbox{where}~~~
   \dot{\cal S}^2 = \frac{2 (M/z^3)}{\cal S} + (f/z^2) + \frac{\Lambda {\cal S}^2}{3} ~,~~~~
   {\cal S} = \frac{R}{z} ~.
 \end{align}

 The accuracy of the series is estimated from the ratio of the last and first terms.  If $\iota$ is the maximum acceptable error (comparable with expected numerical error), then the $z$ value where the program changes from series to numerical integration is given by
 \begin{align}
   \frac{M_6 z^6}{M_3 z^3} < \iota ~~~~\to ~~~~ z < \left( \frac{M_3 \iota}{M_6} \right)^{1/3} ~.
 \end{align}

 \section{The Near-Maximum Series on the PNC}
 \showlabel{MaxSeries}

   Near $z = z_m$, where the maximum $R_m = \Rh(z_m)$ occurs, we can solve the DEs of the PNC by writing the LT arbitrary functions as Taylor series in powers of $\Delta z = z - z_m$:
 \begin{align}
   \Rh & = R_m + \sum_{i=2}^\infty R_i \Delta z^i ~,~~~~~~~~
      \sgh = K_m + \sum_{i=1}^\infty K_i \Delta z^i ~,~~~~~~~~
      \rh = r_m + \sum_{i=1}^\infty r_i \Delta z^i ~, \\
   M & = M_m + \sum_{i=1}^\infty M_i \Delta z^i ~,~~~~~~~~
      \sqrt{1 + f}\; = W = W_m + \sum_{i=1}^\infty W_i \Delta z^i ~.
 \end{align}
 The coefficients of the series for $\rh$, $M$, and $W$ are obtained by substituting these series into the DEs \er{phiDef}, \er{dphidz}, \er{dMdz}, and \er{WPNCz}, again with $\beta = 1$, and the coefficients in the $\Rh$ and $\sgh$ series are found from polynomial fits to the observational data near the maximum in $\Rh$.  Using a Maple program, we find
 \begin{align}
   \varphi_0 = r_1 & = \frac{- 2 R_m R_2}{K_m} ~,   \showlabel{NMSphi0} \\
   \varphi_1 = r_2 & = \Bigg( \Bigg\{ \frac{K_1}{K_m} - \frac{1}{1 + z_m} \Bigg\} R_2 - 3 R_3 \Bigg)
      \frac{R_m}{K_m} ~, \\
   \varphi_2 = r_3 & = \Bigg( \Bigg\{ \frac{2 K_2}{3 K_m} - \frac{K_1^2}{2 K_m^2}
         + \frac{2 K_1}{3 K_m (1 + z_m)} + \frac{1}{2 (1 + z_m)^2} \Bigg\} R_2 \nn \\
      &~~~~ + \Bigg\{ \frac{K_1}{K_m} - \frac{1}{(1 + z_m)} \Bigg\} \frac{3 R_3}{2} - 4R_{4}
         - \frac{2 R_2^2}{3 R_m} \Bigg) \frac{R_m}{K_m} ~, \\
   \varphi_3 = r_4 & = \Bigg( \Bigg\{ \frac{K_3}{2 K_m} - \frac{2 K_1 K_2}{3 K_m^2}
         + \frac{K_1^3}{4 K_m^3} + \frac{K_2}{2 K_m (1 + z_m)} \nn \\
      &~~~~ - \frac{5 K_1^2}{12 K_m^2 (1 + z_m)} - \frac{K_1}{4 K_m (1 + z_m)^2}
         - \frac{1}{4 (1 + z_m)^3} \Bigg\} R_2 \nn \\
      &~~~~ + \Bigg\{ \frac{K_2}{K_m} - \frac{3 K_1^2}{4 K_m^2}
         + \frac{K_1}{K_m (1 + z_m)} + \frac{3}{4 (1 + z_m)^2} \Bigg\} R_3 \nn \\
      &~~~~ + \Bigg\{ \frac{K_1}{K_m} - \frac{1}{1 + z_m} \Bigg\} (2 R_4) - 5 R_5 \nn \\
      &~~~~ + \Bigg\{ \frac{K_1}{6 K_m} - \frac{1}{2 (1 + z_m)} \Bigg\} \frac{R_2^2}{R_m}
         - \frac{3 R_2 R_3}{2 R_m}
         \Bigg) \frac{R_m}{K_m} ~, \\
   \varphi_4 = r_5 & = \Bigg( \Bigg\{ \frac{2 K_4}{5 K_m} - \frac{K_1 K_3}{2 K_m^2}
         - \frac{2 K_2^2}{9 K_m^2} + \frac{K_2 K_1^2}{2 K_m^3} - \frac{K_1^4}{8 K_m^4}
         + \frac{2 K_3}{5 K_m (1 + z_m)} \nn \\
      &~~~~ - \frac{11 K_1 K_2}{18 K_m^2 (1 + z_m)}
         + \frac{K_1^3}{4 K_m^3 (1 + z_m)} + \frac{K_1^2}{9 K_m^2 (1 + z_m)^2} \nn \\
      &~~~~ - \frac{K_2}{6 K_m (1 + z_m)^2} + \frac{K_1}{12 K_m (1 + z_m)^3}
         + \frac{1}{8 (1 + z_m)^4} \Bigg\} R_2 \nn \\
      &~~~~ + \Bigg\{ \frac{3 K_3}{4 K_m} - \frac{K_1 K_2}{K_m^2} + \frac{3 K_1^3}{8 K_m^3}
         + \frac{3 K_2}{4 K_m (1 + z_m)} \nn \\
      &~~~~ - \frac{5 K_1^2}{8 K_m^2 (1 + z_m)} - \frac{3 K_1}{8 K_m (1 + z_m)^2}
         - \frac{3}{8 (1 + z_m)^3} \Bigg\} R_3 \nn \\
      &~~~~ + \Bigg\{ \frac{4 K_2}{3 K_m} - \frac{K_1^2}{K_m^2} + \frac{4 K_1}{3 K_m (1 + z_m)}
         + \frac{1}{(1 + z_m)^2} \Bigg\} R_4 \nn \\
      &~~~~ + \Bigg\{ \frac{K_1}{K_m} - \frac{1}{(1 + z_m)} \Bigg\} \frac{5 R_5}{2}
         - 6 R_6 \nn \\
      &~~~~ + \Bigg\{ \frac{2 K_2}{45 K_m} + \frac{19 K_1}{90 K_m (1 + z_m)}
         + \frac{1}{6 (1 + z_m)^2} \Bigg\} \frac{R_2^2}{R_m} \nn \\
      &~~~~ + \Bigg\{ \frac{7 K_1}{20 K_m} - \frac{23 }{20 (1 + z_m)} \Bigg\}
         \frac{R_2 R_3}{R_m} \nn \\
      &~~~~ - \frac{3 R_3^2}{4 R_m} - \frac{26 R_2 R_4}{15 R_m}
         + \frac{8 R_2^3}{45 R_m^2} \Bigg) \frac{R_m}{K_m} ~,
 \end{align}
 \begin{align}
   M_m & = \Bigg\{ 1 - \frac{\Lambda R_m^2}{3} \Bigg\} \frac{R_m}{2} ~, \\
   M_1 & = M_1 ~,~~~~~~~~~~~~~~ \mbox{i.e. undetermined} \\
   M_2 & = \Bigg\{ \frac{K_1}{K_m} + \frac{1}{1 + z_m} \Bigg\} \frac{M_1}{2} 
         - \frac{\lambda_m R_2}{2} - \frac{K_m^2}{2 R_m} ~, \\
   M_3 & = \Bigg\{ \frac{K_2}{K_m} + \frac{K_1}{K_m (1 + z_m)}
         - \frac{R_2}{R_m} \Bigg\} \frac{M_1}{3} \nn \\
      &~~~~ - \Bigg\{ \frac{K_1}{K_m} + \frac{1}{1 + z_m} \Bigg\} \frac{\lambda_m R_2}{4}
         - \frac{\lambda_m R_3}{4} - \frac{K_m K_1}{2 R_m} ~, \\
   M_4 & = \Bigg\{ \frac{K_3}{K_m} + \frac{K_2}{K_m (1 + z_m)} - \frac{K_1 R_2}{K_m R_m}
         - \frac{R_3}{R_m} - \frac{R_2}{R_m (1 + z_m)} \Bigg\} \frac{M_1}{4} \nn \\
      &~~~~ - \Bigg\{ \frac{5 \lambda_m K_1}{36 K_m (1 + z_m)} + \frac{2 \lambda_m K_2}{9 K_m}
         - \frac{\lambda_m K_1^2}{24 K_m^2} - \frac{K_m^2}{6 R_m^2}
         - \frac{\lambda_m}{24 (1 + z_m)^2} \Bigg\} R_2
         + \Bigg\{ 1 - \frac{\Lambda R_m^2}{4} \Bigg\} \frac{2 R_2^2}{9 R_m} \nn \\
      &~~~~ - \Bigg\{ \frac{K_1}{8 K_m} + \frac{1}{8 (1 + z_m)} \Bigg\} \lambda_m R_3
         - \frac{\lambda_m R_4}{6} - \frac{K_1^2}{8 R_m} - \frac{K_2 K_m}{3 R_m}
         - \frac{K_m^2}{24 R_m (1 + z_m)^2} ~;
 \end{align}
 where
 \begin{align}
   \lambda_m & = 1 - \Lambda R_m^2 ~,
 \end{align}
 and 
 \begin{align}
   W_m & = \frac{M_1}{K_m} ~, \\
   W_1 & = \frac{M_1}{K_m (1 + z_m)} - \frac{\lambda_m R_2}{K_m} - \frac{K_m}{R_m} ~, \\
   W_2 & = - \frac{R_2 M_1}{R_m K_m} + \Bigg\{ \frac{K_1}{4 K_m} - \frac{3}{4 (1 + z_m)}
         \Bigg\} \frac{\lambda_m R_2}{K_m} \nn \\
      &~~~~ - \frac{3 \lambda_m R_3}{4 K_m} - \frac{K_1}{2 R_m} ~, \\
   W_3 & = - \Bigg\{ R_3 + \frac{R_2}{(1 + z_m)} \Bigg\} \frac{M_1}{R_m K_m} \nn \\
      &~~~~ + \Bigg\{ \frac{\lambda_m K_2}{9 K_m} - \frac{\lambda_m K_1^2}{12 K_m^2}
         + \frac{2 K_m^2}{3 R_m^2} + \frac{7 \lambda_m K_1}{36 K_m (1 + z_m)} \nn \\
      &~~~~ + \frac{\lambda_m}{6 (1 + z_m)^2} \Bigg\} \frac{R_2}{K_m}
         + \frac{8 \lambda_m R_2^2}{9 R_m K_m} \nn \\
      &~~~~ + \Bigg\{ \frac{K_1}{4 K_m} - \frac{1}{2 (1 + z_m)} \Bigg\} \frac{\lambda_m R_3}{K_m}
         - \frac{2 \lambda_m R_4}{3 K_m} \nn \\
      &~~~~ - \frac{K_2}{3 R_m} - \frac{K_m}{6 R_m (1 + z_m)^2} ~.
 \end{align}
  These are generalisations to $\Lambda \neq 0$ of the results in \cite{McCHel08}, including a small correction in the expression for $M_4$.  Note that the coefficients of the $\rh(z)$ and $\varphi(z)$ series are fully determined from the data, but in the $M(z)$ and $W(z)$ series, the coefficient $M_1$ remains undetermined.  In other words, its value is fixed by data elsewhere, not by data at $R_m$.

 \section{Conditions for No Shell Crossings or Surface Layers}
 \showlabel{NoShCr}

 The following table presents the conditions that will ensure a model has no shell crossings or surface layers at any time in its evolution.  For $\epsilon = +1$, the first group of conditions are those that apply to the LT model, and the second group are the extra conditions needed in the S model.  The Ellis models \cite{Elli67} are the $\epsilon = 0, -1$, equivalent of LT models.  Although the no-shell-crossing conditions have not been explicitly studied for them, they can be deduced by setting $S$, $P$ \& $Q$ constant.

 \begin{center}
 \begin{tabular}{|l|l|l|l|l|}
 \hline
 \hline
 $\epsilon$ & $R'$  & $f$      & $M',~f',~a'$ & $S',~P',~Q'$ \\
 \hline
 \hline
 $+1$       & $> 0$ & all      & $M' \geq 0$  & \rule[-4mm]{0mm}{12mm}
      $\frac{\sqrt{(S')^2 + (P')^2 + (Q')^2}}{S} \leq \frac{M'}{3M}$ \\
 \cline{3-5}
            &       & $\geq 0$ & \rule[-7mm]{0mm}{17mm}
    \parbox{50mm}{$f' \geq 0$ \\ $a' \leq 0$ \\ but not all 3 equalities at once}
                                              &
       \parbox{45mm}{$\frac{\sqrt{(S')^2 + (P')^2 + (Q')^2}}{S} \leq \frac{f'}{2f}$ \\[0.5mm]
             (no condition where $f = 0$)} \\
 \cline{3-5}
            &       & $< 0$    & \rule[-7mm]{0mm}{17mm}
    \parbox{50mm}{$\tilde{T}' + a' \geq 0$ \\ $a' \leq 0$ \\ but not all 3 equalities at once}
                                              & \\
 \cline{2-5} \\[-4.5mm]
 \cline{2-5}
            & \parbox{15mm}{$= 0$ \\ $R'' > 0$ \\ neck}
                    & $-1$
                               & \rule[-10mm]{0mm}{23mm}
    \parbox{50mm}{$M' = 0$, $f' = 0$, $a' = 0$ \\ $f = -1$ for no surface layer \\
                  $\tilde{T}'' + a'' \geq 0$ \\ $a'' \leq 0$}
                                              &
       \parbox{45mm}{$S' = 0$, $P' = 0$, $Q' = 0$ \\[1mm]
                     $\frac{\sqrt{(S'')^2 + (P'')^2 + (Q'')^2}}{S} \leq \frac{M''}{3M}$} \\
 \cline{2-2}
 \cline{4-5}
            & \parbox{15mm}{$= 0$ \\ $R'' < 0$ \\ belly}
                    & 
                               & \rule[-10mm]{0mm}{23mm}
    \parbox{50mm}{$M' = 0$, $f' = 0$, $a' = 0$ \\ $f = -1$ for no surface layer \\
                  $\tilde{T}'' + a'' \leq 0$ \\ $a'' \geq 0$}
                                              &
       \parbox{45mm}{$S' = 0$, $P' = 0$, $Q' = 0$ \\[1mm]
                     $-\frac{\sqrt{(S'')^2 + (P'')^2 + (Q'')^2}}{S} \geq \frac{M''}{3M}$} \\
 \cline{2-5} \\[-4.5mm]
 \cline{2-5}
            & $< 0$ & all      & $M' \leq 0$  & \rule[-4mm]{0mm}{12mm}
      $-\frac{\sqrt{(S')^2 + (P')^2 + (Q')^2}}{S} \geq \frac{M'}{3M}$ \\
 \cline{3-5}
            &       & $\geq 0$ & \rule[-7mm]{0mm}{17mm}
    \parbox{50mm}{$f' \leq 0$ \\ $a' \geq 0$ \\ but not all 3 equalities at once}
                                              &
       \parbox{45mm}{$-\frac{\sqrt{(S')^2 + (P')^2 + (Q')^2}}{S} \geq \frac{f'}{2f}$ \\[0.5mm]
             (no condition where $f = 0$)} \\
 \cline{3-5}
            &       & $< 0$    & \rule[-7mm]{0mm}{17mm}
    \parbox{50mm}{$\tilde{T}' + a' \leq 0$ \\ $a' \geq 0$ \\ but not all 3 equalities at once}
                                              & \\
 \hline
 \hline
 \end{tabular}

 \begin{tabular}{|l|l|l|l|l|}
 \hline
 \hline
 $\epsilon$ & $R'$ & $f$ & $S'$ & $M'$~,~~$f'$~,~~$a'$~,~~$P'$~,~~$Q'$ \\
 \hline
 \hline
   $= -1$ & $> 0$ & $\geq 1$ & $E S' >0$ &
    \parbox{50mm}{${}$ \\[0.5mm]
                  $(S')^2 > (P')^2 + (Q')^2$ \\[1mm]
                  $\frac{M'}{3 M} \geq - \frac{\sqrt{(S')^2 - (P')^2 - (Q')^2}\;}{S}$ \\[1mm]
                  $\frac{f'}{2 f} \geq - \frac{\sqrt{(S')^2 - (P')^2 - (Q')^2}\;}{S}$ \\[1mm]
                  $a' \leq 0$} \\[14mm]
 \cline{2-5}
   & $= 0$ & $= 1$    & $S' = 0$ &
    \parbox{50mm}{${}$ \\[0.5mm]
                  $M' = 0$~,~~ $f' = 0$~,~~ $a' = 0$~, \\
                  $P' = 0$~,~~ $Q' = 0$} \\[5mm]
 \cline{2-5}
   & $< 0$ & $\geq 1$ & $E S' < 0$ &
    \parbox{50mm}{${}$ \\[0.5mm]
                  $(S')^2 > (P')^2 + (Q')^2$ \\[1mm]
                  $\frac{M'}{3 M} \leq + \frac{\sqrt{(S')^2 - (P')^2 - (Q')^2}\;}{S}$ \\[1mm]
                  $\frac{f'}{2 f} \leq + \frac{\sqrt{(S')^2 - (P')^2 - (Q')^2}\;}{S}$ \\[1mm]
                  $a' \geq 0$} \\[14mm]
 \hline \hline
   $= 0$ & $> 0$ & $\geq 0$ & $= 0$ &
    \parbox{50mm}{${}$ \\[1mm]
                  $M' \geq 0$~,~~ $f' \geq 0$~,~~ $a' \leq 0$~, \\[1mm]
                  $P' = 0$~,~~ $Q' = 0$} \\[5mm]
 \cline{2-5}
   & $= 0$ & $= 0$    & $= 0$ &
    \parbox{50mm}{${}$ \\[0.5mm]
                  $M' = 0$~,~~ $f' = 0$~,~~ $a' = 0$~, \\
                  $P' = 0$~,~~ $Q' = 0$} \\[5mm]
 \cline{2-5}
   & $< 0$ & $\geq 0$ & $= 0$ &
    \parbox{50mm}{${}$ \\[0.5mm]
                  $M' \leq 0$~,~~ $f' \leq 0$~,~~ $a' \geq 0$~, \\[1mm]
                  $P' = 0$~,~~ $Q' = 0$} \\[5mm]
 \hline
 \end{tabular}
 \end{center}

 \end{document}